\documentclass[english]{JHEP3}
\usepackage{graphicx}
\usepackage{amssymb,amsmath,cite}
\usepackage[mathscr]{euscript}
\newcommand{\HWPP}{\textsf{Herwig++}}

\title{A modified CKKW matrix element merging approach to angular-ordered parton showers}
\author{Keith Hamilton \\ Centre for Particle Physics and Phenomenology, \\ Universit\'{e} Catholique de Louvain, \\ Chemin du Cyclotron 2, 1348 Louvain-la-Neuve, Belgium. \\ Email: \email{keith.hamilton@uclouvain.be}}
\author{Peter Richardson\\
   Institute of Particle Physics Phenomenology, Department of Physics, \\ University of Durham,  Durham, DH1 3LE, UK. \\ Email: \email{peter.richardson@durham.ac.uk}}

\author{Jon Tully \\ Institute of Particle Physics Phenomenology, Department of Physics, \\ University of Durham,  Durham, DH1 3LE, UK. \\ Email: \email{j.m.tully@durham.ac.uk}} 

\preprint{ HERWIG/09/03 \\ MCnet/09/09 \\ IPPP/09/41 \\ DCPT/09/82 \\ CP3-09-09 }
\maketitle

\abstract{A modified version of the CKKW matrix element merging algorithm is presented, suitable
for use in an angular-ordered parton shower, using truncated showers and forced 
splittings.  The algorithm is implemented in the \HWPP\ Monte Carlo 
event generator for the benchmark process 
\mbox{$e^{+}e^{-} \to \mathrm{hadrons}$}.   
Results are presented at parton and hadron levels,
demonstrating a smooth merging between the matrix elements and parton shower and an improved description of LEP data.}


\preprint{}

\begin{document}

\section{Introduction}
\label{intro}
Monte Carlo event generators have been successful in providing a full simulation of the physics at collider
experiments and have proven to be invaluable tools in both planning future 
experiments and analysing data from current experiments.
They use a flexible and convenient event-by-event description which allows a range of physics models to be implemented
and as such they describe a wide variety of phenomena.
In particular, they provide a means, via the parton shower, of evolving from hard scales, where partons 
are produced in fixed-order perturbation theory, to soft scales where non-perturbative models must be applied.

Besides the simple fixed-order matrix elements which describe the hard processes, the key perturbative element of 
Monte Carlo event generators is the parton shower.  In this phase of the simulation, partons 
undergo DGLAP evolution from some initial scale, characteristic of the underlying hard scattering, 
down to the hadronization scale.  
In \textsf{Herwig++}\cite{Hw_release,HerwigMan} the evolution variable of the parton shower 
is chosen such that branchings are ordered in their opening angles, accounting for the effects of 
QCD coherence\cite{Corcella:2000bw}. 
This corresponds to resumming next-to-leading logarithm (NLL) terms to all orders in perturbation theory.

While the parton shower accurately simulates soft and collinear radiation, it does not provide a 
reliable description of hard (high transverse momentum) emissions.  
In fact, requiring that the regions of phase space into which each \emph{shower progenitor} 
(the partons which initiate the shower) can emit do not overlap, 
to avoid double counting, gives rise to a \emph{dead zone}: 
a region of phase space corresponding to high transverse momentum, wide angle, emissions which none of the 
showering particles can emit into. Even within the accessible \emph{shower regions} of phase
space, the distribution of radiation involves some degree of approximation, since at any given fixed 
order in perturbation theory, the parton shower effectively approximates the real emission 
corrections to the hard scattering process by a product of splitting functions and Sudakov form factors, 
summed over all combinations of branchings which give rise to the same final state. 
These approximations account for the NLL corrections associated 
with soft and collinear radiation in the perturbative series. 

In order to improve the description in the large transverse momentum region, the parton shower can be
combined with exact fixed-order matrix elements. 
The earliest and simplest means of forming this combination is known as
the matrix element correction method\cite{Seymour:1994df}. This corrects
the hardest emission generated by the parton shower such that it is distributed
according to the real single emission matrix element squared. This technique 
has been successfully applied to important processes
in a number of generators\cite{Norrbin:2000zc,Corcella:1998rs}, including \HWPP\cite{HerwigMan}.

In recent years more general matrix element-parton shower merging 
algorithms have been introduced. These combine tree-level matrix elements with 
parton showers, for a given process, for all parton multiplicities below some 
maximum $N$. Hence these algorithms correct all distributions involving up to 
$N$ external partons, instead of just that of the hardest emission. Several 
schemes of this type have been developed and successfully implemented in event 
generators. The most well known of these are the CKKW\cite{CKKW,CKKW_had,CKKW_imp}, 
CKKW-L\cite{CKKWL}, MLM\cite{MLM} and pseudo-shower\cite{RichMrenna} methods. 
All these methods have the same general approach\cite{MergeComp, nils_leif} 
whereby the phase space for parton emissions is divided into two regions by a merging scale 
$y_{_{MS}}$, defined in some jet measure.  Above the merging scale, emissions are described by
exact matrix elements while below it emissions are produced by the parton shower.

In this article we present a matrix element merging scheme based on 
the CKKW algorithm. A fundamental ingredient in the CKKW method is the 
association of a \emph{pseudo-shower history} to the configurations 
generated according to the fixed-order matrix elements. Each shower history is 
constructed by clustering the two most closely separated partons,
according to the transverse momentum measure defining the merging scale, 
until a leading-order parton configuration is obtained.
The resulting branchings in the shower history 
are therefore ordered according to the jet measure, which may not 
equate to the ordering variable of the parton shower, as is the case 
for the angular-ordered parton shower of \textsf{Herwig++}.
This discrepancy is understood to give rise to serious problems, in particular
it spoils the colour coherence properties of these shower algorithms. 
Although this was already noted, and 
an attempt made to address it, in the original CKKW paper, realisations 
of the method highlight the fact that the colour structure in the events is 
nevertheless in conflict with that expected on the grounds of colour coherence, 
moreover, they show that this is not simply an esoteric consideration but a 
cause of significant practical problems, including a dependence on the 
unphysical merging scale \cite{nils_leif,RichMrenna}. 

We shall present and validate a modified version of the CKKW method, intended to 
optimise the implementation of these colour coherence effects, by a fully 
consistent merging of an angular-ordered parton shower with fixed-order matrix 
elements. The idea behind this method was originally proposed by 
Nason in Ref.~\cite{PowhegOriginal}. The central result of that theoretical work 
is the observation that the parton shower may be formally decomposed in terms of 
\emph{truncated showers}, \emph{hard emissions}, and \emph{vetoed showers}. 
Reference~\cite{PowhegOriginal} advocates that the CKKW algorithm may then best 
model the coherent emission of radiation by including these truncated showers, 
consisting of only soft emissions, prior to, and between, the hard emissions 
in the shower history, thereby rendering it angular-ordered. 
In the following we will develop the full details necessary for our practical 
implementation of this idea for the process \mbox{$e^{+}e^{-} \to \mathrm{hadrons}$} 
and compare the results of it to LEP data.

This work builds on the infra-structure introduced in Refs.~\cite{HRT_drell_yan, HRT_higgs} 
for the \textsf{Herwig++} angular-ordered parton shower, where the truncated 
shower was initially developed for use in the POWHEG\cite{PowhegOriginal, PowhegMan, Frixione:2007vw} 
next-to-leading order matching scheme applied to Drell-Yan vector boson production. This was the 
first time a full implementation of the truncated shower had been 
developed\footnote{An approximate, single emission truncated shower is also described in 
Refs.~\cite{LatundeDada:2006gx,LatundeDada:2008bv}.}.

While this work was in production, Ref.~\cite{Hoeche:2009rj} appeared on the arXiv,
employing similar techniques to those described in the present work
for matrix element merging with a transverse-momentum-ordered dipole shower.

The paper is set out as follows. In Sect.~\ref{CKKW_merging} we review the 
original CKKW merging prescription. In Sect.~\ref{Shower_reorganisation} 
we go on to describe the way in which the angular-ordered parton shower may be 
decomposed into hard emissions, truncated showers, and vetoed showers. Having 
introduced the relevant conceptual ingredients we then give a more detailed 
technical description of our modified CKKW algorithm in Sect.~\ref{algorithm}. 
In Sect.~\ref{results} we present a validation of our algorithm by comparing to
LEP data for \mbox{$e^{+}e^{-} \to \mathrm{hadrons}$}, before giving our 
conclusions in Sect.~\ref{conclusion}.

\section{CKKW merging}
\label{CKKW_merging}

In this section we present an overview and discussion of the original CKKW algorithm
for the process \mbox{$e^{+}e^{-}\to \mathrm{hadrons}$}.  We first describe the algorithm for 
the case where the parton shower evolution variable is identical to the merging variable before
describing the adaptations which must be made for the \HWPP\ angular-ordered parton shower.

\subsection{Transverse-momentum-ordered CKKW merging}
\label{transCKKW}

The algorithm is simplest if the merging variable is the same as the ordering variable 
of the parton shower.  We therefore consider first the case where we have a single transverse 
momentum variable $q$ as the parton shower evolution and merging variables.

The basic principle underlying the CKKW approach, is that the distribution of 
radiation in the region of phase space where all partons are separated by an 
amount $q$ greater than the \emph{merging scale} $q_{MS}$, should be given by 
tree-level matrix elements, while for $q\le q_{MS}$ it should be given by the 
parton shower. The algorithm then requires, as input, samples of events of the 
process with up to $N$ partons in the final state. These input samples are easily obtained using fully 
automated tree-level event generators such as \textsf{Madgraph}/\textsf{MadEvent} 
\cite{Maltoni:2002qb,Alwall:2007st}\footnote{Currently, computational efficiency
limits the total number of final-state particles to around six.}. As 
well as producing the events themselves, for each sample with $n$ partons the 
generator will provide a finite, tree-level, jet cross section 
$\sigma_{n}^{(ME)}\left(q_{_{MS}}\right)$. 

Na\"{\i}vely, with the input events in hand, one might then consider filling the
remaining phase space by selecting events from each sample with $n$ partons, 
with a probability proportional to $\sigma_{n}^{(ME)}$, and simply invoking 
the parton shower on each of the external legs, starting from the scale $q_{_{MS}}$. 
However, the merging scale $q_{_{MS}}$ is not a physical parameter and so all 
distributions of partons should be insensitive to its value. This would certainly 
not be the case for such a na\"{\i}ve procedure, since the distribution of 
radiation from the parton shower and the fixed-order matrix elements are known 
to differ, especially in the regions corresponding to high and low $q$ emissions. 
The great success of the CKKW algorithm is in its ability to correct for the 
mismatch at the phase space partition $q_{_{MS}}$ by providing a smooth, physical, 
interpolation between the matrix element distribution at high $q$ values and 
that of the parton shower in the low $q$ region. 

To illustrate how this works, consider the simplified case of merging only 
samples of 2- and 3-parton events, with $q\ge q_{_{MS}}$, for 
\mbox{$e^{+}e^{-} \to \mathrm{hadrons}$}, with a $q$-ordered parton shower. 
In general, the parton shower cross section analogous to $\sigma_{n}^{(ME)}\left(q_{_{MS}}\right)$, 
with $n$ partons resolved at the merging scale, may be written as the product 
of the leading-order cross section together with a set of Sudakov form factors 
and splitting functions. The product of these splitting functions and the 
leading-order cross section approximate the exact tree-level $n$-jet cross 
section. For the case of three partons this cross section is
\begin{equation}
\label{R3_PS_simple}
  \sigma_{3}^{(PS)}\left(q_{_{MS}}\right)= \sigma_{2} \times 
  2\left[\Delta_{q}\left(q_{I},q_{_{MS}}\right)\right]^{2}
  \int^{q_{I}}_{q_{_{MS}}} \mathrm{d}q \,
  \alpha_{S}\left(q\right)\,\Gamma_{q\to qg}\, \left(q\right)\Delta_{g}(q,q_{_{MS}}),
\end{equation}
where $q_{I}$ is the scale at which the parton shower is initiated and 
$\alpha_{S}(q)\Gamma_{\widetilde{ij} \to ij}(q)$ 
is the probability for a parent parton $\widetilde{ij}$ to branch into two 
daughter partons $i$ and $j$, in the interval $\left[q,q+dq\right]$ \footnote{
The dependence on auxiliary splitting variables has been suppressed.}. The 
overall normalisation factor $\sigma_{2}$ is simply the leading-order cross 
section. Finally, in Eq.~(\ref{R3_PS_simple}), $\Delta_{\widetilde{ij}}(q,q_{_{MS}})$ is 
the Sudakov form factor, which can be interpreted as the probability for the 
parent parton $\widetilde{ij}$ to evolve from a scale $q$ down to the scale 
$q_{_{MS}}$ without undergoing a resolvable branching,
\begin{equation}
\label{simpleSudDef}
 \Delta_{\widetilde{ij}}(q,q_{_{MS}})=\exp\left[ -\sum_{\widetilde{ij} \to ij} 
    \int^{q}_{q_{_{MS}}}\mathrm{d}q^{\prime} \,
    \alpha_{S}\left(q^{\prime}\right)
    \Gamma_{\widetilde{ij} \to ij}\left(q^{\prime}\right)
    \right].
\end{equation} 

In parton shower (NLL) expansions of the jet cross sections, such as that in Eq.~(\ref{R3_PS_simple}), the exact
tree-level matrix elements are approximated by the product of the leading-order cross section and
splitting functions.  In order to improve the parton shower with exact tree-level matrix elements,
this product should be replaced by the corresponding exact, tree-level jet cross section.

The CKKW merging should not affect the NLL expansion of the jet cross section therefore the 
NLL expansion of the matrix element contribution should give the result in Eq.~(\ref{R3_PS_simple}).
Since a NLL expansion of the tree-level matrix elements yields a corresponding product 
of parton shower splitting functions, it is clear that in order to retain the NLL form of
Eq.~(\ref{R3_PS_simple}), the matrix element contribution above $q_{_{MS}}$ should be given
by configurations generated according to the tree-level jet cross sections reweighted by
appropriate Sudakov and running $\alpha_{S}$ factors.

In order to determine appropriate reweighting factors for events from the tree-level generator, a
pseudo-shower history must be assigned to each event.  This shower history interprets the set of external
parton momenta as a set of branchings originating from a leading-order configuration.  This procedure gives rise to
 a set of \emph{nodal values}, $q_{i}$, for the scales at which each pseudo-branching
occurred. These scales provide the arguments for the Sudakov form factors and $\alpha_{S}$
factors with which the configuration should be reweighted.  In the original CKKW publication,
this pseudo-shower history is assigned by repeatedly clustering the pair of 
partons\footnote{Only pairs of partons whose flavours correspond to allowed branchings are considered.}
with the smallest separation according to the jet resolution variable, until only the particles of the leading-order process remain.

In the case being considered, where the evolution variable has been taken to match the merging scale variable,
combining the matrix elements with the parton shower is straightforward.  The parton shower evolution can 
be split into two parts: first an evolution from the initial scale down to the merging scale $q_{_{MS}}$; then an evolution
from the merging scale down to the hadronization scale $q_{0}$.  This results in a simple procedure for attaching
the parton shower to the reweighted matrix elements, where each external parton produces a shower line 
evolving from the merging scale.

The full CKKW algorithm then proceeds as follows:
\begin{enumerate}
\item a jet multiplicity $n$ is generated with probability
  \begin{equation}
    P_{n}=\frac{\sigma_{n}^{(ME)}(q_{_{MS}})}{\sum_{N}\sigma_{i}^{(ME)}(q_{_{MS} })},
  \end{equation}
  where all cross sections are evaluated at a fixed strong coupling $\alpha_{S_{ME}}$;
\item  \label{step_ME}a configuration of $n$ parton momenta is generated 
  according to $\mathrm{d}\sigma_{n}^{(ME)}(q_{_{MS}})$;
\item \label{step_cluster}external partons are clustered, defining a pseudo-shower history with 
  a set of nodal scales $q_{i}$;
\item  \label{step_reweight}the configuration is reweighted by Sudakov and $\alpha_{S}$ factors: 
  each internal line between two nodes at
  $q_{i}$ and $q_{i+1}$ contributes a factor of $\Delta_{f}(q_{i},q_{_{MS}}) / \Delta_{f}(q_{i+1},q_{_{MS}})$, each external line emanating from a node with scale $q_{i}$ contributes  $\Delta_{f}(q_{i},q_{_{MS}})$, while each node itself contributes
  $\frac{\alpha_{S}(q_{i})}{\alpha_{S_{ME}}}$;
\item  \label{step_invoke}the parton shower is invoked on each external parton from a starting scale of $q_{_{MS}}$.
\end{enumerate}

This scheme is independent of the merging scale to NLL order\cite{CKKW}.  We have reweighted configurations
such that the NLL three-jet cross section resolved at the merging scale is given by Eq.~(\ref{R3_PS_simple}).
This NLL cancellation of merging scale dependence can be seen by considering the cross section for three jets 
resolved at the hadronization scale.  This cross section is given by the sum of the probability of generating
a single emission in the matrix element region and none in the parton shower, together with the probability of generating
no emissions in the matrix element region and a single emission in the parton shower region.
The cross section is
\begin{align}
\label{R3_PS_merge}
  \sigma_{3}^{(PS+ME)}\left(q_{0}\right) & = 
  \bar{\sigma}_{3}^{(ME)}\left(q_{_{MS}}\right)
  \left[\Delta_{q}\left(q_{_{MS}},q_{0}\right)
    \right]^{2}\Delta_{g}\left(q_{_{MS}},q_{0}\right) \\
  & +\sigma_{2} \times 
  2\left[\Delta_{q}\left(q_{I},q_{0}\right)\right]^{2}
  \int^{q_{_{MS}}}_{q_{0}} \mathrm{d}q \,
  \alpha_{S}(q) \, \Gamma_{q \to qg}( q ) \, \Delta_{g}(q,q_{0}) \notag ,
\end{align}
where $\bar{\sigma}_{3}^{(ME)}\left(q_{_{MS}}\right)$ is the reweighted matrix element contribution for three jets
resolved at the merging scale. The first term in Eq.~(\ref{R3_PS_merge}) corresponds to a single emission 
above the merging scale followed by parton shower evolution from the merging scale down to the hadronization
scale with no resolvable emissions.  The second term corresponds to no emissions above the
merging scale followed by a single parton shower emission below the merging scale.  In the NLL expansion 
of Eq.~(\ref{R3_PS_merge}), we replace $\bar{\sigma}_{3}^{(ME)}\left(q_{_{MS}}\right)$ by the NLL parton shower approximation in Eq.~(\ref{R3_PS_simple}).
This results in a simplification of Eq.~(\ref{R3_PS_merge})
\begin{equation}
\label{R3_PS_cancel}
  \sigma_{3}\left(q_{0}\right)= \sigma_{2} \times 
  2\left[\Delta_{q}\left(q_{I},q_{0}\right)\right]^{2}
  \int^{q_{I}}_{q_{0}} \mathrm{d}q \,
  \alpha_{S}(q) \, \Gamma_{q \to qg}( q ) \, \Delta_{g}(q,q_{0}),
\end{equation}
yielding the expected NLL parton shower cross section for a single resolved emission which is independent
of the merging scale.

\subsection{Angular-ordered CKKW merging}
\label{standard_ang_CKKW}

The merging variable used to define the jet cross sections must regulate both
soft and collinear singularities, so it must be a transverse momentum measure.
The merging variable in the original CKKW publication is defined in terms of the Durham 
jet measure\cite{Durham} for two partons $i$ and $j$,
\begin{equation}
  y_{\mathrm{dur}_{ij}}=\frac{2\mathrm{min}\left(E_{i}^{2},E_{j}^{2}\right)}
  {s}\left(1-\cos{\theta_{ij}}\right),
\end{equation} 
where $E_{i,j}$ are the energies of the two partons, $\theta_{ij}$ is the angle between
the two partons and $s$ is the centre-of-mass-energy squared.
The merging transverse momentum variable is defined by 
\begin{equation}
  \label{simpleScale}
  k_{\perp}=\sqrt{y_{ij}s}.
\end{equation}

The parton shower with which we wish to merge the matrix elements may not be ordered in transverse momentum,
in which case the merging variable cannot be chosen to be the same as the evolution variable, as was assumed in 
Sect.~\ref{transCKKW}.

In the \HWPP\ parton shower, splittings are described by the variables $(\tilde{q}, z, \phi)$, where 
$\tilde{q}$ is an angular-ordered evolution variable, $z$ is the momentum fraction
of the emitted parton and $\phi$ is the azimuthal angle of the branching.
The evolution variable is defined by
\begin{equation}
\label{hw_scale}
  z^{2}(1-z)^{2}\tilde{q}^{2}=
  p_{\perp}^2+(1-z)m_{i}^{2}+zm_{j}^{2}-z(1-z)m_{\widetilde{ij}}^{2},
\end{equation}
where $p_{\perp}$ is the relative transverse momentum of the branching and $m_{\widetilde{ij}}$ and $m_{i,j}$ 
are the masses of the parent and child partons respectively.

The probability for a branching $\widetilde{ij}\to ij$ in the phase space 
measure $\tilde{q} \to \tilde{q} +\mathrm{d}\tilde{q}$ is given by
\begin{equation}
   \mathrm{d}\mathcal{P}_{\widetilde{ij}\to ij}(\tilde{q},z) =
  \frac{\alpha_{S}(p_{\perp})}{2\pi} \,
   \frac{\mathrm{d} \tilde{q}^{2}}{\tilde{q}^{2}} \,
   \mathrm{d}z \, P_{\widetilde{ij}\to ij}(z,\tilde{q}),
\end{equation}
where $P_{\widetilde{ij}\to ij}(z,\tilde{q})$ is the corresponding quasi-collinear 
splitting function\cite{Catani:2000ef}.
The Sudakov form factor, giving the probability of a parton shower
line of flavour $\widetilde{ij}$ evolving from a scale $\tilde{q}_{1}$ down to $\tilde{q}_{2}$ undergoing no resolvable
emissions, is given by
\begin{equation}
\Delta_{\widetilde{ij}}(\tilde{q}_{1},\tilde{q}_{2})=
\exp \left[-\sum_{\widetilde{ij}\to ij} \int^{\tilde{q}_{1}}_{\tilde{q}_{2}} 
  \mathrm{d}\mathcal{P}_{\widetilde{ij}\to ij}(\tilde{q},z) \right].
\end{equation} 

In order to accommodate the fact that the evolution and merging variables are not identical, 
the CKKW algorithm must include some additional features to that outlined
in Sect.~\ref{transCKKW}. Changes must be made to the Sudakov form factors with which the
matrix elements are reweighted and the initial conditions with which the parton shower is invoked,
furthermore, when the shower is invoked, a veto must be applied to prevent it
generating emissions with $k_{\perp}(\tilde{q},z)>k_{\perp_{MS}}$.

The Sudakov form factor used for the matrix element reweighting, 
defined in Eq.~(\ref{simpleSudDef}), corresponds to the probability of evolving from
a scale $q$ down to the hadronization scale with no emissions resolvable at the merging scale.  
In the case of Sect.~\ref{transCKKW}, this was achieved by setting the lower limit on
the integral to $q_{_{MS}}$, however, now this cut, defining what is meant by a \textit{resolvable emission},
must be implemented as a $\theta$-function in the Sudakov form factors used in step \ref{step_reweight}. 
The Sudakov form factors for the reweighting are then given by
\begin{equation}
\label{sud_w_theta}
\Delta_{\widetilde{ij}}^{R}(\tilde{q};k_{\perp_{MS}})=
\exp \left[-\sum_{\widetilde{ij}\to ij} \int^{\tilde{q}}_{\tilde{q}_{2}}
  \mathrm{d}\mathcal{P}_{\widetilde{ij}\to ij}(\tilde{q}^{\prime},z)
  \theta\left(k_{\perp}(\tilde{q}^{\prime},z)-k_{\perp_{MS}}\right)
 \right].
\end{equation} 
The prescription for constructing the Sudakov weights is then identical
to that in Sect.~\ref{transCKKW} except for factors of $z$ in the scale from which each child 
evolves, which are required for the angular-ordered evolution.
Each intermediate line, connecting branchings at $(\tilde{q}_{1},z_{1})$
and $(\tilde{q}_{2},z_{2})$ in the pseudo-shower history, contributes a factor
\begin{equation}
  \Delta_{\widetilde{ij}}^{R}(z_{1}\tilde{q}_{1};k_{\perp_{MS}}) /
  \Delta_{\widetilde{ij}}^{R}(\tilde{q}_{2};k_{\perp_{MS}}).
\end{equation}
Each external line, from a branching at $(\tilde{q},z)$ in the pseudo-shower history,
contributes a factor
\begin{equation}
  \Delta_{\widetilde{ij}}^{R}(z\tilde{q};k_{\perp_{MS}}). 
\end{equation}

\subsection{Highest multiplicity treatment}

The original CKKW publication did not treat the highest multiplicity matrix element contribution any differently
to the other multiplicities. In Refs.~\cite{CKKW_imp, RichMrenna } it was noted that a different treatment
of highest multiplicities must be employed in order to fill the phase space in the matrix element region to all
orders in $\alpha_{S}$.   Since computational limits mean that only matrix elements with up to a
maximum of $N$ 
partons can be calculated, the standard approach leads to a maximum of $N$ partons being generated above the 
merging scale.  The parton shower generates to all orders in $\alpha_{S}$ and therefore we should also
let the matrix element region generate to all orders.  This can be achieved by allowing the highest
multiplicity channel parton shower to generate emissions in the region with $k_{\perp}$ less than 
that of the lowest transverse momentum of the matrix element emissions, $k_{\perp_{L}}$.  This is achieved
by changing the scale of the parton shower vetoes and Sudakov form factor cuts from $k_{\perp_{MS}}$ to
$k_{\perp_{L}}$.

\subsection{Problems with the algorithm}

The above procedure is heavily reliant on having an exact mapping between the shower variables
and the merging measure $k_{\perp_{MS}}$ so that the parton shower vetoes and
Sudakov cuts can be correctly applied.  
A mapping from the momentum clustered in step \ref{step_cluster}
to the corresponding shower variables is also required, so that the correct scales for the
Sudakov reweighting and initial shower conditions are obtained.
In practice obtaining such mappings may be difficult due to the complexity of the shower kinematics.

The initial scale at which the parton shower is invoked is vital to the algorithm.  
Initiating the parton shower directly from the merging scale would result
in a radiation gap, where emissions with transverse momentum less than the merging scale but 
evolution scale greater than the merging scale are missed.  In the angular-ordered shower, this
radiation corresponds to soft, wide-angle emissions.  The original CKKW publication attempts to
resolve this by invoking the parton shower from each external parton at a scale 
corresponding to the node at which it was `created' in the pseudo-shower history.
Although adopting this maximal initial scale helps fill the radiation gap, the
extra soft, wide-angle radiation that results, is emitted from the external
parton in the pseudo-shower history, rather than the intermediates, as implied by colour coherence \cite{PowhegOriginal}.  The original CKKW publication argued 
that this should be a sub-leading effect, however, it will certainly change the colour
structure of the configuration, which may cause problems when non-perturbative 
hadronization models are applied.

The original CKKW algorithm also assumes that the clustering of momentum in step \ref{step_cluster}
and subsequent mapping to parton shower variables results in a set of emission scales that 
respect the ordering of the parton shower, \emph{i.e.}
\begin{equation}
  \tilde{q}_{I}>\tilde{q}_{1}>...>\tilde{q}_{n}>\tilde{q}_{0}.  
\end{equation}
The clustering scheme of the original algorithm does not guarantee this.

These problems were studied in Ref.~\cite{nils_leif} and found to result in discontinuities
around the merging scale and a false dependence on the merging scale when the parton shower
was not ordered in transverse momentum. In Ref.~\cite{RichMrenna}, a study of the algorithm
with angular and virtuality-ordered parton showers was presented.  In that work, a number of \textit{ad hoc} 
adaptations were applied and tuned in order to achieve a reasonably
smooth merging at the parton level, nevertheless, some problems remained at the hadron level.  
In this article we aim to overcome these problems with a set of well motivated modifications based on the
POWHEG shower reorganisation.

\section{Shower reorganisation }
\label{Shower_reorganisation}
In the POWHEG next-to-leading-order matching scheme it was shown that 
a general parton shower may be rearranged such that the 
hardest emission\footnote{The hardest emission here refers to the emission with highest transverse momentum.} 
can be generated first.
The hardest emission can then be generated according to the exact NLO
matrix elements such that inclusive observables are distributed according to the NLO
cross section while the soft/collinear resummation of the shower is undisturbed.
In the POWHEG shower reorganisation, the hardest emission is
then dressed by inserting a truncated shower of wide angle, soft emissions prior to it, 
and a vetoed shower consisting of smaller $p_\perp$, smaller angle emissions, after it.

The CKKW algorithm generates a set of $n$ emissions above the merging scale $y_{_{MS}}$ according to 
exact tree-level matrix elements up to $\mathcal{O}(\alpha_{S}^{N})$.  This defines a set of $n$ hard emissions.
In order to reproduce the full shower around this set of hard emissions we employ 
a generalisation of the POWHEG shower reorganisation. 
In the following a review of the POWHEG reorganisation is presented followed by its extension 
to the CKKW case.  The notation used relates specifically to that of the \textsf{Herwig++} shower, 
however the treatment is independent of the details of the parton shower.
\subsection{POWHEG reorganisation}
\label{powheg_review}

A general \textsf{Herwig++} parton shower, evolving from a parton of flavour $\widetilde{ij}$ and 
initial scale $\tilde{q}_{I}$, can be represented by a generating functional 
$S_{\widetilde{ij}}(\tilde{q}_{I})$.
The evolution of the parton shower may be expressed by the recursive equation,
\begin{equation}
  \label{shower_recursion}
  S_{\widetilde{ij}}(\tilde{q}_{I})=\Delta_{\widetilde{ij}}(\tilde{q}_{I},\tilde{q}_{0})S_{\widetilde{ij}}(\tilde{q}_{0})
  + \int^{\tilde{q}_{I}}_{\tilde{q}_{0}}
  \Delta_{\widetilde{ij}}(\tilde{q}_{I},\tilde{q}) 
  \mathrm{d}\mathcal{P}_{\widetilde{ij}\to ij}(\tilde{q},z) 
  S_{i}\left(z\tilde{q}\right)
  S_{j}\left((1-z)\tilde{q}\right).
\end{equation}

The first term in Eq.~(\ref{shower_recursion}) is a no emission term corresponding
to the probability of evolving from the initial scale to hadronization scale with no resolvable emissions
generated.  The second term represents the probability of producing at least one emission, with the first
generated at a scale $\tilde{q}$ and further showers evolving down to the hadronization scale.

It is possible to expand Eq.~(\ref{shower_recursion}) to explicitly show the hardest 
emission of the shower.
The hardest emission is described by the shower variables $(\tilde{q}_{h},z_{h},\phi_{h})$ and has 
an associated transverse momentum $p_{\perp_{h}}$.  The shower may produce any number of other emissions before the 
hardest emission and any number of emissions after it but all of these must have
$p_{\perp}<p_{\perp_h}$.  The shower can therefore be written as
\begin{equation}
  \label{shower_decomp}
  S_{\widetilde{ij}}(\tilde{q}_{I})=\Delta_{\widetilde{ij}}(\tilde{q}_{I},\tilde{q}_{0})S_{\widetilde{ij}}(\tilde{q}_{0})  +
  \int^{\tilde{q}_{I}}_{\tilde{q}_{0}} 
  \bar{S}^{T}_{\widetilde{ij}}\left(\tilde{q}_{I},\tilde{q}_{h};p_{\perp_h}\right) 
  \mathrm{d}\mathcal{P}_{\widetilde{ij}\to ij}(\tilde{q}_{h},z_{h})
  \bar{S}_{i}^{V}\left(z_{h}\tilde{q}_{h};p_{\perp_h}\right)
  \bar{S}_{j}^{V}\left((1-z_{h})\tilde{q}_{h};p_{\perp_h}\right)
\end{equation}
where $\bar{S}^{T}$ refers to a \textit{truncated shower} and $\bar{S}^{V}$ refers to a
\textit{vetoed shower}.  The truncated shower in Eq.~(\ref{shower_decomp}) is responsible
for evolving from the initial scale down to the scale of the hardest emission producing 
any number of emissions with transverse momentum less than $p_{\perp_{h}}$.  
The emissions within the truncated shower correspond to soft,
wide angle gluon emissions, and so do not change the flavour of the shower line.
The vetoed shower evolves from the scale of the hardest emission down to the hadronization
scale also generating only emissions with transverse momentum less than $p_{\perp_{h}}$, its
evolution is defined by
\begin{align}
  \label{vetoed_shower}
   \bar{S}_{\widetilde{ij}}^{V}\left(\tilde{q}_{h};p_{\perp_h}\right)
   =\Delta_{\widetilde{ij}}(\tilde{q}_{h},\tilde{q}_{0})S_{\widetilde{ij}}(\tilde{q}_{0})
  + & \int^{\tilde{q}_{h}}_{\tilde{q}_{0}}
  \Delta_{\widetilde{ij}}(\tilde{q}_{h},\tilde{q}) 
  \mathrm{d}\mathcal{P}_{\widetilde{ij}\to ij}(\tilde{q},z)
  \theta(p_{\perp_h}-p_{\perp}(\tilde{q},z)) \\
  \times & \bar{S}_{i}^{V}\left(z\tilde{q};p_{\perp_h}\right)
  \bar{S}_{j}^{V}\left((1-z)\tilde{q};p_{\perp_h}\right) \notag.
\end{align}
The recursive equation describing the evolution of the truncated shower is given by
\begin{align}
  \label{trunc_shower}
   \bar{S}_{\widetilde{ij}}^{T}\left(\tilde{q}_{I},\tilde{q}_{h};p_{\perp_h}\right)
   =\Delta_{\widetilde{ij}}(\tilde{q}_{I},\tilde{q}_{h})S_{\widetilde{ij}}(\tilde{q}_{h})
  + & \int^{\tilde{q}_{I}}_{\tilde{q}_{h}}
  \Delta_{\widetilde{ij}}(\tilde{q}_{I},\tilde{q}) 
  \mathrm{d}\mathcal{P}_{\widetilde{ij}\to ij}(\tilde{q},z)
  \theta(p_{\perp_h}-p_{\perp}(\tilde{q},z)) \\
  \times &
  \bar{S}_{\widetilde{ij}}^{T}\left(z\tilde{q},\tilde{q}_{h};p_{\perp_h}\right)
  \bar{S}_{g}^{V}\left((1-z)\tilde{q};p_{\perp_h}\right) \notag.
\end{align} 

The Sudakov form factors and splitting functions appearing in the Eqs.~({\ref{trunc_shower}, \ref{vetoed_shower})
are identical to those in the standard shower equation of Eq.~(\ref{shower_recursion}) with the exception
that the splitting functions in both new showers have an additional $\theta$-function.  This $\theta$-function
guarantees that no emissions with transverse momentum greater than that of the hardest emission are generated.
Standard Monte Carlo techniques require that the splitting functions of a parton shower match those appearing
in the Sudakov form factors.  The introduction of the $\theta$-functions mean that this is not the case for
the vetoed and truncated showers in Eq.~(\ref{shower_decomp}), we highlight this in our notation with a bar.

In order to make the truncated and vetoed showers suitable for a Monte Carlo treatment, the original
POWHEG publication\cite{PowhegOriginal} splits the Sudakov form factor appearing in 
Eqs.~({\ref{trunc_shower}, \ref{vetoed_shower})
into two parts according to
\begin{equation}
\label{SudSplit}
  \Delta_{f}(z_{i}\tilde{q}_{i},\tilde{q}_{i+1})
  =\Delta_{f}^{V}(z_{i}\tilde{q}_{i},\tilde{q}_{i+1};p_{\perp_h}) 
  \bar{\Delta}_{f}^{R}(z_{i}\tilde{q}_{i},\tilde{q}_{i+1};p_{\perp_h}).
\end{equation}
Here, $\Delta_{f}^{V}$ refers to a \textit{vetoed Sudakov} in which the exponent contains a $\theta$-function,
which matches that in the splitting function of Eqs.(\ref{trunc_shower}, \ref{vetoed_shower}) and is given by 
\begin{equation}
  \Delta_{\widetilde{ij}}^{V}(z_{i}\tilde{q}_{i},\tilde{q}_{i+1};p_{\perp_h})
   =\exp\left[ -\sum_{\widetilde{ij}\to ij} \int^{z_{i}\tilde{q}_{i}}_{\tilde{q}_{i+1}}
    \mathrm{d}\mathcal{P}_{\widetilde{ij}\to ij}(\tilde{q},z) 
    \theta\left(p_{\perp_{h}}-p_{\perp}(\tilde{q},z)\right)
    \right].
\end{equation}
The other factor, $\bar{\Delta}_{f}^{R}$,
contains the opposite $\theta$-function and is referred to as a \textit{remnant Sudakov} given by 
\begin{equation}
  \label{sud_rem}
  \bar{\Delta}_{\widetilde{ij}}^{R}(z_{i}\tilde{q}_{i},\tilde{q}_{i+1};p_{\perp_h})=
  \exp\left[ -\int^{z_{i}\tilde{q}_{i}}_{\tilde{q}_{i+1}}
    \sum_{\widetilde{ij}\to ij}
    \mathrm{d}\mathcal{P}_{\widetilde{ij}\to ij}(\tilde{q},z)
    \theta\left(p_{\perp}(\tilde{q},z)-p_{\perp_{h}}\right)
    \right].
\end{equation}
The combination of the splitting functions in Eqs.~(\ref{trunc_shower}, \ref{vetoed_shower}) and the vetoed  Sudakov form
factors result in a parton shower that may be generated with standard vetoes allowing only emissions with 
$p_{\perp}<p_{\perp_h}$, 
however, the presence of the remnant Sudakov form factors appears to spoil this picture.  
On the contrary, it turns out that the seemingly awkward remnant 
factors have a key role to play in formalising how to generate the 
hardest emission first. In Ref.~\cite{PowhegOriginal} it was proven 
that the hardest emission in the shower is generated along the hardest 
line, the line for which all $z_{i}>\frac{1}{2}$ and, moreover, all 
non-soft emissions preceding it give rise to subleading collinear contributions. 
This means that the truncated shower may be considered as comprising solely of 
soft gluon emissions and that $z_{i}$ can be effectively replaced by 
one in all of the associated remnant factors. It is also shown that 
the $\theta$-function in the remnant Sudakov form factor exponent is 
zero for scales less than $\tilde{q}_{h}$ and so the replacement 
$z_{i}\rightarrow 1$ also holds for the vetoed showers. The net result 
of these replacements is that the product of all remnant Sudakov form 
factors combine as a single remnant Sudakov factor:
\begin{equation}
  \label{sud_rem_tot}
  \Delta_{\widetilde{ij}}^{R}(\tilde{q}_{I},\tilde{q}_{0};p_{\perp_h})=
 \exp\left[ -\int^{\tilde{q}_{I}}_{\tilde{q}_{0}}
   \mathrm{d}\mathcal{P}_{\widetilde{ij}\to ij}(\tilde{q},z)
    \theta\left(p_{\perp}(\tilde{q},z)-p_{\perp_{h}}\right)
   \right].
\end{equation}

The POWHEG treatment which has been outlined, results in a reorganisation of the shower such 
that the hardest emission may be generated first.  The Monte Carlo interpretation of this 
reorganisation is:
\begin{enumerate}
\item the hardest emission $(q_{h},z_{h},\phi_{h})$ is generated according to the appropriate splitting function 
  reweighted with the remnant Sudakov factor of Eq.~(\ref{sud_rem_tot});
\item a truncated shower, allowing only non-flavour-changing emissions with $p_{\perp}<p_{\perp_h}$ is initiated,
evolving the shower from $\tilde{q}_{I}$ down to $\tilde{q}_{h}$;
\item the hardest emission is forced with shower variables $(q_{h},z_{h},\phi_{h})$;
\item showers with a veto, allowing only emissions with $p_{\perp}<p_{\perp_h}$, 
  evolve all external lines down to the hadronization scale.
\end{enumerate}
A shower generated in this way should differ from the standard shower by only sub-leading terms.

The vetoed shower is defined by
\begin{align}
  \label{vetoed_shower_actual}
   S_{\widetilde{ij}}^{V}\left(\tilde{q}_{h};p_{\perp_h}\right)
   =\Delta_{\widetilde{ij}}^{V}(\tilde{q}_{h},\tilde{q}_{0})S_{\widetilde{ij}}^{V}(\tilde{q}_{0})
  + & \int^{\tilde{q}_{h}}_{\tilde{q}_{0}}
  \Delta_{\widetilde{ij}}^{V}(\tilde{q}_{h},\tilde{q}) 
  \mathrm{d}\mathcal{P}_{\widetilde{ij} \to ij}(\tilde{q},z)
  \theta(p_{\perp_h}-p_{\perp}(\tilde{q},z) ) \\
  \times & S_{i}^{V}\left(z\tilde{q};p_{\perp_h}\right)
  S_{j}^{V}\left((1-z)\tilde{q};p_{\perp_h}\right) \notag .
\end{align}
and corresponds to a standard shower with vetoes applied such that only emissions 
with $p_{\perp}<p_{\perp_{h}}$ are generated.
The truncated shower is defined by
\begin{align}
  \label{trunc_shower_actual}
   S_{i}^{T}\left(\tilde{q}_{I},\tilde{q}_{h};p_{\perp_h}\right)
   =\Delta_{i}^{V}(\tilde{q}_{I},\tilde{q}_{h})S_{\widetilde{ij}}^{V}(\tilde{q}_{0})
  + & \int^{\tilde{q}_{I}}_{\tilde{q}_{h}}
  \Delta_{i}^{V}(\tilde{q}_{I},\tilde{q}) 
  \mathrm{d}\mathcal{P}_{i \to ig}(\tilde{q},z)
  \theta(p_{\perp_h}-p_{\perp}(\tilde{q},z)) \\
  \times & S_{i}^{T}\left(z\tilde{q},\tilde{q}_{h};p_{\perp_h}\right)
  S_{g}^{V}\left((1-z)\tilde{q};p_{\perp_h}\right) \notag .
\end{align} 
and corresponds to a standard vetoed parton shower line, constrained not to produce any flavour
changing emissions, that is stopped once the truncated line 
has evolved down to the scale $\tilde{q}_{h}$.

\subsection{CKKW shower reorganisation}
\label{CKKW_recon}

In the POWHEG treatment reviewed in Sect.~\ref{powheg_review} a single hardest emission is
separated such that it may be corrected with matrix elements.  In the CKKW algorithm we aim
to improve the parton shower with tree-level matrix elements for all parton multiplicities 
resolved at the merging scale $k_{\perp_{MS}}$.  We perform a reorganisation of the 
parton shower, analogous to the POWHEG reorganisation, splitting the shower into two parts: 
a \textit{hard shower} describing emissions resolved above the merging scale; and another shower 
producing the rest of the shower emissions around this hard shower.  
The hard shower can then
be generated according to the tree-level matrix elements as required by the CKKW algorithm.

The result of this generalisation of the POWHEG reconstruction
is a set of truncated and vetoed showers which fill in the radiation between the hard emissions 
defined by the hard shower history.
In order to see how this works we first consider the next step
up from the POWHEG case of a single hard emission, where we have exactly two hard emissions
along the hard shower line, generated at scales $q_{1}$ and $q_{2}$.  One possible configuration of this 
hard shower line is given in Fig.~\ref{2_emission_line}.
\begin{figure}
\begin{centering}
\includegraphics[width=0.4\textwidth]{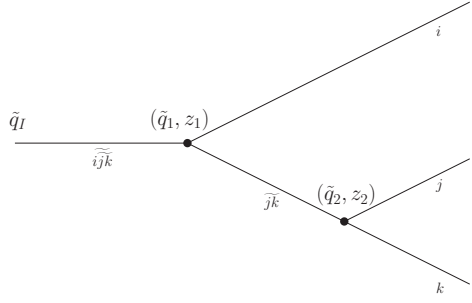}
\caption{An example of a hard shower line configuration where two emissions are generated above $k_{\perp_{MS}}$.}
\label{2_emission_line} 
\end{centering}
\end{figure}
As was done in formulating the POWHEG scheme, the full parton shower can be constructed around this hard
shower line by constructing an equation analogous to Eq.~(\ref{shower_decomp}), 
\begin{align}
  \label{CKKW_decomp}
  S_{\widetilde{i\widetilde{jk}}}^{(2)}(\tilde{q}_{I})= &
  \int^{\tilde{q}_{I}}_{\tilde{q}_{0}}
  \bar{S}_{\widetilde{i\widetilde{jk}}}^{T}\left(\tilde{q}_{I},\tilde{q}_{1};k_{\perp_{MS}}\right) 
  \mathrm{d}\mathcal{P}_{\widetilde{i\widetilde{jk}} \to i\widetilde{jk}}(\tilde{q}_{1},z_{1})
  \bar{S}_{i}^{V}\left((1-z_{1})\tilde{q}_{1};k_{\perp_{MS}}\right) \\
  \times & \int^{\tilde{q}_{1}}_{\tilde{q}_{0}}
  \bar{S}_{\widetilde{jk}}^{T}\left(z_{1}\tilde{q}_{1},\tilde{q}_{2};k_{\perp_{MS}}\right) 
  \mathrm{d}\mathcal{P}_{\widetilde{jk} \to jk}(\tilde{q}_{2},z_{2})
  \bar{S}_{j}^{V}\left(z_{2}\tilde{q}_{2};k_{\perp_{MS}}\right)
  \bar{S}_{k}^{V}\left((1-z_{2})\tilde{q}_{2};k_{\perp_{MS}}\right) \notag.
\end{align}
The superscript $(2)$ on $S$ denotes that this does not describe a general shower line, 
but the subset of shower lines with exactly two emissions above the merging scale. 
Eq.~(\ref{CKKW_decomp}) contains two truncated showers, 
one containing parton shower emissions with $k_{\perp}<k_{\perp_{MS}}$ before the 
hard emission $(\tilde{q}_{1},z_{1})$ and the other containing emissions with 
$k_{\perp}<k_{\perp_{MS}}$ between the hard emissions at $(\tilde{q}_{1},z_{1})$
and $(\tilde{q}_{2},z_{2})$.  The truncated and vetoed showers are
given by Eq.~(\ref{trunc_shower}) and 
Eq.~(\ref{vetoed_shower})\footnote{With the replacement $p_{\perp} \to k_{\perp}$. 
The variable $p_{\perp}$ is the shower definition of transverse momentum 
and  $k_{\perp}$ is the merging variable transverse momentum measure.}, respectively.

As in the POWHEG case, the splitting functions and Sudakov form factors for the 
truncated and vetoed showers in Eq.~(\ref{CKKW_decomp}) do not match each other and are therefore not suitable 
for a standard Monte Carlo treatment.  However, we can use the same manipulations as in the POWHEG formulation
to split the Sudakov form factors into a product of a Sudakov form factor that matches the vetoed
splitting functions and a remnant Sudakov form factor, as in Eq.~(\ref{SudSplit}).
The truncated showers contain only soft radiation so again we may set
$z_{i} \to 1$ in the remnant Sudakov form factor of Eq.~(\ref{sud_rem}) with only sub-leading differences.  
The result of this is that the product of remnant Sudakov form factors for a particular truncated 
or vetoed line combine to give a remnant Sudakov factor.  Rather than resulting in a single remnant
Sudakov factor as in the POWHEG scheme, we now get a product of remnant Sudakov factors.
The product of remnant Sudakov factors for the hard shower configuration 
of Fig.~\ref{2_emission_line} is given by
\begin{align}
 & \frac{ \Delta_{\widetilde{i\widetilde{jk}}}^{R}(\tilde{q}_{I};k_{\perp_{MS}}) }
 { \Delta_{\widetilde{i\widetilde{jk}}}^{R}(\tilde{q}_{1};k_{\perp_{MS}}) }
  \Delta_{i}^{R}((1-z_{1})\tilde{q}_{1};k_{\perp_{MS}})
  \frac{ \Delta_{\widetilde{jk}}^{R}(z_{1}\tilde{q}_{1};k_{\perp_{MS}}) }
 { \Delta_{\widetilde{jk}}^{R}(\tilde{q}_{2};k_{\perp_{MS}}) } \\
 & \times \Delta_{j}^{R}(z_{2}\tilde{q}_{2};k_{\perp_{MS}})
   \Delta_{k}^{R}((1-z_{2})\tilde{q}_{2};k_{\perp_{MS}}) \notag ,
\end{align}
where the remnant Sudakov factor is given by
\begin{equation}
  \Delta_{\widetilde{ij}}^{R}(\tilde{q};k_{\perp_{MS}})=
  \exp\left[ -\int^{\tilde{q}}_{\tilde{q}_{0}}
    \sum_{\widetilde{ij}\to ij}
    \mathrm{d}\mathcal{P}_{\widetilde{ij}\to ij}(\tilde{q},z)
    \theta\left(k_{\perp}(\tilde{q},z)-k_{\perp_{MS}}\right)
    \right].
\end{equation}

In the CKKW algorithm the hard shower is generated by choosing a jet multiplicity $n$ as described in
Sect.~\ref{transCKKW} and generating $n$ parton momenta according to the appropriate jet cross section.
A pseudo-shower history and corresponding shower variables are then assigned by applying a clustering algorithm
to the $n$ parton momenta, until they are clustered back to a leading order configuration.
The shower reorganisation presented here results in a product of remnant Sudakov factors, with which these
hard shower configurations should be reweighted.  
These remnant Sudakov factors can generally be found
from the pseudo-shower history by applying the following prescription:
\begin{itemize}
  \item each internal line from a branching at $(q_{1},z_{1})$ to $(q_{2},z_{2})$ contributes a factor
    \begin{equation}
      \frac{\Delta_{f}^{R}(z_{1}\tilde{q}_{1};k_{\perp_{MS}})}
      {\Delta_{f}^{R}(\tilde{q}_{2};k_{\perp_{MS}})};
    \end{equation}
  \item each external line from a branching at $(q,z)$ contributes a factor
    \begin{equation}
      \Delta_{f}^{R}(z\tilde{q};k_{\perp_{MS}}).
    \end{equation}
\end{itemize}
These remnant Sudakov factors match the Sudakov factors in Eq.(\ref{sud_w_theta}) that 
we argued should be introduced in order to extend the CKKW procedure for transverse momentum showers to
the angular-ordered shower.

The parton shower for emissions below the cut is generated by producing truncated and vetoed
showers around the hard shower according to the following prescription:
\begin{itemize}
  \item each internal line from a branching at $(q_{1},z_{1})$ to $(q_{2},z_{2})$ results in a truncated shower
    \begin{equation}
      S_{f}^{T}\left(z_{1}\tilde{q}_{1},\tilde{q}_{2};k_{\perp_{MS}}\right);
    \end{equation}
  \item each external line from a branching at $(q,z)$ results in a vetoed shower
    \begin{equation}
      S_{f}^{V}\left(z\tilde{q};k_{\perp_{MS}}\right).
    \end{equation}
\end{itemize}

\section{The algorithm}
\label{algorithm}

In order to implement the procedure described in Sect.~\ref{CKKW_recon} we employ the strategy of
Ref.~\cite{HRT_drell_yan}, where the hardest emissions, or set of hard emissions in this case,
are interpreted as parton shower emissions.  This approach leads to a straightforward 
implementation of the truncated showers, where a truncated shower, evolving between hard emissions
at $(\tilde{q}_{1},z_{1},\phi_{1})$ and $(\tilde{q}_{2},z_{2},\phi_{2})$, is generated by initiating a standard parton 
shower at $z_{1}\tilde{q}_{1}$ with vetoes allowing only non-flavour-changing emissions with $k_{\perp}<k_{\perp_{MS}}$
and stopping the truncated shower once it has evolved beyond $\tilde{q}_{2}$, at which point the 
second hard emission is forced with splitting variables $(\tilde{q}_{2},z_{2},\phi_{2})$.
This allows the full shower of truncated showers, hard emissions and vetoed showers to be generated
as a single shower evolution from the leading-order configuration. 
This results in a substantial improvement over earlier
CKKW implementations with angular-ordered showers
\cite{CKKW,RichMrenna}, since now the \emph{colour structure}
in the event is plainly equivalent to that which the
shower would have produced by default \emph{i.e.}
it respects colour coherence.

In order to interpret the matrix element emissions as shower emissions, we require
an exact mapping from the set of $n$ external parton momentum and assigned pseudo-shower history
to a set of shower splitting variables, $(\tilde{q},z,\phi)$, describing each emission.
Obtaining such a mapping equates to inverting the momentum reconstruction, which is performed at the end of
the standard parton shower to translate the set of shower variables into the parton momenta, and is described in
Sect.~\ref{IMR}.  Having such a mapping also provides the exact shower variables that are to used 
for the Sudakov and $\alpha_{S}$ reweighting.

The full modified CKKW algorithm is described below.
\begin{enumerate}
\item
  \label{step_start2}
  The jet multiplicity $n$ is generated with probability   
  \begin{equation}
    P_{n}=\frac{\sigma_{n}(k_{\perp_{MS}})}{\sum_{N}\sigma_{i}(k_{\perp_{MS }})},
  \end{equation}
  where cross sections are evaluated at a fixed strong coupling $\alpha_{S_{ME}}$.
\item 
  \label{dist_mom}
  The $n$ external parton momenta are  generated according to
  $\mathrm{d}\sigma ( k_{\perp_{MS}} )$.
\item
  \label{step_cluster2}
  Pairs of external parton momenta are clustered\footnote{The clustering procedure is 
    discussed in Sect.~\ref{clustering}.} down to a leading-order configuration, 
  assigning a pseudo-shower history.
\item The inverse momentum reconstruction is applied to the external momenta and shower history such that a set of 
  shower splitting variables $(\tilde{q},z,\phi)$ are found, describing $n-2$ hard branchings.
\item The configuration is reweighted to include the Sudakov form factors and running $\alpha_{S}$.  
  This corresponds to assigning the configuration a weight $W$ and rejecting the configuration if
  $W<\mathcal{R}$\footnote{$\mathcal{R}$ refers to a random number, generated in the interval $[0,1]$.}.
  The weight is constructed from the pseudo-shower history, according to the following prescription:
  \begin{itemize}
  \item each hard emission at $(\tilde{q},z)$ contributes a running $\alpha_{S}$ factor
    \begin{equation}
      \frac{\alpha_{S}\left(p_{\perp}(\tilde{q},z)\right)}{\alpha_{S_{ME}}};
    \end{equation}
  \item each internal line between hard emissions at $(\tilde{q}_{1},z_{1})$ to $(\tilde{q}_{2},z_{2})$ contributes a Sudakov factor
    \begin{equation}
      \frac{\Delta_{f}^{R}(z_{1}\tilde{q}_{1};k_{\perp_{MS}})}
           {\Delta_{f}^{R}(\tilde{q}_{2};k_{\perp_{MS}})};
    \end{equation}
  \item each external line from a hard emission at $(\tilde{q},z)$ contributes a Sudakov factor
    \begin{equation}
      \Delta_{f}^{R}(z\tilde{q};k_{\perp_{MS}}).
    \end{equation}
  \end{itemize}
  If the configuration is rejected\footnote{This reweighting procedure relies on the weight 
    generated in this step satisfying $W<1$.  The fixed strong coupling used in the matrix 
    elements $\alpha_{_{ME}}$ can be chosen to be large enough 
    that the $\alpha_{S}$ weight is always less than one. 
    Individual Sudakov form factors are also guaranteed to be less than one
    while the ratio of Sudakov form factors contributed by intermediate lines must be less 
    than one due to the angular ordering condition
    $z_{i}\tilde{q}_{i}>\tilde{q}_{i+1}$.} return to step \ref{step_start2}.
\item Parton shower lines are initiated from the leading-order configuration which are to be evolved
  according to the procedure:
  \begin{enumerate}
    \item
      \label{step_dec}
      If a hard emission exists at a lower scale on the shower line, 
      then the shower is evolved as a truncated shower otherwise proceed
      with step~(\ref{step_vet}).  The truncated showers evolve
      as the standard parton shower with vetoes allowing only non-flavour-changing-emissions with 
      $k_{\perp}<k_{\perp_{MS}}$. Each truncated 
      emission generates a soft gluon which should be evolved according to step~(\ref{step_vet}).
    \item Once the scale of the next hard emission is reached, the hard emission is forced
      creating  two further shower lines, each of which should be evolved according to
      step~(\ref{step_dec}).
    \item
      \label{step_vet}
      Vetoed showers evolve all external shower lines down to the hadronization scale, 
      with vetoes allowing only emissions with $k_{\perp}<k_{\perp_{MS}}$.
  \end{enumerate}
\end{enumerate}
The above scheme is adapted for the highest multiplicity channel, where $n=N$, by the replacement 
$k_{\perp_{MS}} \to k_{\perp_{l}}$ in the shower vetoes and Sudakov form factors.
\subsection{Shower kinematics}

In order to implement the inverse momentum reconstruction and parton shower vetoes,
the kinematics of the \textsf{Herwig++} final-state parton shower must be understood.  
The full details of this are given in 
Ref.~\cite{HerwigMan}
and we present a review here for completeness and to introduce our notation.

Each external parton from the hard sub-process with momentum $p_{J}$ is interpreted as a progenitor for a
parton shower jet. 
Each parton shower jet evolves from the initial scale $\tilde{q}_{I}=\sqrt{s}$ down to the hadronization scale
$\tilde{q}_{0}$ undergoing a series of branchings, each described by the shower variables $(\tilde{q},z,\phi)$.

Once all of the shower lines have evolved down to the hadronization scale, 
the shower evolution is stopped and the momentum
of all external and intermediate partons are reconstructed from the shower variables.
This is done in the centre-of-mass frame via the Sudakov decomposition, 
where the momentum of a parton in the shower is written
\begin{equation}
  q_{i}=\alpha_{i}p_{J}+\beta_{i}n_{J}+q_{\perp i},
\label{sudakovDecomp}
\end{equation}
where the reference vector $n_{J}$ is taken to be a light-like vector with three-momentum equal to that of
the colour partner to the jet progenitor and $q_{\perp}$ is the component of momentum transverse to
both $p_{J}$ and $n_{J}$.  The momentum fraction $z$ of each branching is defined by
\begin{equation}
  z=\frac{\alpha_{i}}{\alpha_{\widetilde{ij}}},
  \label{zdef}
\end{equation}
the scale of the emission $\tilde{q}$ is defined in Eq.~(\ref{hw_scale}).
The relative transverse momentum of the branching is defined by
\begin{equation}
  p_{\perp i}=q_{\perp i}-zq_{\perp \widetilde{ij} }.
\label{relPt}
\end{equation}
The $p_{\perp i}$ vector is written in terms of the azimuthal angle $\phi$
\begin{equation}
  p_{\perp}=\left(
  \left|p_{\perp}\right|\cos{\phi},
  \left|p_{\perp}\right|\sin{\phi},
  0;0\right).
\label{ptDef}
\end{equation}
The Sudakov variables $\alpha_{i}$, $\beta_{i}$ and $q_{\perp i}$ are 
calculated from the shower
variables recursively for all partons in the shower jet and the momentum are constructed according to
Eq.~(\ref{sudakovDecomp}).

The momentum reconstruction procedure results in the progenitor
momenta $q_{J}$ gaining some off-shell momentum and leads to the loss of global
momentum conservation.
Longitudinal boosts are applied to each shower jet to restore momentum conservation
while disturbing the internal structure of each jet as little as possible.  These 
reshuffling boosts are defined for each jet by the transformation
\begin{equation}
  \left(\vec{q}_{J};\sqrt{\vec{q}_{J}^{2}+q_{J}^{2}}\right)\rightarrow
  \left(k\vec{p}_{J};\sqrt{k^{2}\vec{p}_{J}^{2}+q_{J}^{2}}\right).
  \label{reshuffle}
\end{equation} 
Momentum conservation is ensured by requiring that the rescaling parameter $k$ satisfies
\begin{equation}
  \sum_{J}\sqrt{k^{2}\vec{p}_{J}^{2}+q_{J}^{2}}=\sqrt{s}.
\label{boostParam}
\end{equation}

\subsection{Inverse momentum reconstruction}

\label{IMR}
In the CKKW algorithm a clustering procedure is applied to a configuration of external parton momenta,
defining a pseudo-shower history for producing that configuration as a set of branchings from a
leading-order parton configuration, which in this case would be $q\bar{q}$.
We aim to interpret this branching history as a set of shower branchings
such that when this set of branchings are forced, the momentum reconstruction will reconstruct the
original parton momenta.  In order to map the shower history to a set of shower variables 
the momentum reconstruction procedure must be inverted; this requires two steps.  
First, the boost applied to each shower jet in order to conserve global momentum must be found and its 
inverse applied to the momenta of the shower jet.  Second, the resulting momenta are decomposed into
the shower variables according to Eq.~(\ref{sudakovDecomp}).

The momenta of the set of progenitors defined by the branching $q_{J}^{\prime}$ correspond to the momenta on the right 
hand side of Eq.~(\ref{reshuffle}).  
The original on-shell progenitors $p_{J}$ are related to $q_{J}^{\prime}$ by
\begin{equation}
  p_{J}=\left(\frac{\vec{q}_{J}^{\prime}}{k};
  \sqrt{\frac{\vec{q}_{J}^{\prime 2}}{k^{2}}+m_{J}^{2}}\right),
\label{refP}
\end{equation} 
where $m_{J}$ is the on-shell mass of the jet progenitor.
The set of on-shell progenitors respect global momentum conservation therefore we can find the 
boost parameter $k$ by solving
\begin{equation}
  \sum_{J}\sqrt{\frac{\vec{q}_{J}^{\prime 2}}{k^{2}}+m_{J}^{2}}=\sqrt{s}.
\end{equation}
Once $k$ is found, the reference vector $p_{J}$ is given by Eq.~(\ref{refP}); similarly $n_{J}$ is given by
\begin{equation}
 n_{J}=\left(
 \frac{\vec{q}^{\prime}_{\bar{J}}}{k};
 \sqrt{\frac{\vec{q}_{\bar{J}}^{\prime 2}}{k^{2}}} \right),
\label{refN}
\end{equation} 
where $\bar{J}$ refers to the colour partner jet of the shower jet $J$.
In the reconstruction procedure, the Sudakov parameters of the progenitor partons 
are set to $\alpha_{0}=1$ and $q_{\perp 0}=\boldmath{0}$.  
Furthermore, Eq.~(\ref{sudakovDecomp}) implies that 
\begin{equation}
  \beta_{0}=\frac{q_{J}^{2}-m_{J}^{2}}
       {2p \cdot n}.
\end{equation}
Since $q_{J}^{\prime}$ and $q_{J}$ are related by a boost, we also have
$q_{J}^{2}=q_{J}^{\prime 2}$.  The momentum of the reconstructed progenitors $q_{J}$
can then be constructed according to Eq.~(\ref{sudakovDecomp}).  
This defines the
reshuffling boost as in Eq.~(\ref{reshuffle}).  The boosts for all shower jets
can then be calculated, inverted and applied to all momenta in each jet.
The momentum can then be decomposed into Sudakov parameters and the shower variables 
$(\tilde{q},z,\phi)$ for each branching calculated from
Eqs.~(\ref{zdef}--\ref{ptDef}).

\subsection{Shower vetoes}

The vetoes that are applied to the truncated and vetoed showers and the cuts applied to the remnant Sudakov form
factors require a mapping between the shower variables, $(\tilde{q},z)$ 
and the merging scale transverse momentum measure $k_{\perp}$.
The merging variable, for an emission $\widetilde{ij} \to ij$, is defined in some jet measure according to, 
$k_{\perp}=\sqrt{y_{ij}s}$.
We have implemented the merging algorithm with the Durham\cite{Durham} and LUCLUS\cite{LUCLUS}
jet measures, defined by
\begin{equation}
  \label{dur}
  y_{\mathrm{dur}_{ij}}=\frac{2\mathrm{min}\left(E_{i}^{2},E_{j}^{2}\right)}
  {s}\left(1-\cos{\theta_{ij}}\right),
\end{equation} 
\begin{equation}
  \label{luc}
  y_{\mathrm{luc}_{ij}}=\frac{2\left(E_{i}E_{j}\right)^{2}}
  {s\left(E_{i}+E_{j}\right)^{2}}
  \left(1-\cos{\theta_{ij}}\right).
\end{equation} 

In order to implement these vetoes a mapping between the shower variables $(\tilde{q},z,\phi)$ and $y_{ij}$, 
in the chosen jet measure, must be found.  
The \textsf{Herwig++} shower produces off-shell intermediate states and therefore a set of boosts must be applied
to each shower line in order to ensure momentum conservation.  Since the boosts depend on the full 
shower history, an exact mapping between the shower variables and the merging variable cannot be found.
We use a mapping that is exact for a single shower emission and should give a good approximation for 
larger numbers of emissions.  For clarity in the following, we treat partons as massless while in
our implementation parton masses are retained.

In the case of \mbox{$e^{+}e^{-}\rightarrow\mathrm{hadrons}$} we have two shower
progenitor partons $q(p_{a})$ and $\bar{q}(p_{b})$. 
The momenta of these progenitors, in the centre-of-mass frame, can be written,
\begin{equation}
  p_{a,b}=\frac{\sqrt{s}}{2}\left(0, 0, \pm 1; 1\right).
\end{equation}
We now consider a single gluon emission along the quark line 
such that $q(p_{a}) \rightarrow q(q_{1}) g(q_{2})$.  
This is the first emission so we have $\alpha_{1}=z$ and $\alpha_{2}=1-z$.
The transverse momentum of the emitted partons are given by
$q_{\perp_{1,2}}=\pm p_{\perp}$.  The emitted partons are considered to be external partons
and therefore their momenta should be set on-shell. The $\beta$ parameters 
in Eq.~(\ref{sudakovDecomp}) are therefore given by
\begin{equation}
  \beta_{1}=\frac{p_{\perp}^{2}}{zs},
\end{equation}
\begin{equation}
  \beta_{2}=\frac{p_{\perp}^{2}}{(1-z)s}.
\end{equation}
The vetoes should correspond to vetoes on the reshuffled momenta that have had the boosts, defined
in Eq.~(\ref{reshuffle}), applied to them.  We should therefore solve Eq.~(\ref{boostParam})
and calculate these boosts before applying Eqs.(\ref{dur}, \ref{luc}).
The reconstructed progenitor momenta are given by $q_{a}=q_{1}+q_{2}$ for the quark jet and 
$q_{b}=p_{b}$ for the anti-quark jet.
Inserting these momenta into Eq.~(\ref{boostParam}) yields the solution
\begin{equation}
k=1-\frac{p_{\perp}^{2}}{sz(1-z)},
\end{equation}
for the boost parameter.  The reshuffling boost for the quark line is then defined
by Eq.~(\ref{reshuffle}).  It follows that the three-vector of the shuffled quark progenitor
$q_{a}^{\prime}$ should be given by
\begin{equation}
\label{shuff_three}
\vec{q}_{a} \ ^{\prime}=\frac{\sqrt{s}}{2}\left( 1-\frac{p_{\perp}^{2}}{sz(1-z)} \right).
\end{equation}
The expression in Eq.(\ref{shuff_three}) is exactly the same as $q_{a}$ and therefore the
boost to be applied to the quark jet is the unit matrix.
The shuffled momenta for the emitted partons have now been constructed and
we can apply Eqs.~(\ref{dur}, \ref{luc}) to give expressions
for the jet measures used to define the merging scale.  These give
\begin{equation}
y_{\mathrm{dur}}=\mathrm{min}\left[z+\frac{p_{\perp}^{2}}{sz},(1-z)+\frac{p_{\perp}^{2}}{s(1-z)}\right]^{2}
\frac{(1-\cos{\theta})}{2},
\end{equation}
\begin{equation}
y_{\mathrm{luc}}=\left[ \frac{\left( p_{\perp}^{2}+(1-z)^{2}s \right)\left( p_{\perp}^{2}+z^{2}s \right)}
  { p_{\perp}^{2}s+s^{2}z(1-z) } \right]^{2}
\frac{(1-\cos{\theta})}{2},
\end{equation}
where
\begin{equation} 
\cos{\theta}=1-\frac{2p_{\perp}^{2}s}
{\left( p_{\perp}^{2}+(1-z)^{2}s \right)\left( p_{\perp}^{2} + z^{2}s \right)}.
\end{equation}
These mappings allow a transverse momentum measure, $k_{\perp}=\sqrt{ys}$, to be calculated 
in the merging variable for each parton shower emission.  
Parton shower vetoes and Sudakov cuts can then be applied by comparing this measure to the merging scale.

\subsection{Clustering scheme}
\label{clustering}
The parton shower decomposition presented in Sect.~\ref{CKKW_recon} relied on our ability to interpret
the series of hard branchings, defined by the matrix element momenta and assigned pseudo-shower history,
as a parton shower.  The inverse momentum reconstruction procedure ensures that, given an assigned
pseudo-shower history, a set of parton shower emissions are found that will exactly reproduce 
the matrix element momenta.  

The assignment of a pseudo-shower history is dependent on the clustering 
scheme used and this should be carefully chosen to ensure that the assigned history is similar to that
which the parton shower would have generated.  In particular, Sect.~\ref{CKKW_recon} assumes that the assigned history 
 is angular-ordered.  We therefore assign histories that obey the angular-ordering condition
\begin{equation}
\label{ang_ordering}
\tilde{q}_{i}z_{i}>\tilde{q}_{i+1},
\end{equation}
for all emissions along all shower lines.

The inverse momentum reconstruction allows us to find the shower variables of all branchings
in a particular pseudo-shower history.  We can therefore determine whether a history
is angular-ordered by following all shower lines outwards from the hard process and 
explicitly checking that all of the branchings satisfy Eq.~(\ref{ang_ordering}). 

We employ a clustering procedure that creates all possible angular-ordered pseudo-shower 
histories and chooses the one that the shower was most likely to produce.
This is achieved by the procedure: 
\begin{enumerate}
\item all possible shower histories are created by clustering all pairs of partons 
  whose flavours correspond to allowed branchings;
\item non-angular-ordered histories are discarded;
\item the angular-ordered history for which the scalar sum 
of the transverse momentum of its branchings is smallest is chosen;
\item if no angular-ordered histories could be found, the unordered history
with smallest transverse momentum is chosen.
\end{enumerate}
The procedure of always choosing the 
shower history with smallest transverse momentum is taken rather than a probabilistic choice so
that unnatural clusterings are not assigned, for example in a three jet event a gluon collinear
to the quark line could be clustered to the anti-quark. 

\section{Results}
\label{results}

In this section we present the results of the implementation of the modified CKKW algorithm for the process
\mbox{$e^{+}e^{-}\to\mathrm{hadrons}$} at a centre-of-mass energy of 
$91.2\mathrm{\,GeV}$ at both parton and hadron level.
The parton level results provide a test of the algorithm's ability to provide a
smooth merging between the matrix element and parton shower regions of phase space, 
showing features that may be hidden by the addition of a hadronization model.
The hadron level results provide the ultimate test of the algorithm's ability to describe
data and in particular are sensitive to the parton colour structure assignment which we
expect the modified algorithm to improve with respect to traditional CKKW methods.

A key test of the merging algorithm is its insensitivity to changes in the merging scale and merging
variable.
The algorithm was implemented with two merging variables: the Durham and LUCLUS jet resolution
variables.  
For each merging variable, merging scales of $y_{_{MS}}=5\times10^{-2}$, $y_{_{MS}}=10^{-2}$ and
$y_{_{MS}}=5\times10^{-3}$ were used.  Samples of events with all partons separated by
$y>y_{_{MS}}$ were generated using 
\textsf{MadGraph/MadEvent}\cite{Maltoni:2002qb,Alwall:2007st} for 
the process with up to five partons in the final state\footnote{The maximum multiplicity for each 
merging scale was decided according to the phase space available in the matrix element region.}.

\subsection{Parton level results}
\label{partonRes}

We present the distributions of the merging variable itself since these should be the most sensitive to problems with the 
merging procedure.  In order to provide a direct comparison to Ref.~\cite{nils_leif}, we first present
a systematic look at the algorithm with the maximum multiplicity set to three, so that the matrix element
region is responsible for, at most, a single hard emission. 

Figure \ref{fig:23} shows distributions of the scale at which three jets are resolved in the Durham jet measure 
for the three chosen merging scales with the Durham jet measure as the merging variable.  Jet analyses were performed with the \textsf{KtJet} 
package\cite{Butterworth:2002xg}.   
Each of the merging
scale choices exhibit a smooth transition between the two phase space regions, there also appears
to be little dependence on the choice of merging scale.  

\begin{figure}
\begin{centering}
\includegraphics[width=0.4\textwidth,angle=0]{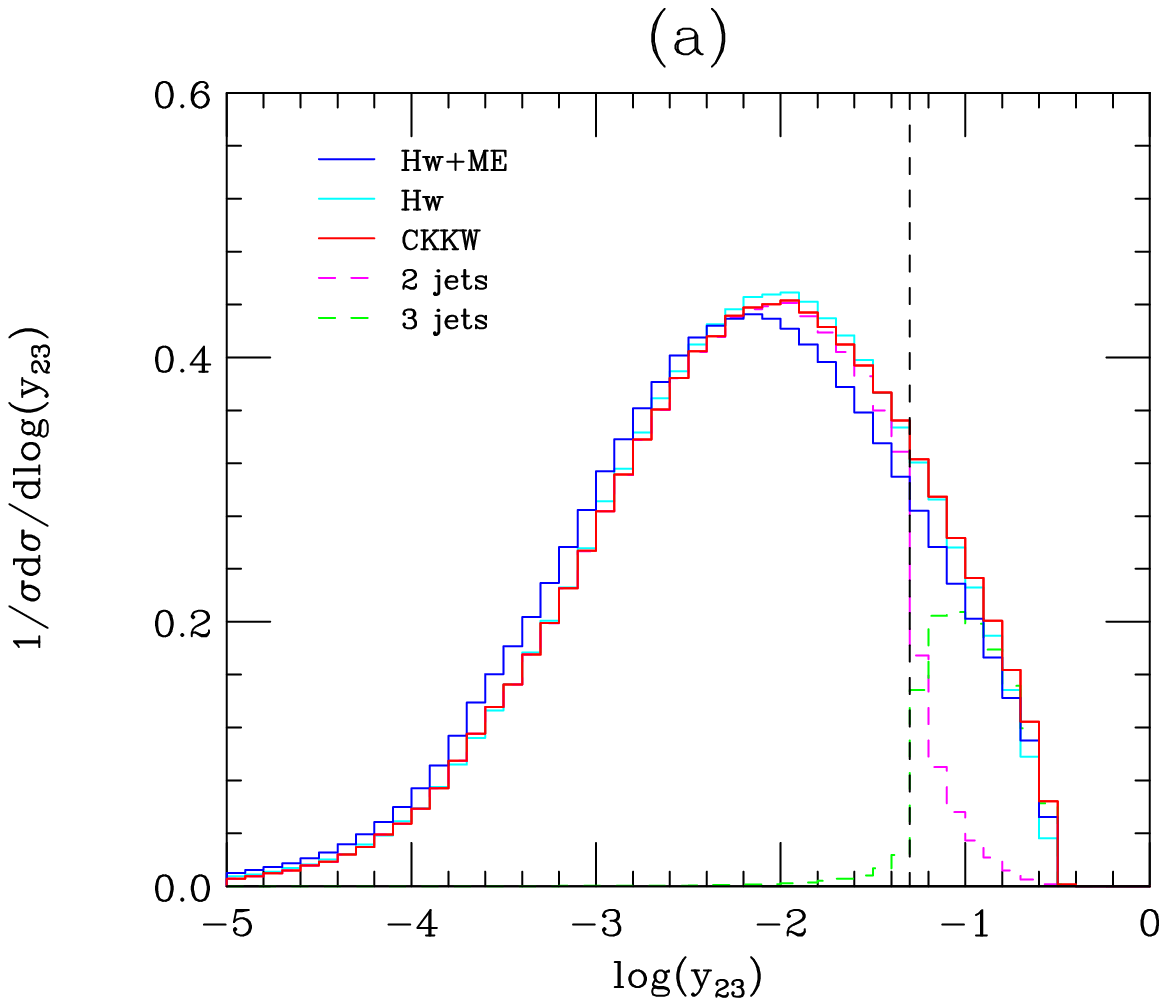}\hspace{3mm}
\includegraphics[width=0.4\textwidth,angle=0]{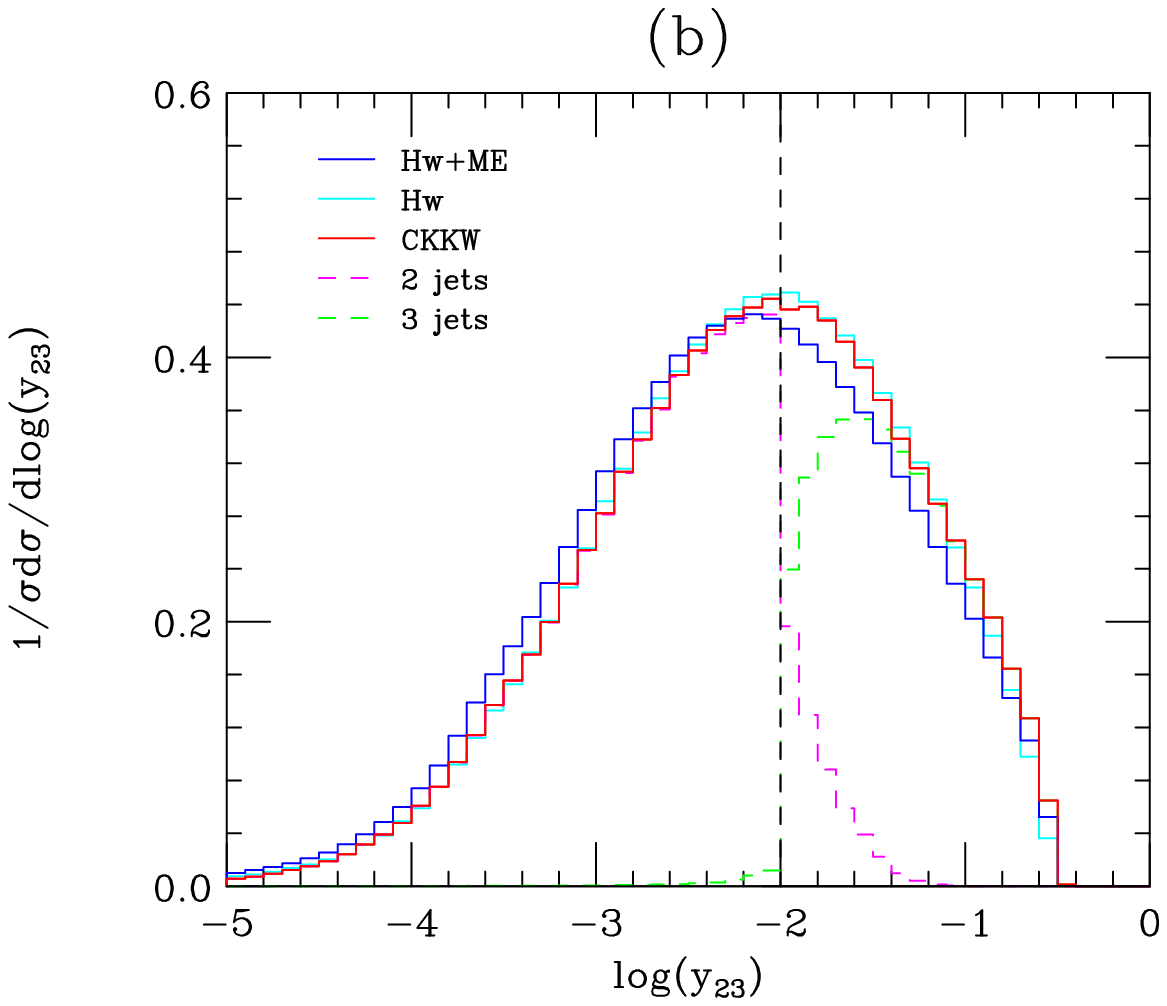}\\
 \vspace{3mm}
 \includegraphics[width=0.4\textwidth,angle=0]{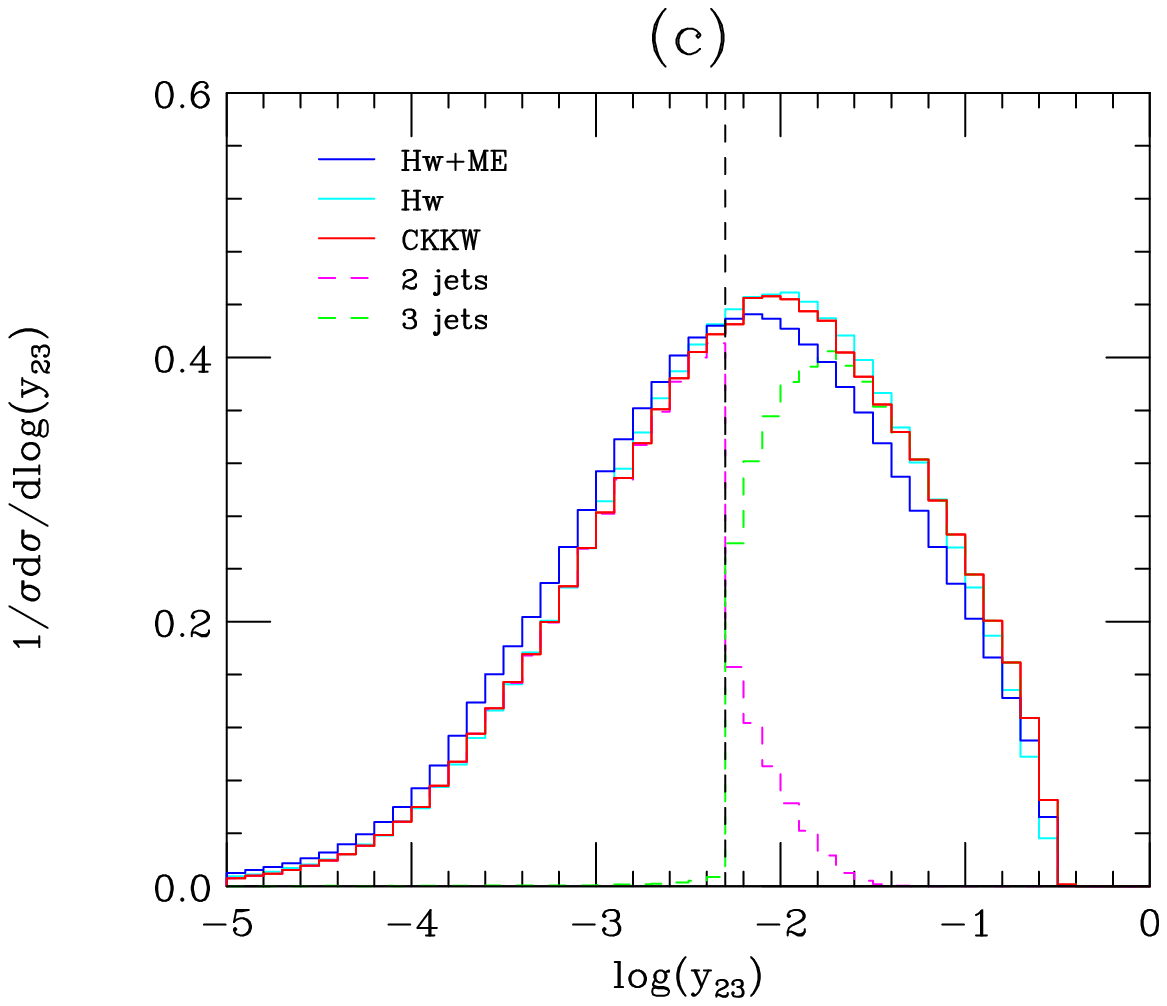}\hspace{3mm}
\includegraphics[width=0.4\textwidth,angle=0]{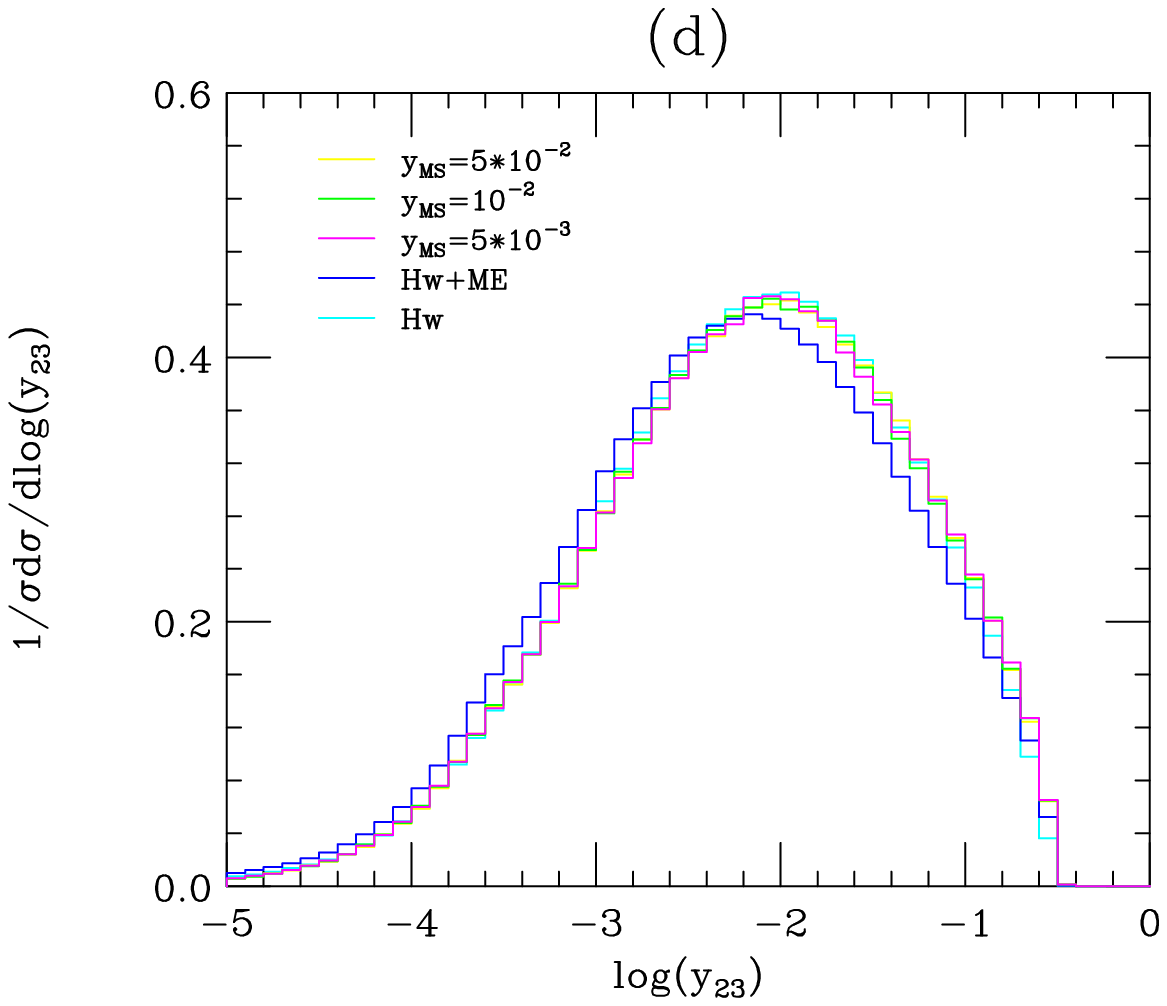} \\
\par\end{centering}
\caption{ Parton level distributions of the scale at which three jets are resolved in the Durham jet measure for \mbox{$e^{+}e^{-} \to \mathrm{hadrons}$} at $\sqrt{s}=91.2\mathrm{\,GeV}$. The red line shows the CKKW distribution with maximum multiplicity set to three, the blue line shows the \HWPP\ parton shower distribution with a matrix element correction applied and the cyan line shows the \HWPP\ parton shower distribution without the matrix element correction.  Plots (a)-(c) show the CKKW distributions with merging scales set to $y_{_{MS}}=5\times10^{-2}$, $y_{_{MS}}=10^{-2}$ and $y_{_{MS}}=5\times10^{-3}$ in the Durham jet measure.  Plot (d) shows the CKKW distributions at all of the
merging scale choices on the same plot. }
\label{fig:23}
\end{figure}

Figure \ref{fig:23_no_trunc} shows the same distributions as Fig.~\ref{fig:23} but with the truncated shower
switched off.  Switching off the truncated shower results in a radiation gap, meaning that emissions that 
would be generated at scales greater than that of the hard emission, 
but with transverse momentum less than that of the hard emission, are never produced.  
This radiation gap corresponds to a deficit in the amount of
soft wide angle emissions produced from the three jet samples.
This problem will only become more serious as higher multiplicity contributions are included 
and underlines the importance of the truncated shower in the merging algorithm.

\begin{figure}
  \begin{centering}
  \includegraphics[width=0.4\textwidth,angle=0]{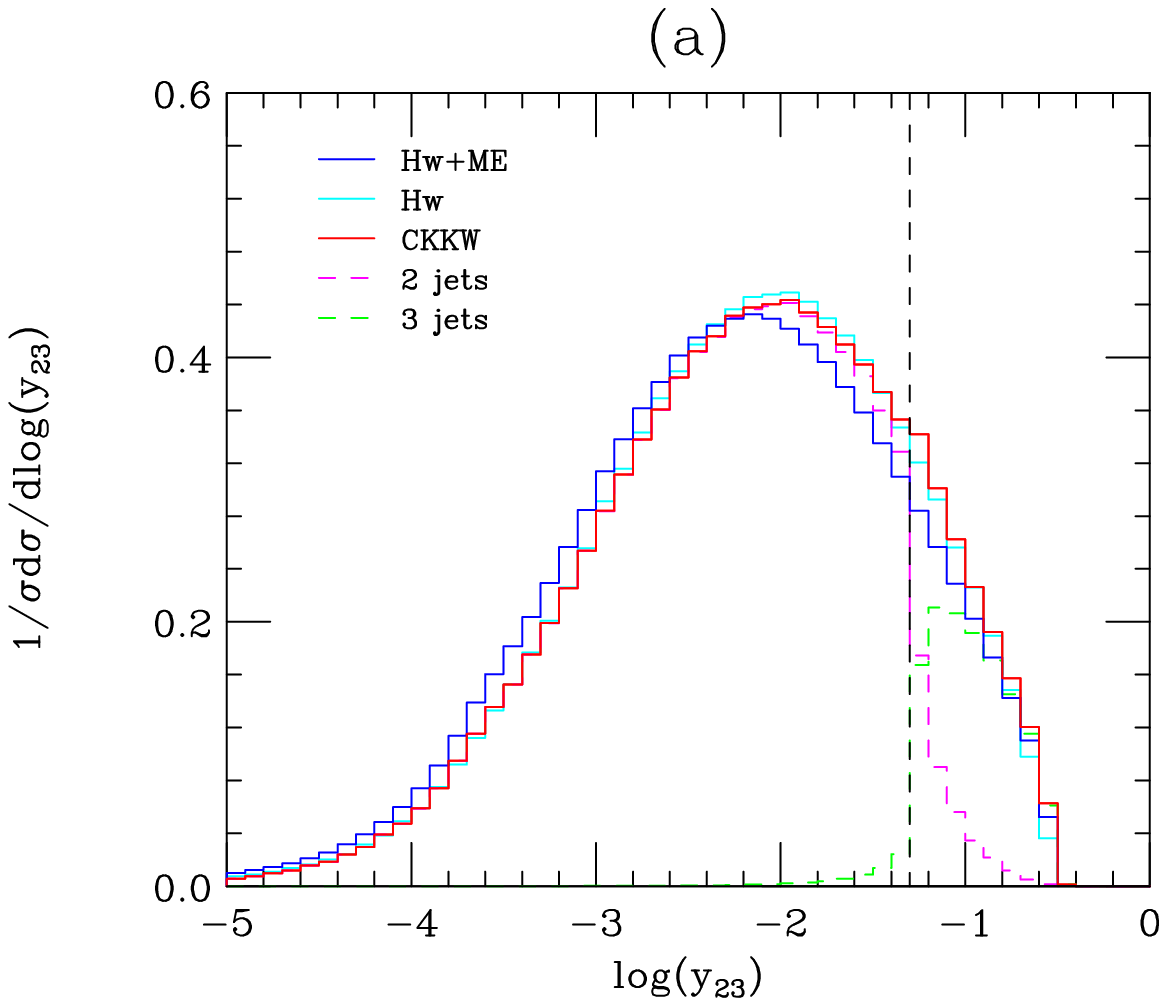}\hspace{3mm}
  \includegraphics[width=0.4\textwidth,angle=0]{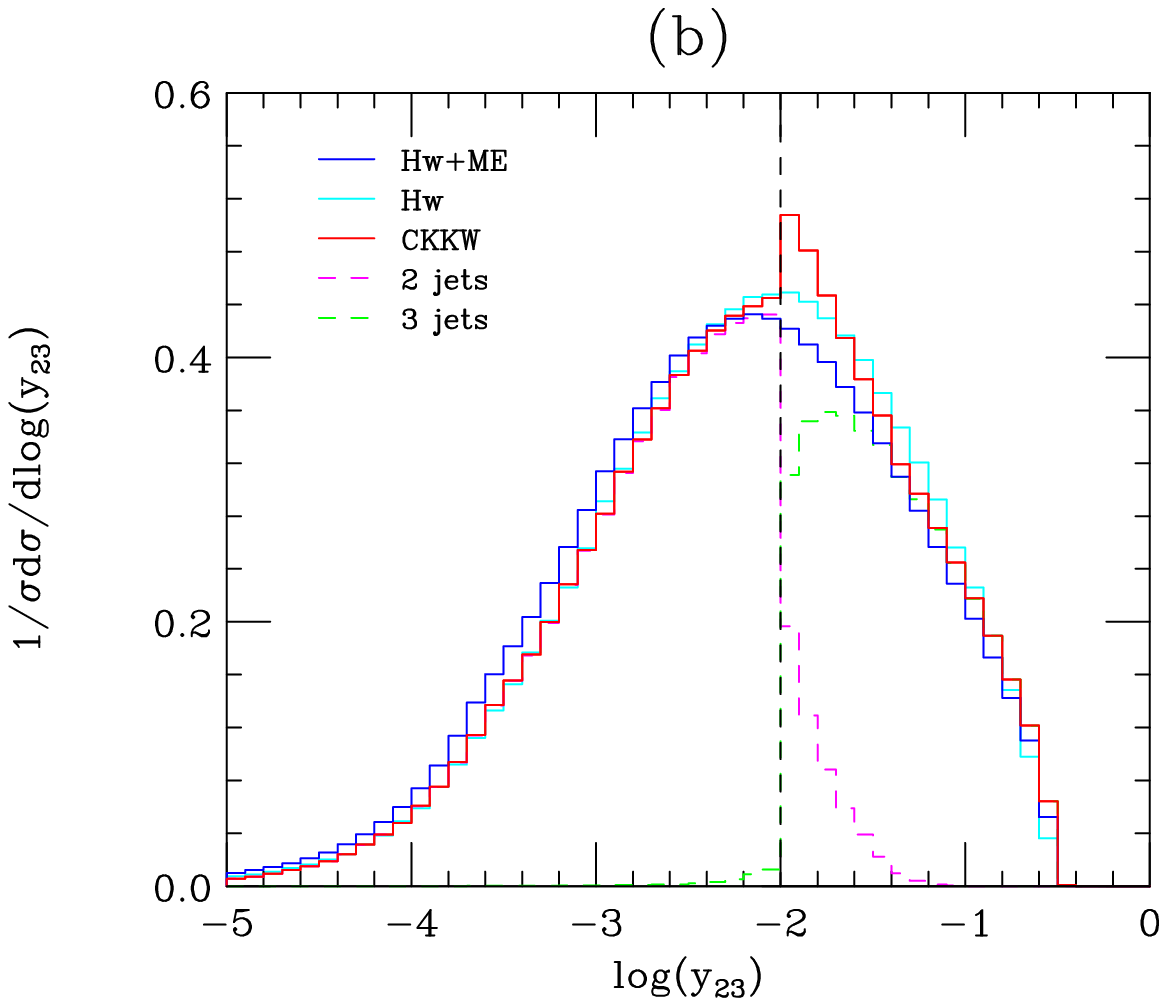}\\
  \par\end{centering}
  \vspace{3mm}
  \begin{centering}
  \includegraphics[width=0.4\textwidth,angle=0]{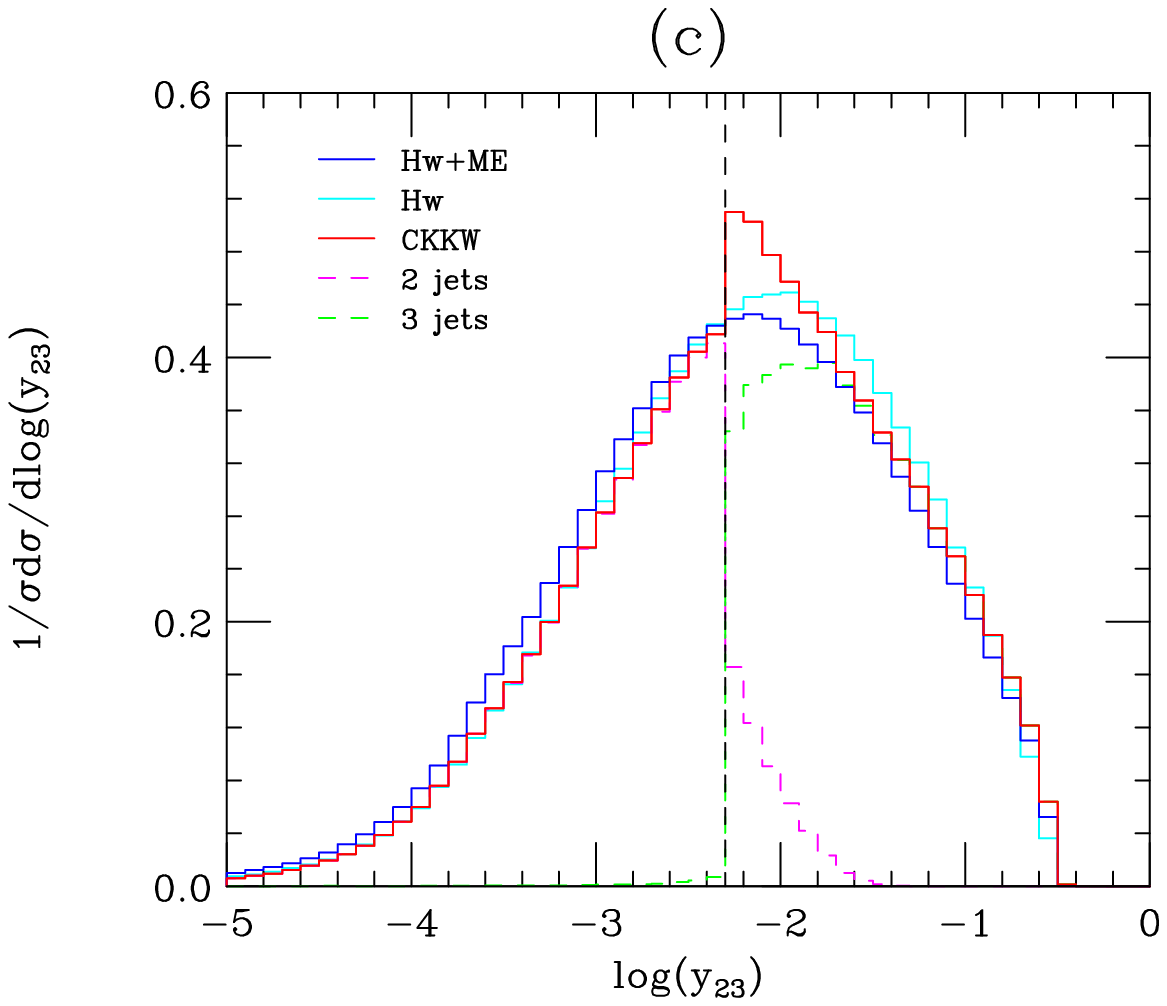}\hspace{3mm}
  \includegraphics[width=0.4\textwidth,angle=0]{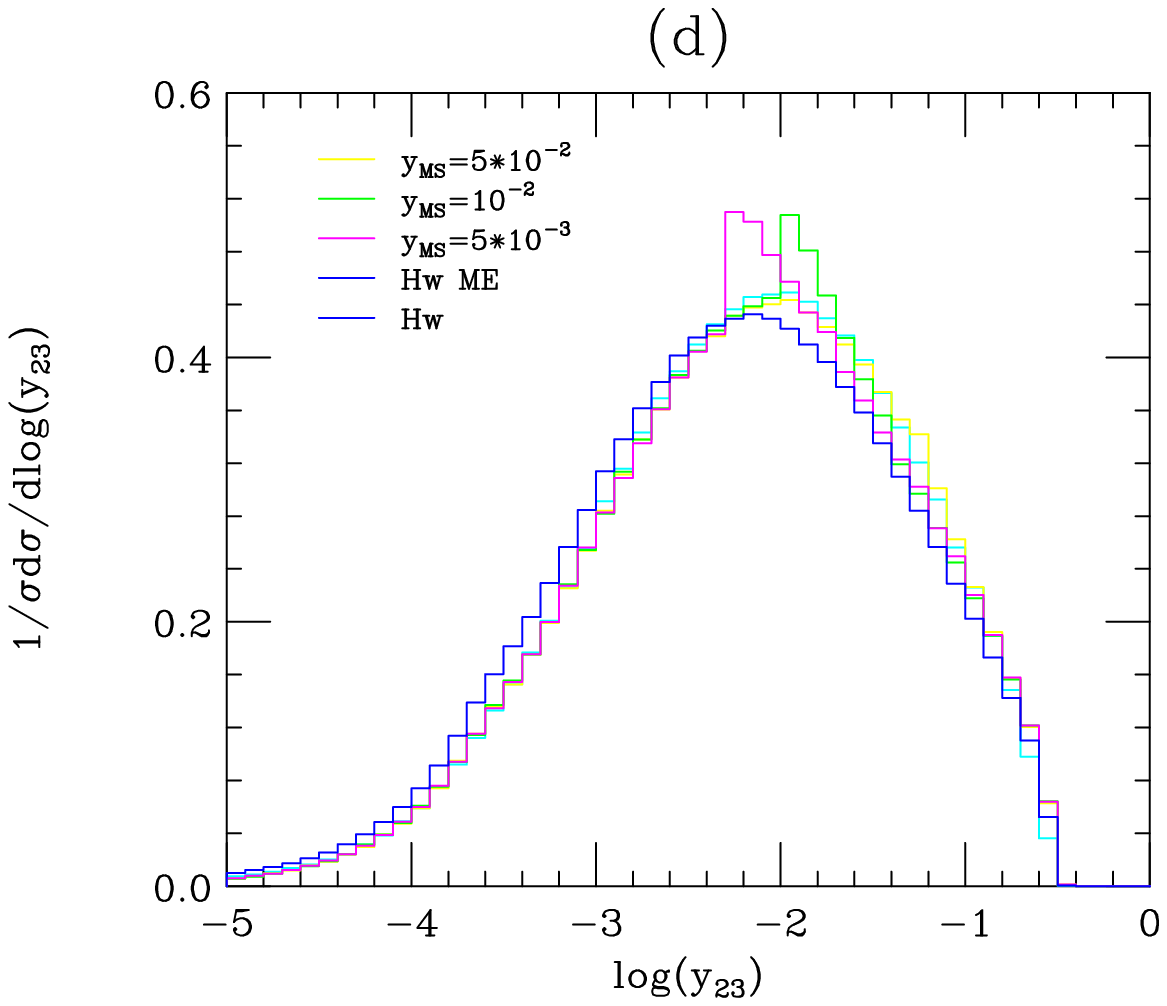} \\
  \par\end{centering}
  \caption{The same distributions as in  Fig.~\protect\ref{fig:23} 
    but with the truncated shower switched off in the CKKW treatment.}
\label{fig:23_no_trunc}
\end{figure}

Figure \ref{fig:23_no_high} shows the same distributions as Fig.~\ref{fig:23} but with the highest multiplicity treatment
switched off.  The result of switching off the highest multiplicity treatment is that a maximum of three emissions
 may be generated in the matrix element region.  This violates the all-orders-in-$\alpha_{S}$ resummation of the parton shower.
The result is a suppression of the three-jet channel and a deficit in the hard radiation generated in 
the three jet channel.
In Fig.~\ref{fig:23_no_high} this presents itself as distributions that are peaked around the merging scale and
have a large dependence on the choice of merging scale.

\begin{figure}
\begin{centering}
\includegraphics[width=0.4\textwidth,angle=0]{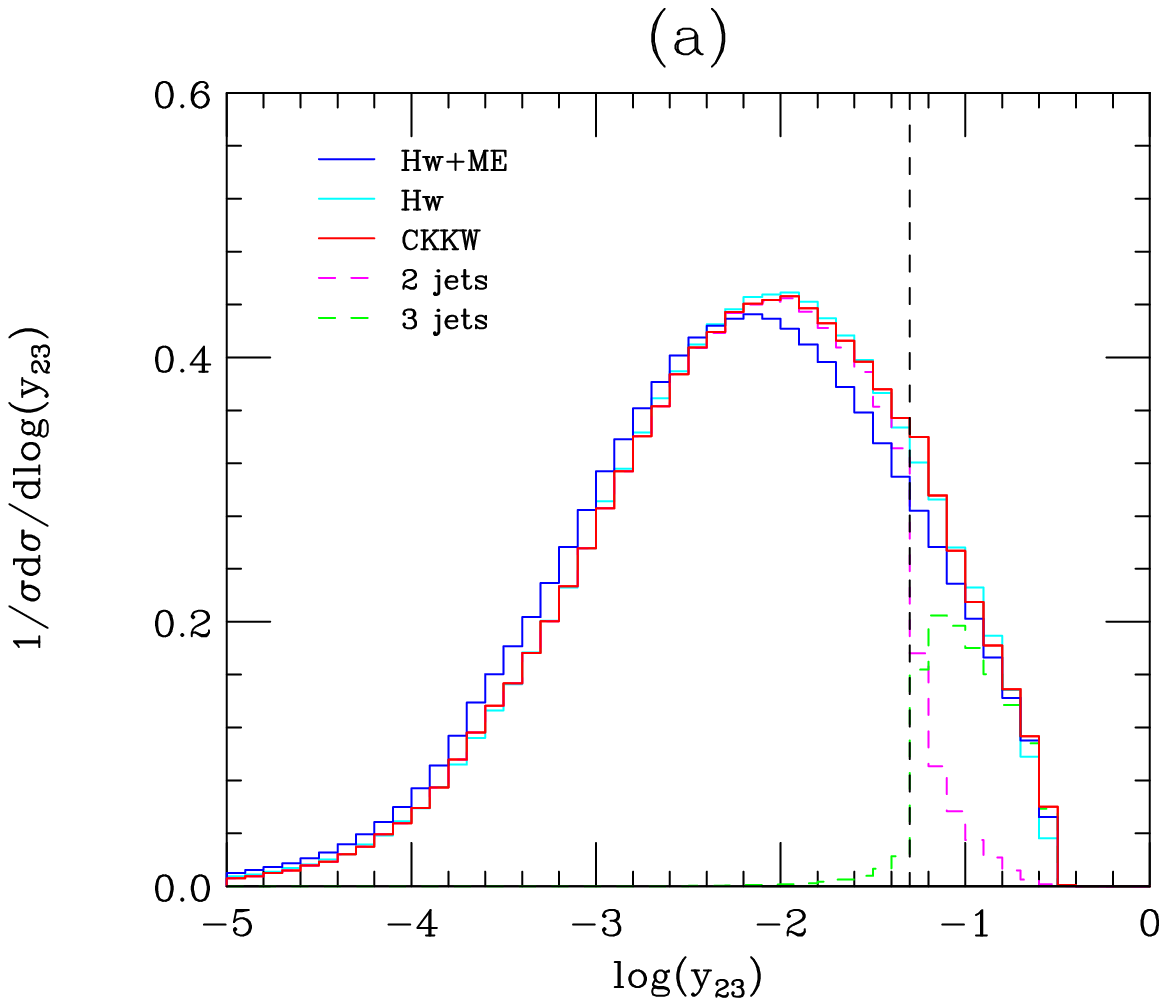}\hspace{3mm}
\includegraphics[width=0.4\textwidth,angle=0]{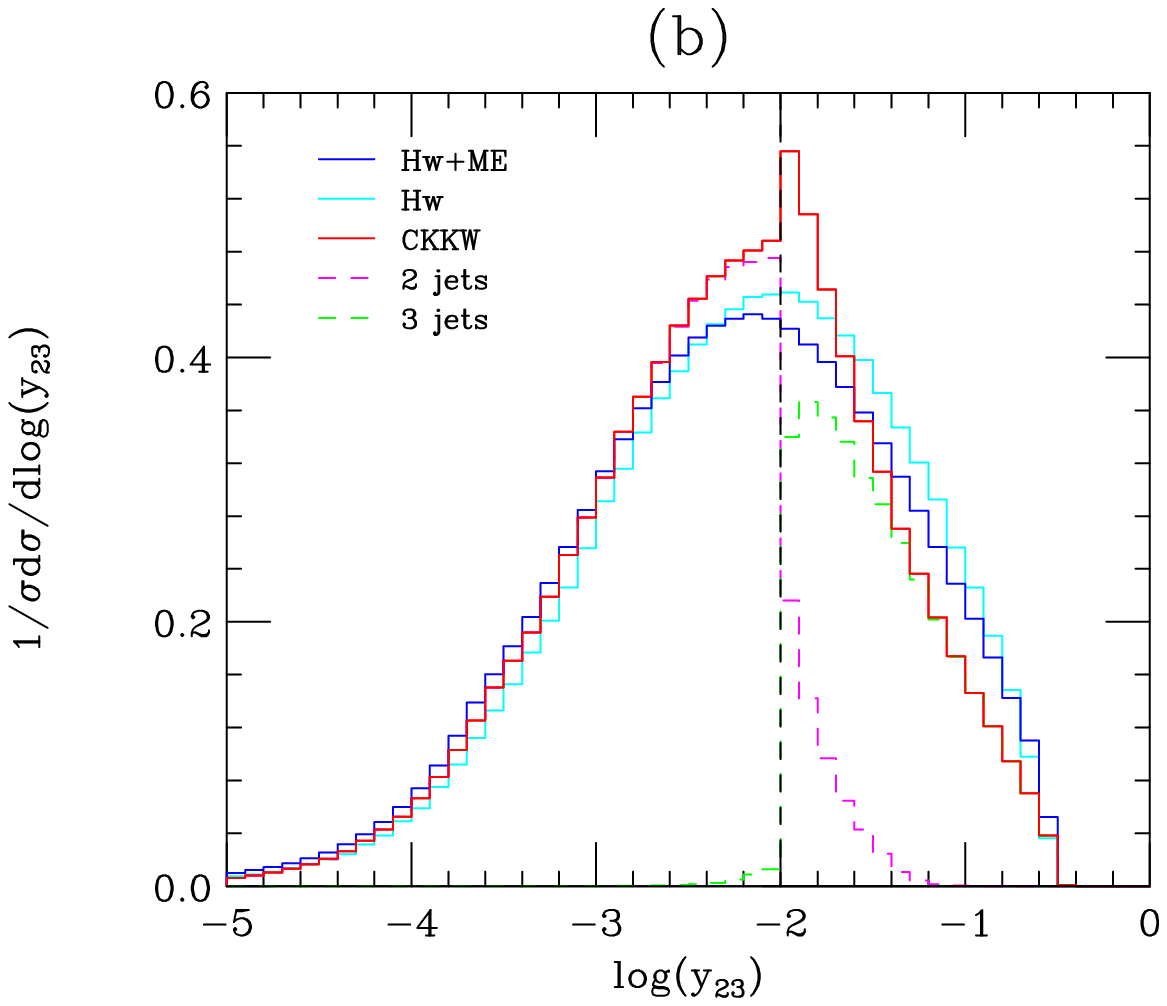}\\
 \vspace{3mm}
 \includegraphics[width=0.4\textwidth,angle=0]{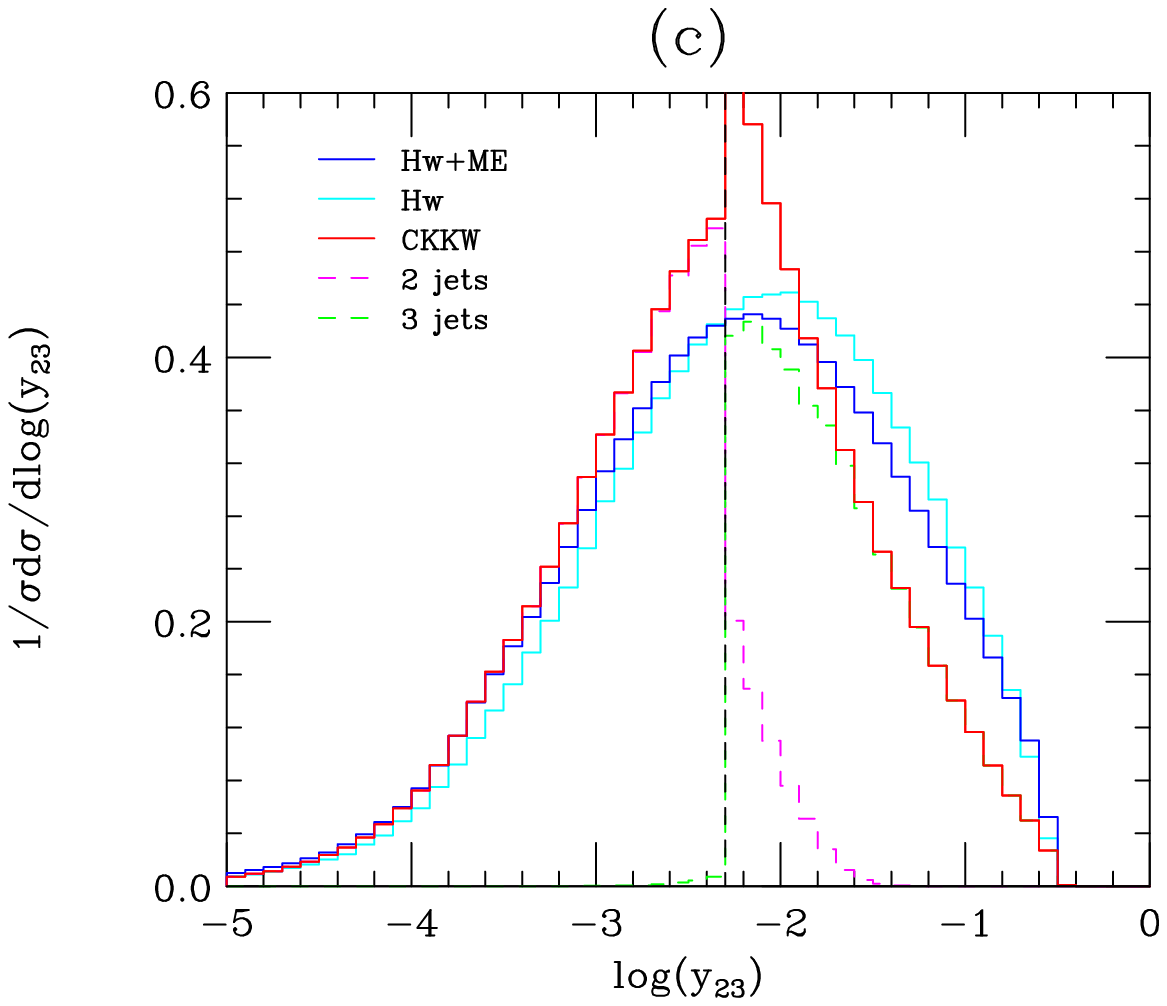}\hspace{3mm}
\includegraphics[width=0.4\textwidth,angle=0]{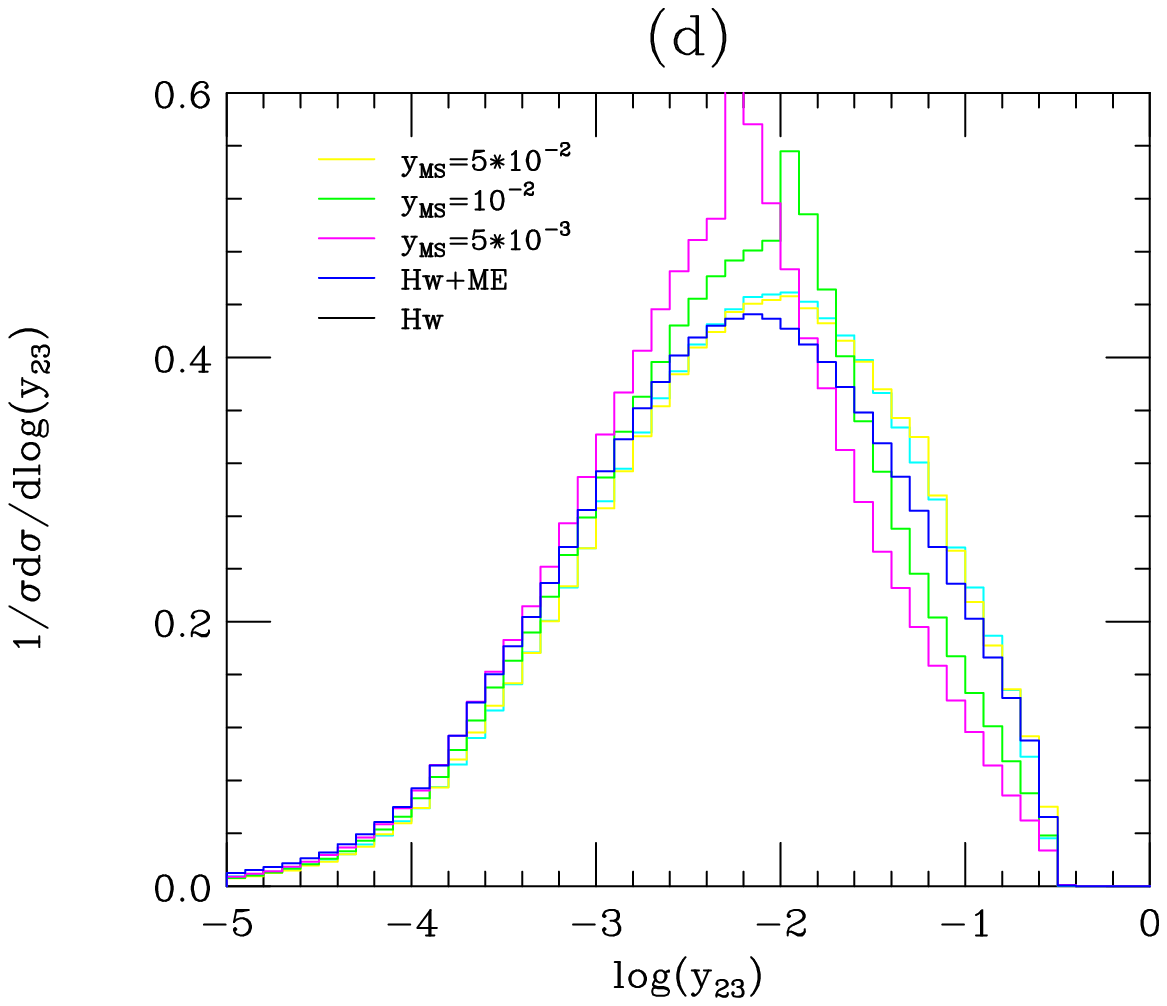} \\
\par\end{centering}
\caption{ The same distributions of Fig.~\protect\ref{fig:23} but with the highest multiplicity treatment switched off in the CKKW treatment. }
\label{fig:23_no_high}
\end{figure}

\begin{figure}
\begin{centering}
\includegraphics[width=0.4\textwidth,angle=0]{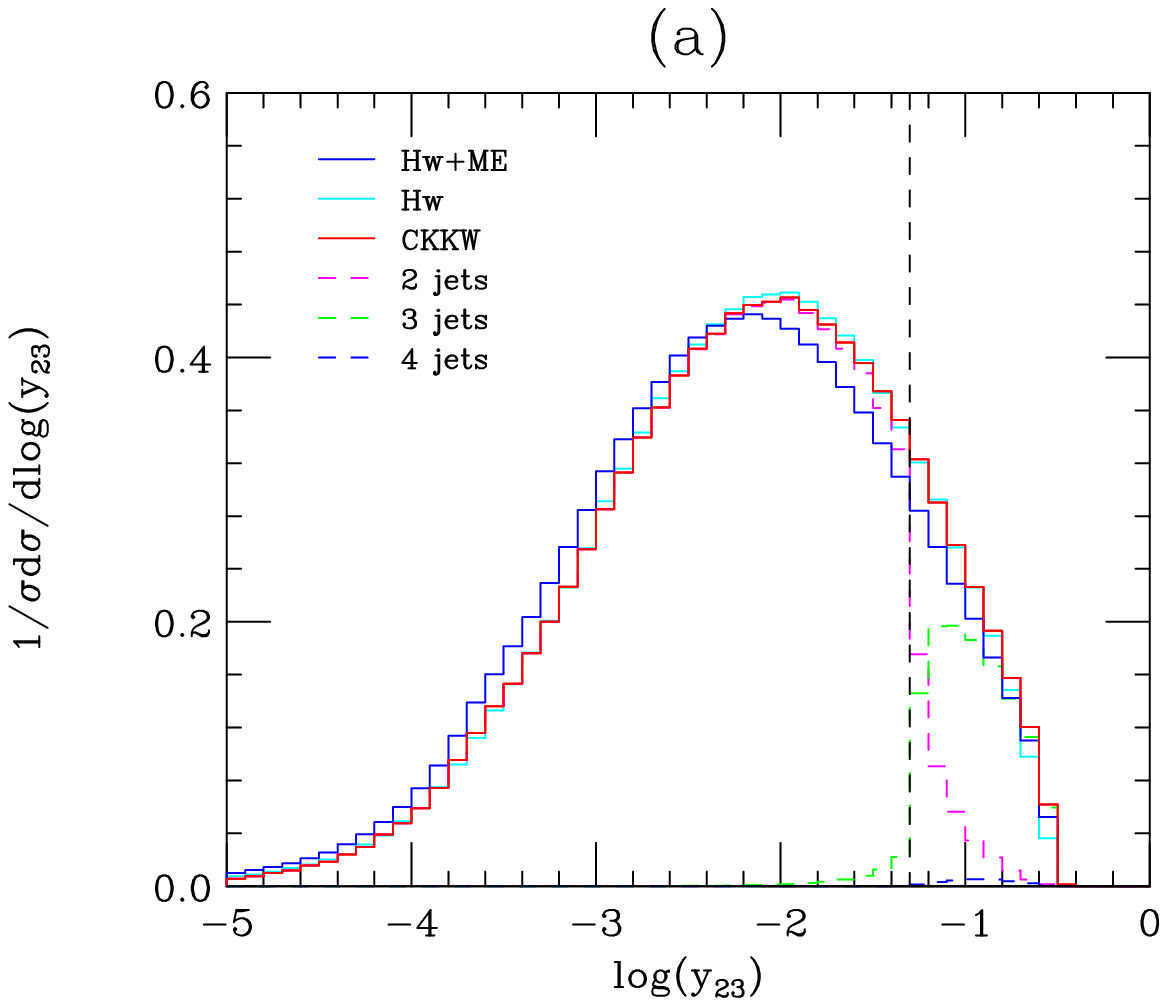}\hspace{3mm}
\includegraphics[width=0.4\textwidth,angle=0]{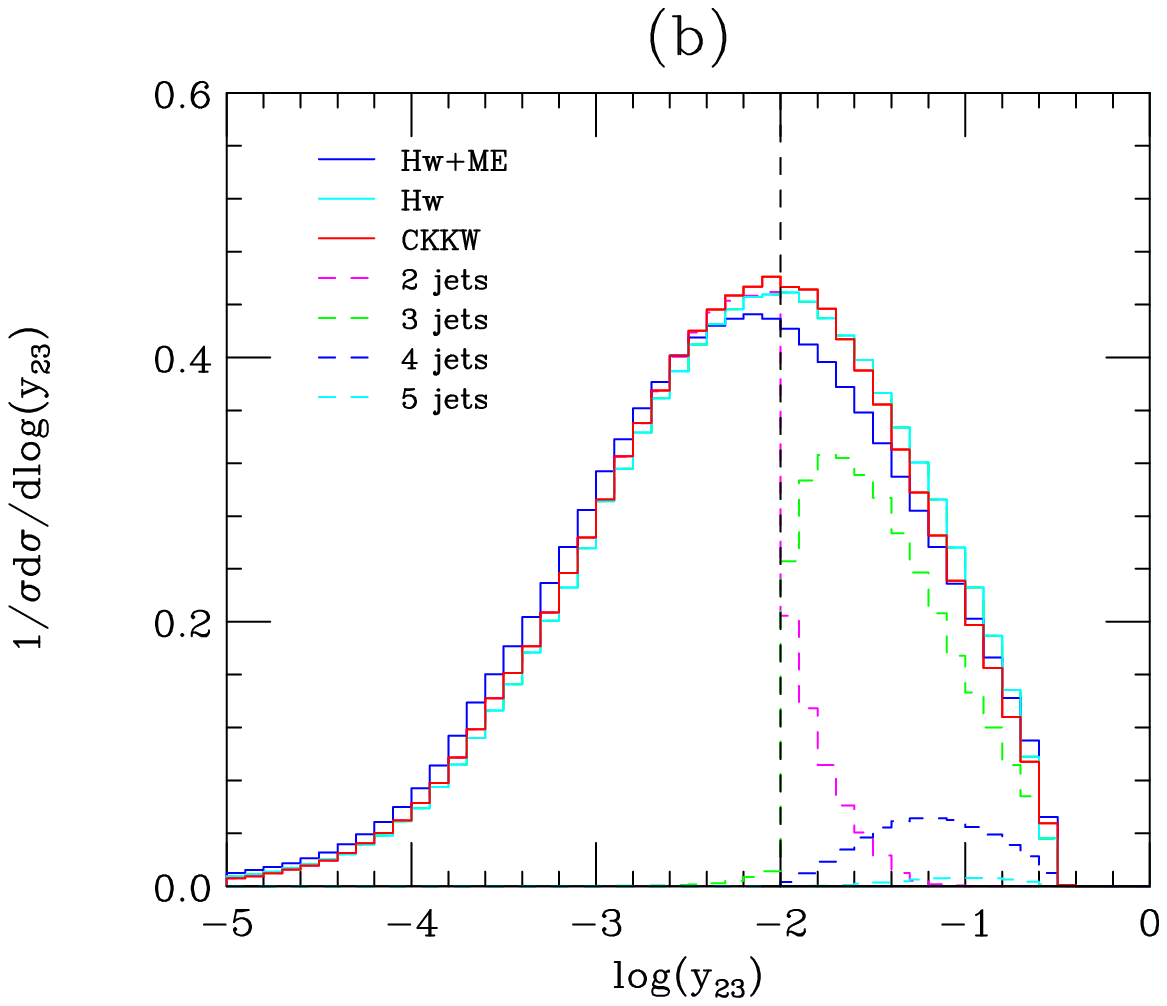}\\
 \vspace{3mm}
 \includegraphics[width=0.4\textwidth,angle=0]{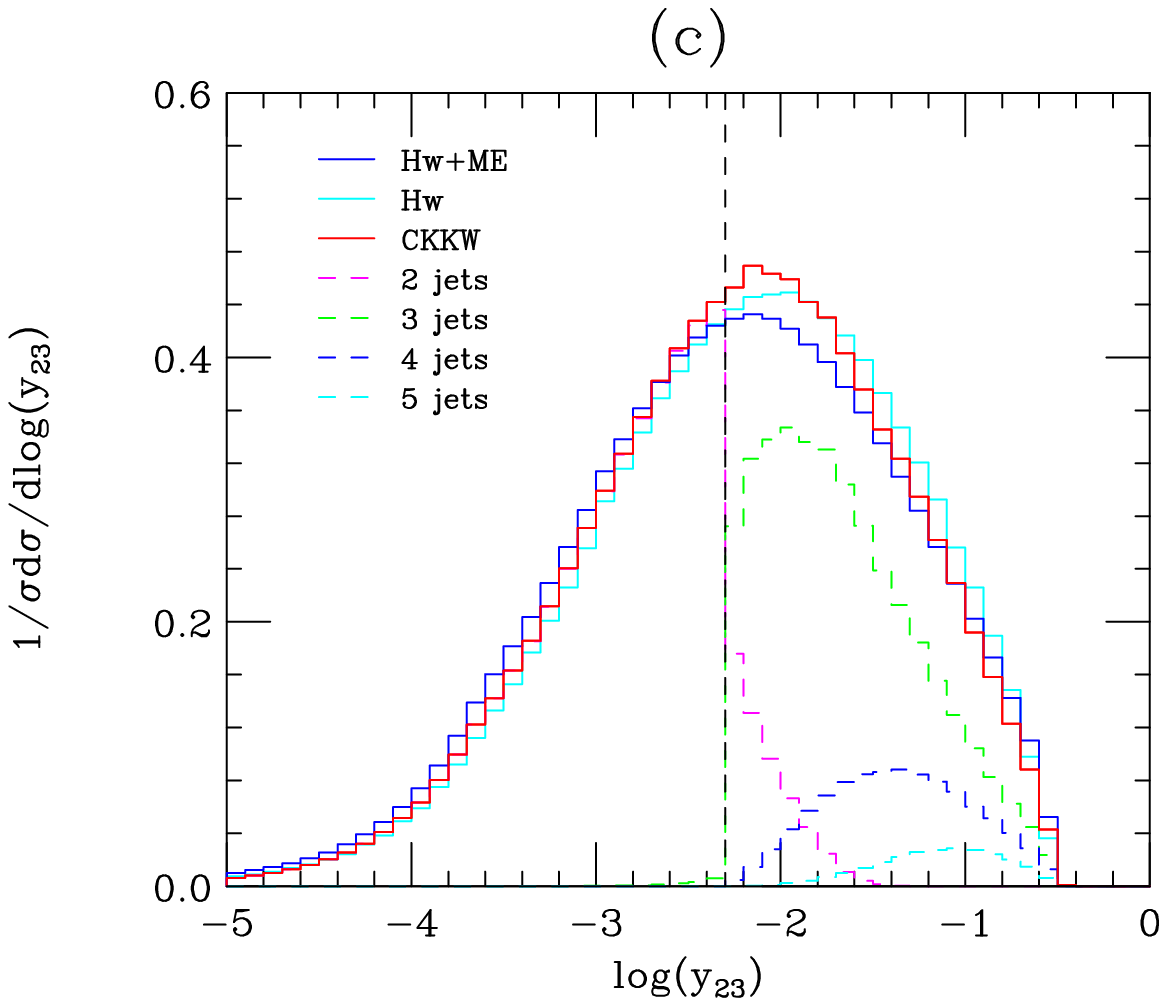}\hspace{3mm}
\includegraphics[width=0.4\textwidth,angle=0]{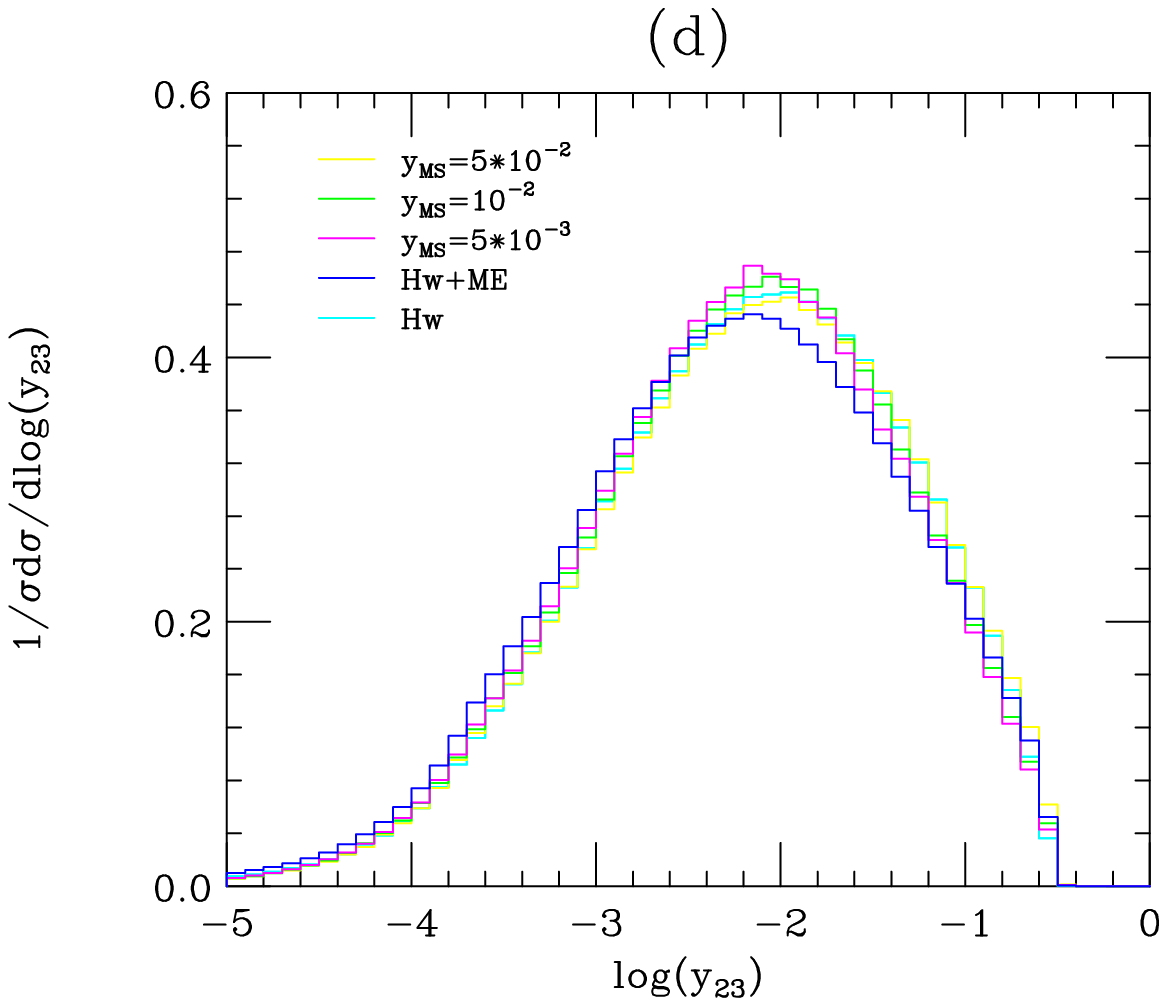} \\
\end{centering}
\caption{Distributions of the scale at which three jets are resolved in the Durham jet measure. The red line shows the distributions for the CKKW treatment with all multiplicity channels (up to a maximum of five jets) included at a set of merging scale choices in the Durham jet measure. }
\label{fig:dur}
\end{figure}

\begin{figure}
\begin{centering}
\includegraphics[width=0.4\textwidth,angle=0]{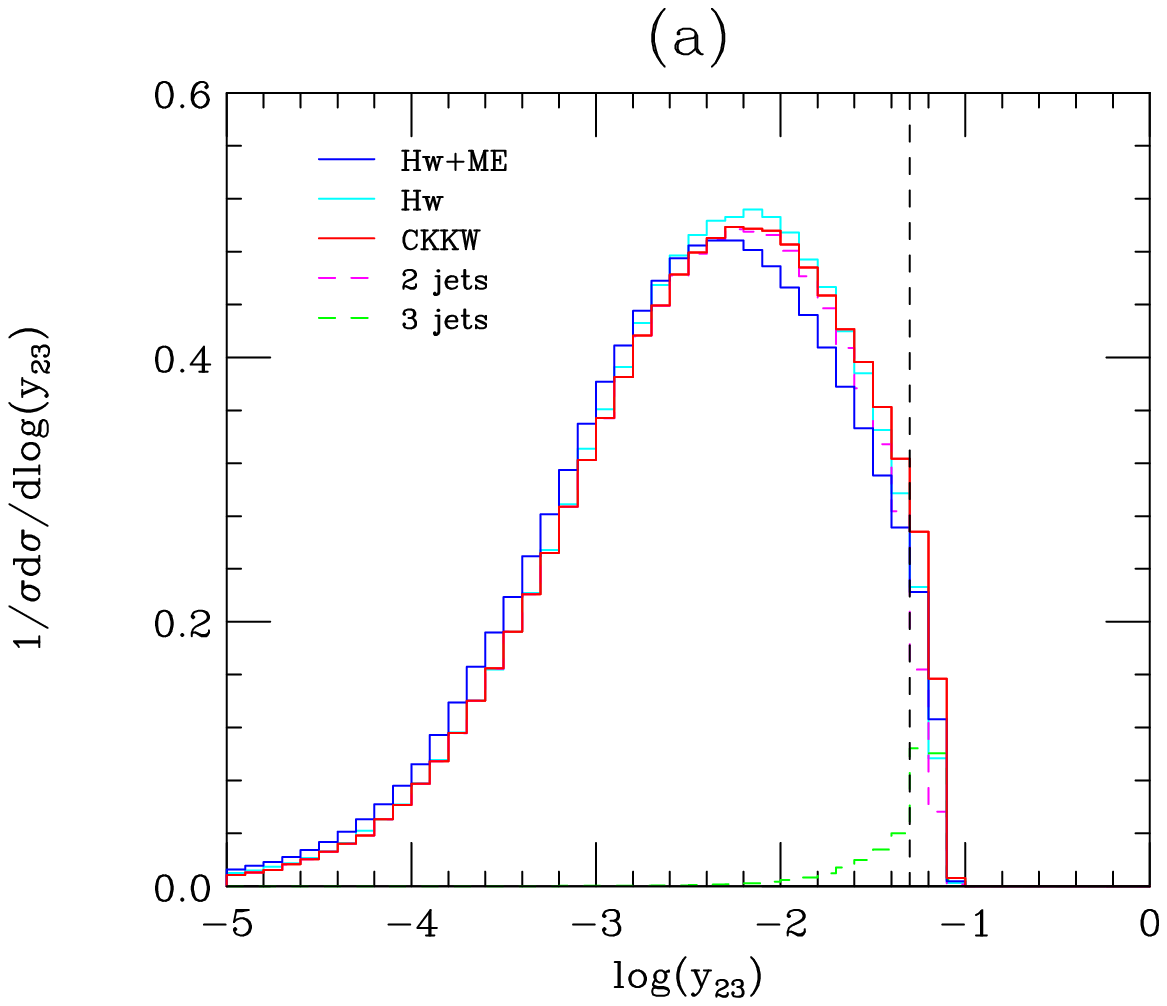}\hspace{3mm}
\includegraphics[width=0.4\textwidth,angle=0]{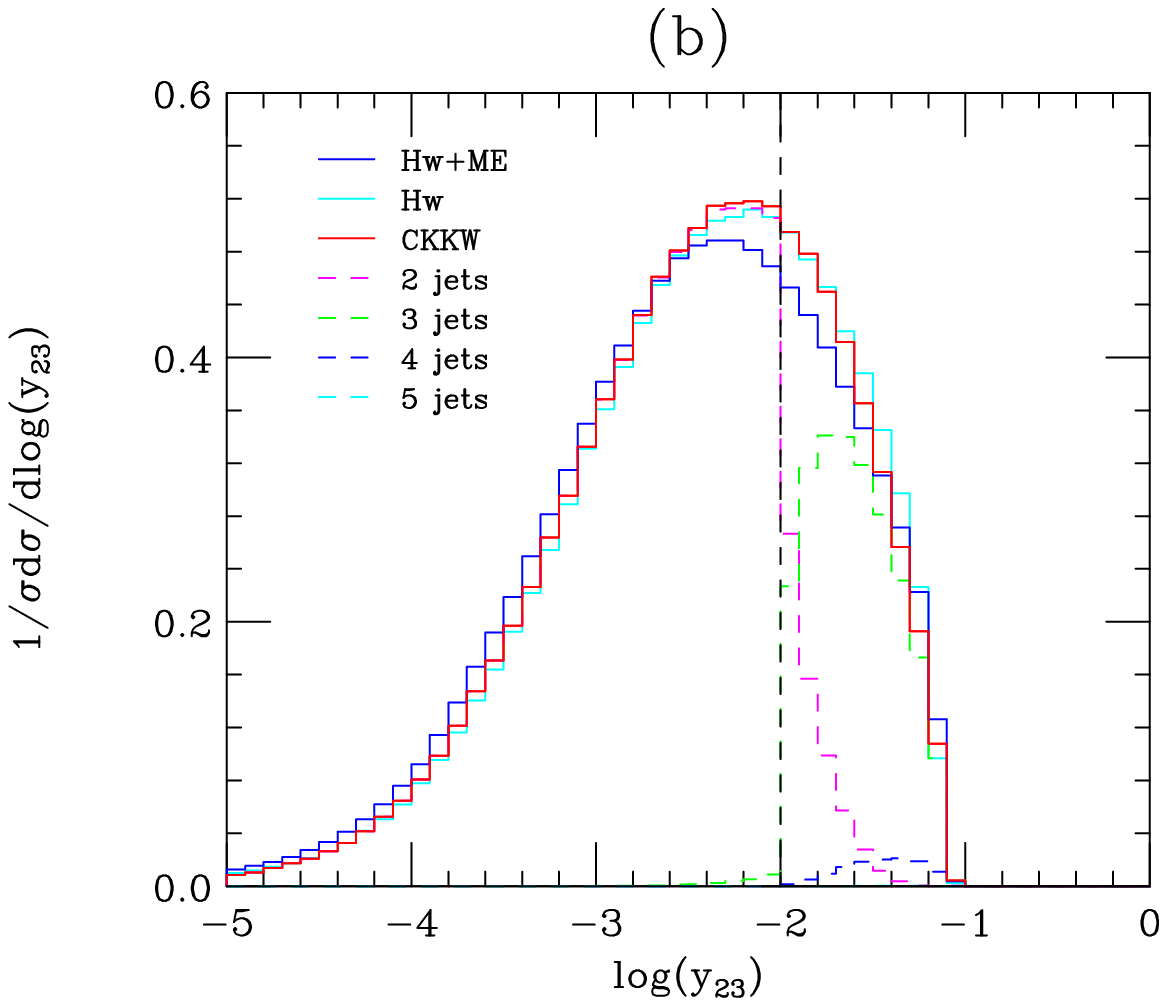}\\
 \vspace{3mm}
 \includegraphics[width=0.4\textwidth,angle=0]{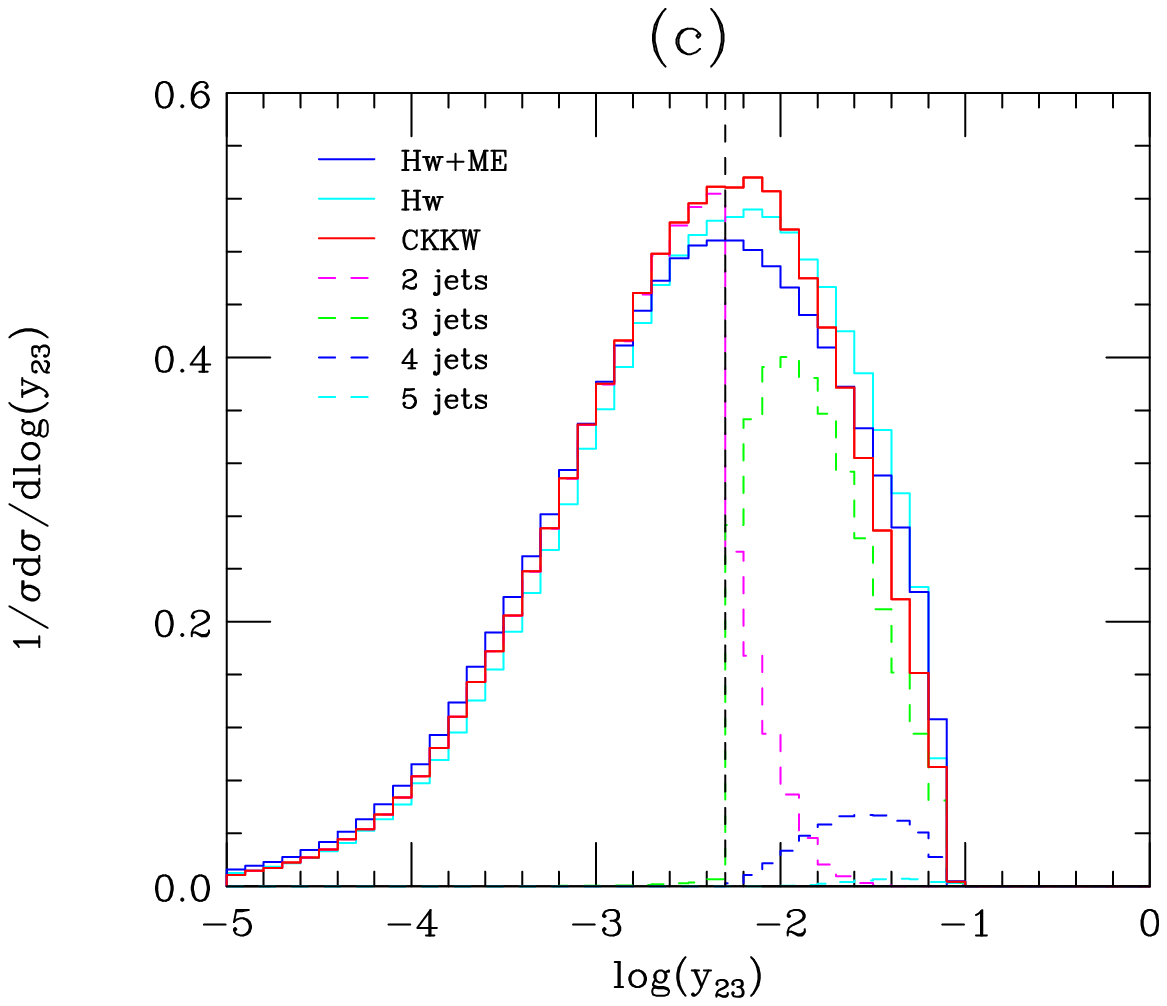}\hspace{3mm}
\includegraphics[width=0.4\textwidth,angle=0]{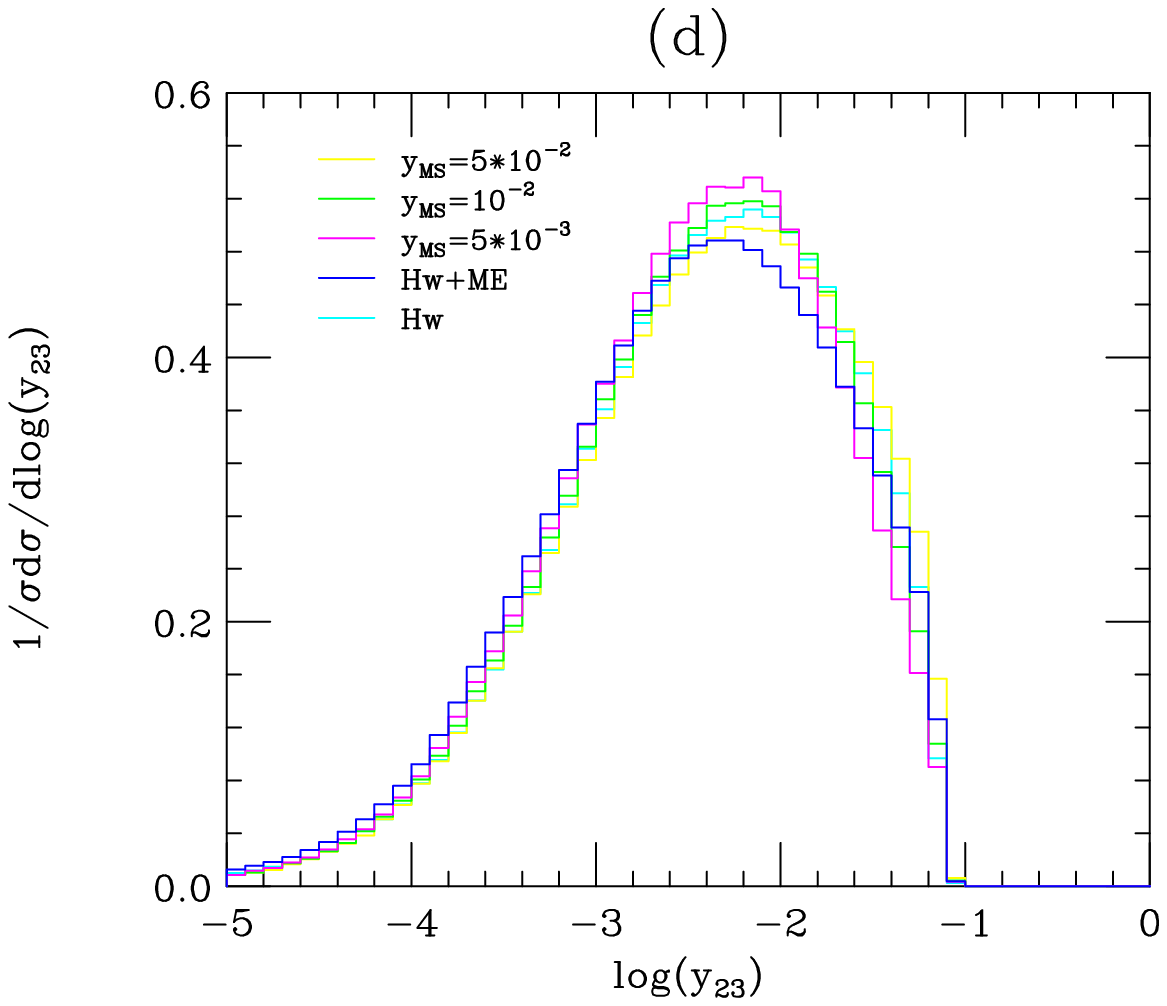} \\
\end{centering}
\caption{Distributions of the scale at which three jets are resolved in the LUCLUS jet measure. The red line shows the distributions for the CKKW treatment with all multiplicity channels (up to a maximum of five jets) included at a set of merging scale choices in the LUCLUS jet measure.}
\label{fig:luc}
\end{figure}

Figures \ref{fig:dur} and \ref{fig:luc} show the distributions of the scale at which three jets are resolved, 
for the  algorithm with maximum multiplicity set to up to five jets, with the merging algorithm defined
in the Durham and LUCLUS jet measures respectively. As in Fig.~\ref{fig:23}, all distributions appear to
be smooth around the merging scale and to be relatively insensitive to the choice of merging scale.  

Since we are now including higher multiplicity channels in our merging algorithm we check
the distributions of scales at which higher numbers of jets are resolved.  This is done in Fig.~\ref{fig:dury45}
for the resolution of four and five jets in the Durham and LUCLUS jet measures.
The merging in these distributions is well behaved.

These distributions demonstrate a degree of insensitivity to the choice of merging scale, which has been varied 
over an order of magnitude, however there is still some residual dependence on this choice.  While the parton
shower and merged matrix element treatments formally have the same large logarithm behavior, there 
are differences between the two.  The degree of these differences will directly influence the amount 
of residual dependence on the merging scale that is observed.  In changing the merging scale we are changing the volume
of the matrix element phase space region and therefore changing the proportion of parton emissions that 
are corrected by exact matrix elements.  

It can be seen in 
Fig.~\ref{fig:dur} and even more so in Fig.~\ref{fig:luc} that when the merging scale is set to its upper value
of $y_{_{MS}}=5\times10^{-2}$, the volume of phase space into which the matrix elements are allowed to emit is small.
This results in the CKKW distributions being very close to those of the \HWPP\ parton shower
with no matrix element correction.  Lowering the merging scale increases the volume of the phase
space region described by exact matrix elements and moves the distributions farther from those of the bare parton shower.
Table \ref{CrossSection} gives the cross sections for the CKKW treatment at different
choices of the merging scale and exhibits variation at the 5\% level.

\begin{figure}
\begin{centering}
 \includegraphics[width=0.4\textwidth,angle=0]{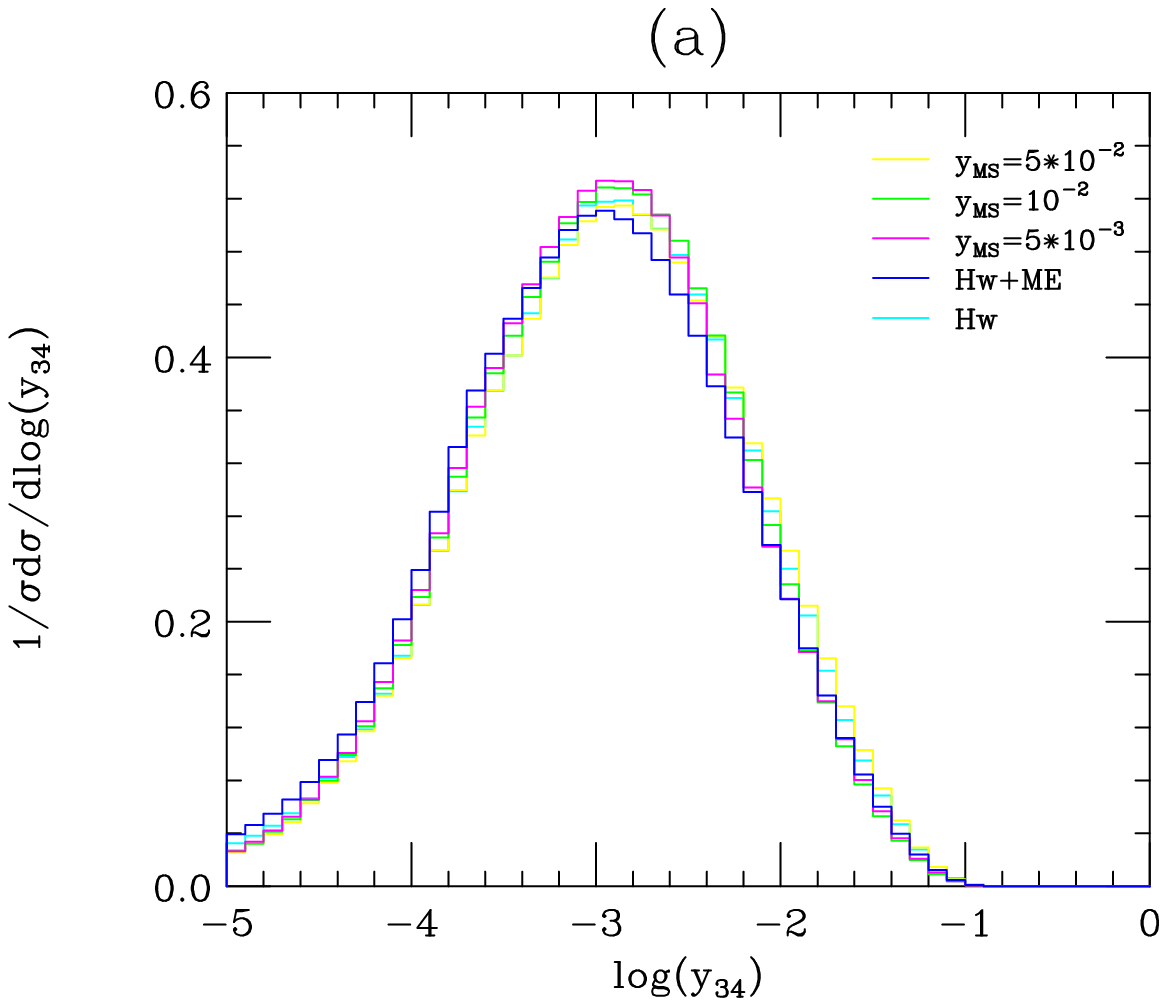}\hspace{3mm}
 \includegraphics[width=0.4\textwidth,angle=0]{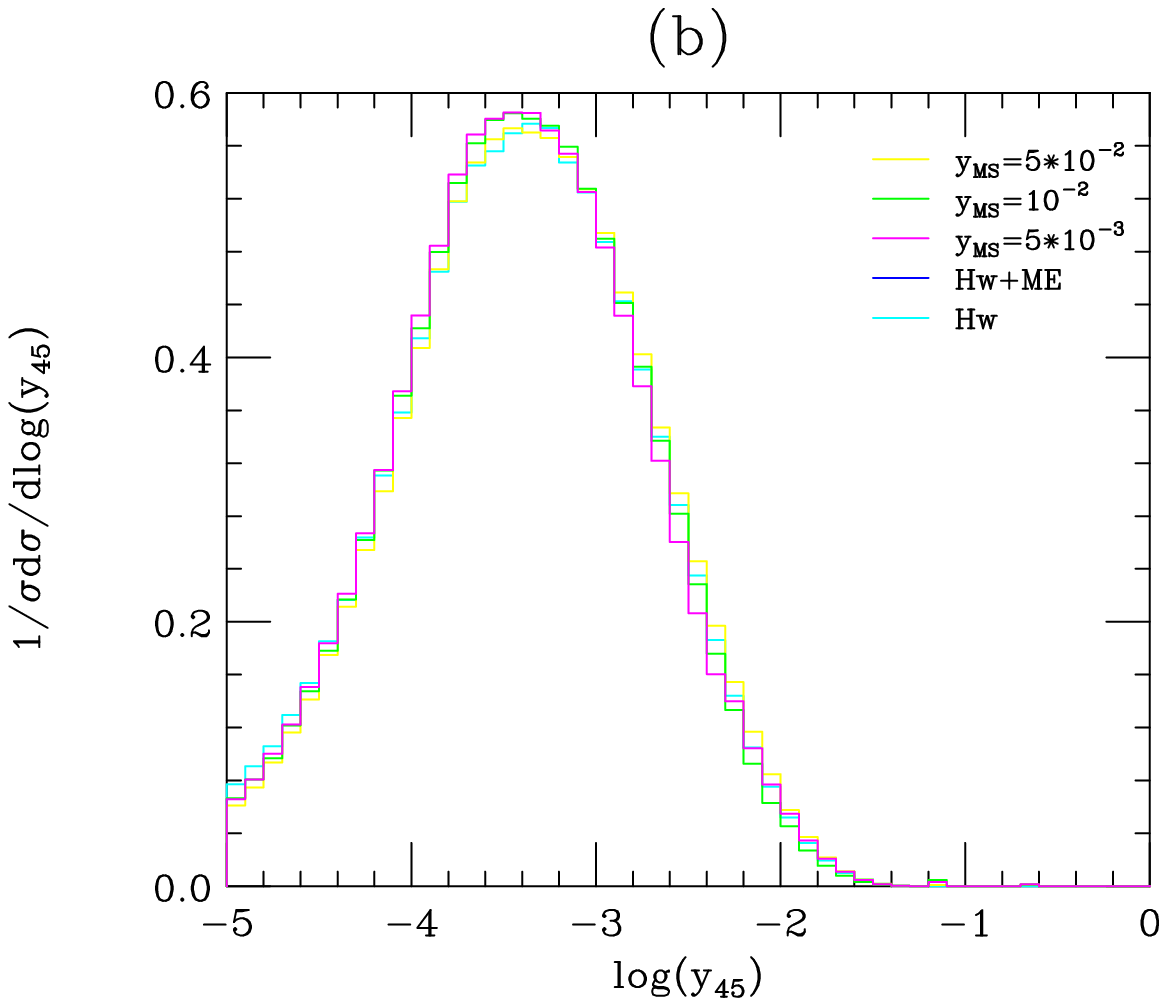}\\
  \vspace{3mm}
 \includegraphics[width=0.4\textwidth,angle=0]{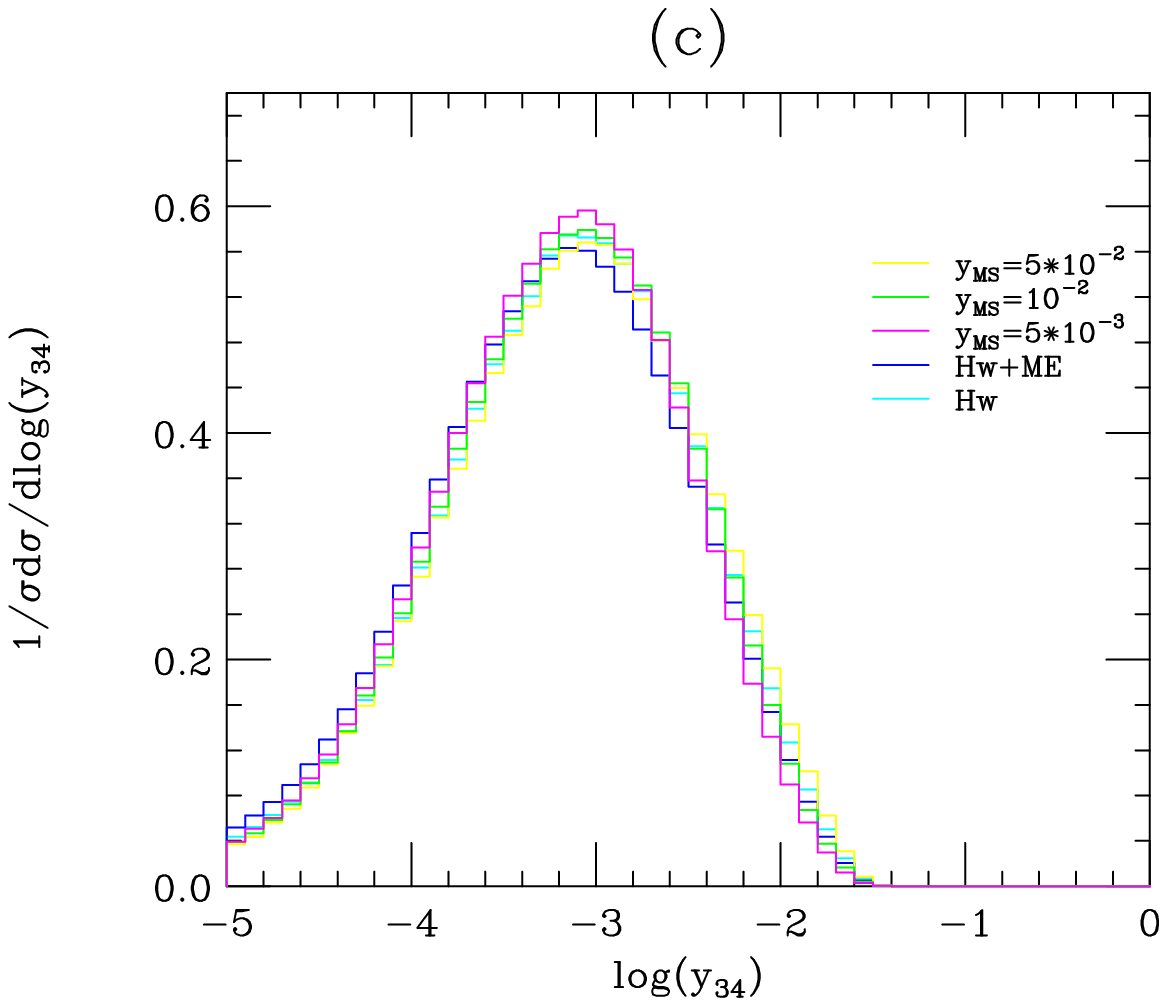}\hspace{3mm}
 \includegraphics[width=0.4\textwidth,angle=0]{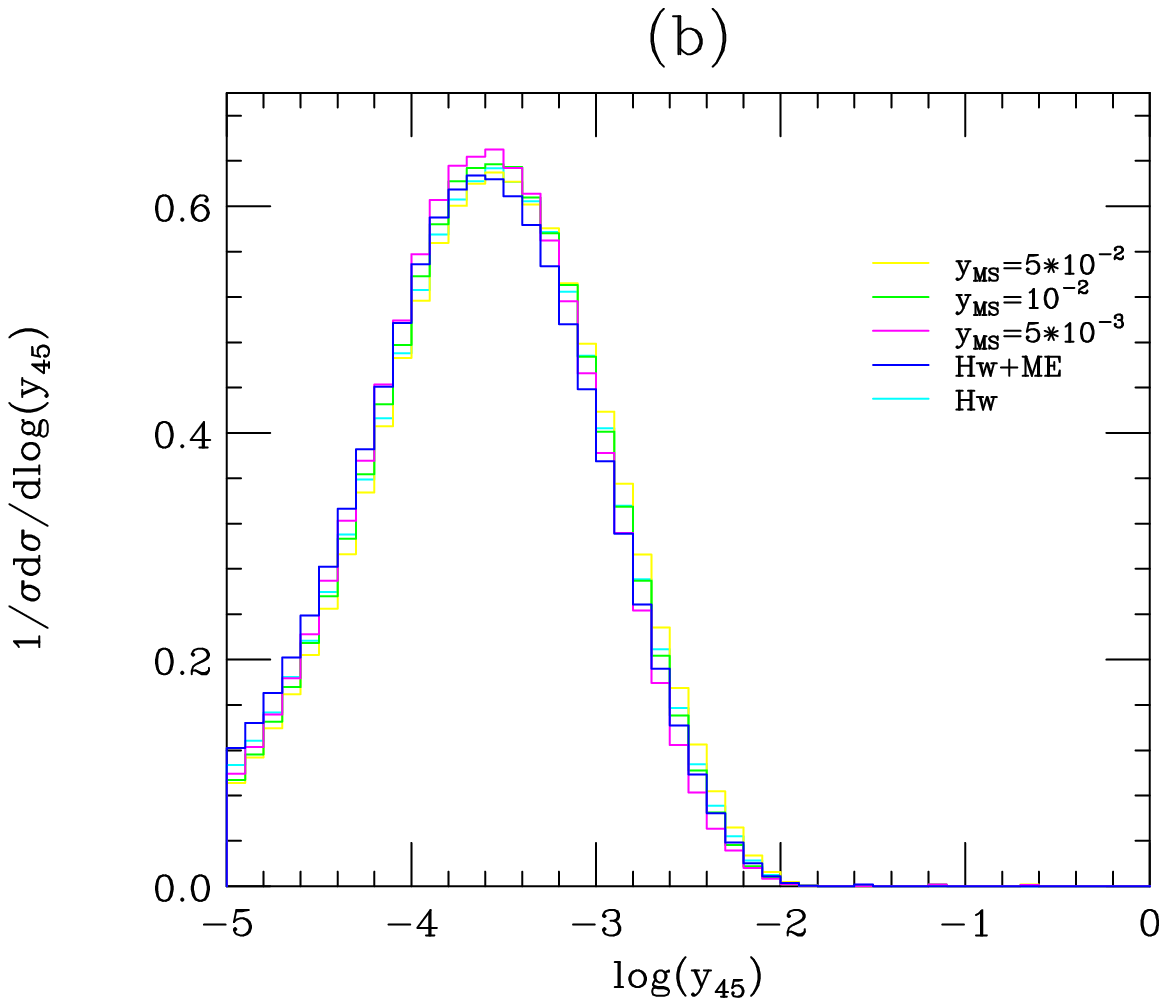} \\
\par\end{centering}
\caption{Distributions of the scale at which (a) four and (b) five jets are resolved in the Durham jet measure 
  and the resolution scales for (c) four and (d) five jets in the LUCLUS jet measure.}
\label{fig:dury45}
\end{figure}

\TABLE{
  \begin{tabular}{|l|c|c|}
    \hline
    $y_{_{MS}}$      &	Durham cross section / nb  & LUCLUS cross section / nb \\ \hline
    $5\times10^{-2}$ &	38.2			   &	38.6                   \\
    $10^{-2}$        &	36.5			   &	37.1                   \\
    $5\times10^{-3}$ &	35.7			   &	35.9                   \\
    \hline
  \end{tabular}
  \caption{Table of cross sections of the process \mbox{$e^{+}e^{-}\to\mathrm{hadrons}$} for different choices of the 
    merging scale in the Durham and LUCLUS jet measures.}
  \label{CrossSection}
}

\subsection{Hadron level results}

We present a comparison of the \HWPP\ CKKW implementation with hadronization switched on to LEP data 
for a variety of event shapes.
It is standard practice to \textit{tune} the free parameters of an event generator to LEP data and this
has been done with the default \HWPP\ parton shower with matrix element corrections.  Since the CKKW
merging algorithm significantly changes the parton shower component of the event generator and in order
to provide a fair comparison with default \HWPP, a new tune was performed on the free parameters for 
\HWPP\ with the CKKW algorithm. This tune was performed with the merging scale set to $y_{_{MS}}=10^{-2}$ in 
the Durham jet measure.
The most notable change to the default \HWPP\ parameter set was a change in the value of the strong coupling
at $Z$ mass from a value of $\alpha_{S}(M_{z})=0.128$ for \HWPP\ with matrix elements corrections down to
$\alpha_{S}(M_{z})=0.120$ for \HWPP\ with CKKW.  This change brings the parameter closer to its measured value
and is indicative of an improved treatment of hard radiation.

Figures \ref{fig:LEPEvent}-\ref{fig:LEPFourJet} show distributions of a range of
event shape, jet resolution and four-jet observables in comparison to LEP data.  The parton level analysis
shows that the merging scale choice of $y_{_{MS}}=5\times10^{-2}$ leaves only a very small region of phase space
that is corrected by the matrix elements.  This very high scale choice will therefore not
 give the improvement expected in introducing the merging algorithm, we therefore omit this merging scale choice
from the hadron level analysis.  In each of the figures the red band shows the variation in distributions
over the four merging scale choices of $y_{_{MS}}=10^{-2}$ and $y_{_{MS}}=5\times10^{-3}$ in the Durham and LUCLUS
jet measures.

The CKKW distributions (red band) in Figs.~\ref{fig:LEPEvent}-\ref{fig:LEPFourJet} all demonstrate 
improved descriptions of the data in comparison to the default \HWPP\ parton 
shower with matrix element corrections.  In particular the tails of the distributions in Fig.~\ref{fig:LEPEvent}, 
corresponding to hard emissions, and the jet resolution distributions of Fig.~\ref{fig:LEPJet} with 
four and five jets are significantly improved as would be expected given the aims of the merging algorithm.  
The four-jet angle distributions of Fig.~\ref{fig:LEPFourJet} are also all improved, with the exception
of the $\alpha_{34}$ angle, which was already well described by the default \HWPP\ parton shower.  The
$\theta_{NR}$ distribution provides the most notable improvement in its description of the data in comparison
to the default \HWPP\ parton shower.

The width of the red band on the distributions shows that there is some residual dependence on the merging scale 
however it does not appear to be too serious and is at a similar level to that observed at parton level.  
This shows that the problems with colour structure, that appear in the standard CKKW algorithm, 
are not present here and that the truncated shower is working as intended.  It should be noted that a fixed
set of \HWPP\ shower and hadronization parameters was used for each of the four merging scale choices; 
the variation would be reduced further if 
a tune of the parameters was performed for each merging scale choice.

The $\chi^{2}$ per degree of freedom values for the distributions in Figs.~\ref{fig:LEPEvent}-\ref{fig:LEPFourJet} 
are given in Table~\ref{Chis} for the merging scale choice of $y_{_{MS}}=10^{-2}$ in the Durham jet measure, which 
was used in the tune.  The CKKW values are lower than those of the default \HWPP\ shower in all cases except for 
the $\alpha_{34}$ angle, where the default implementation already gave an accurate description, and in many cases the CKKW 
values are significantly lower.

\begin{figure}
\begin{centering}
\includegraphics[width=0.4\textwidth,angle=0]{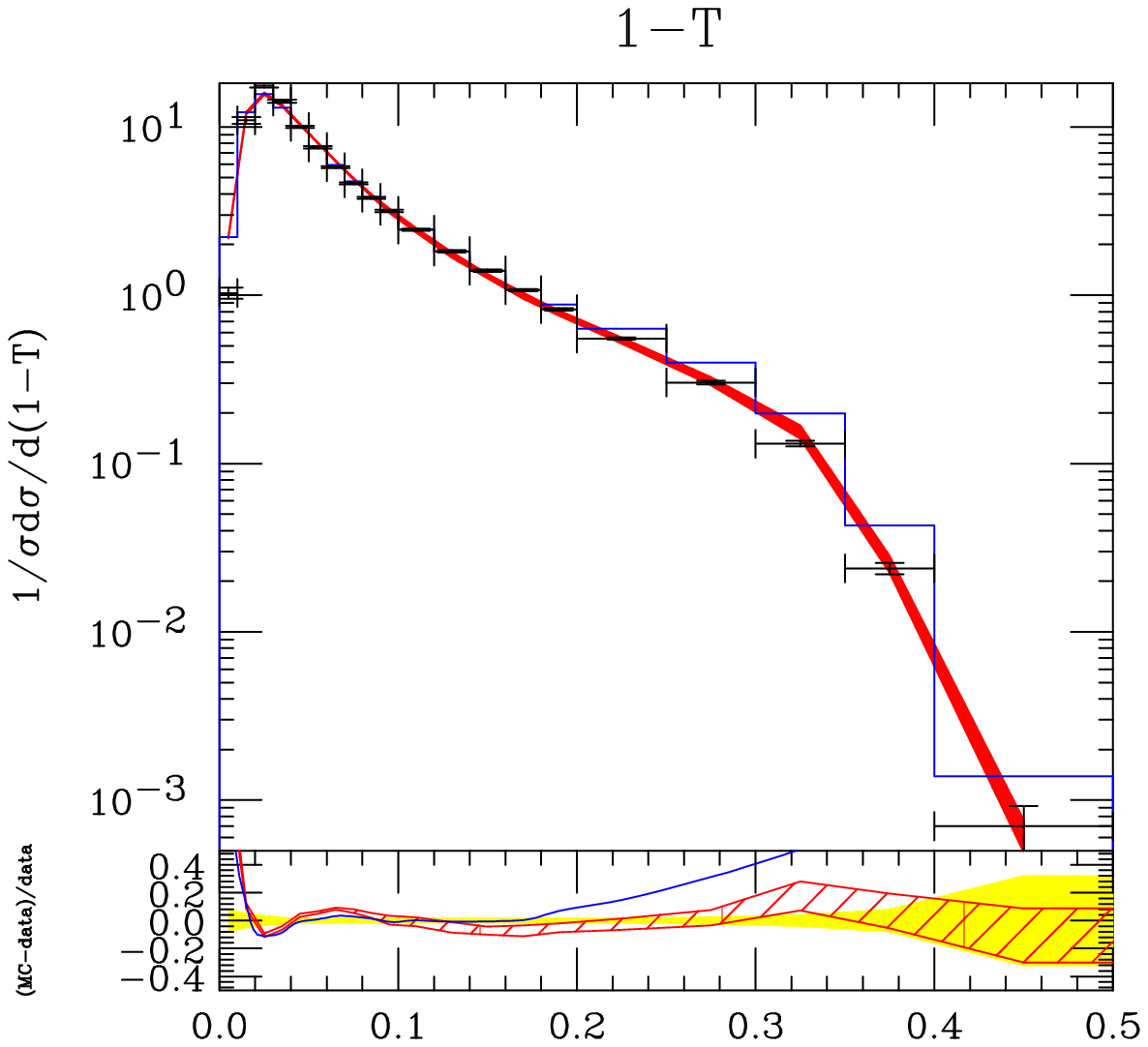}\hspace{3mm}
\includegraphics[width=0.4\textwidth,angle=0]{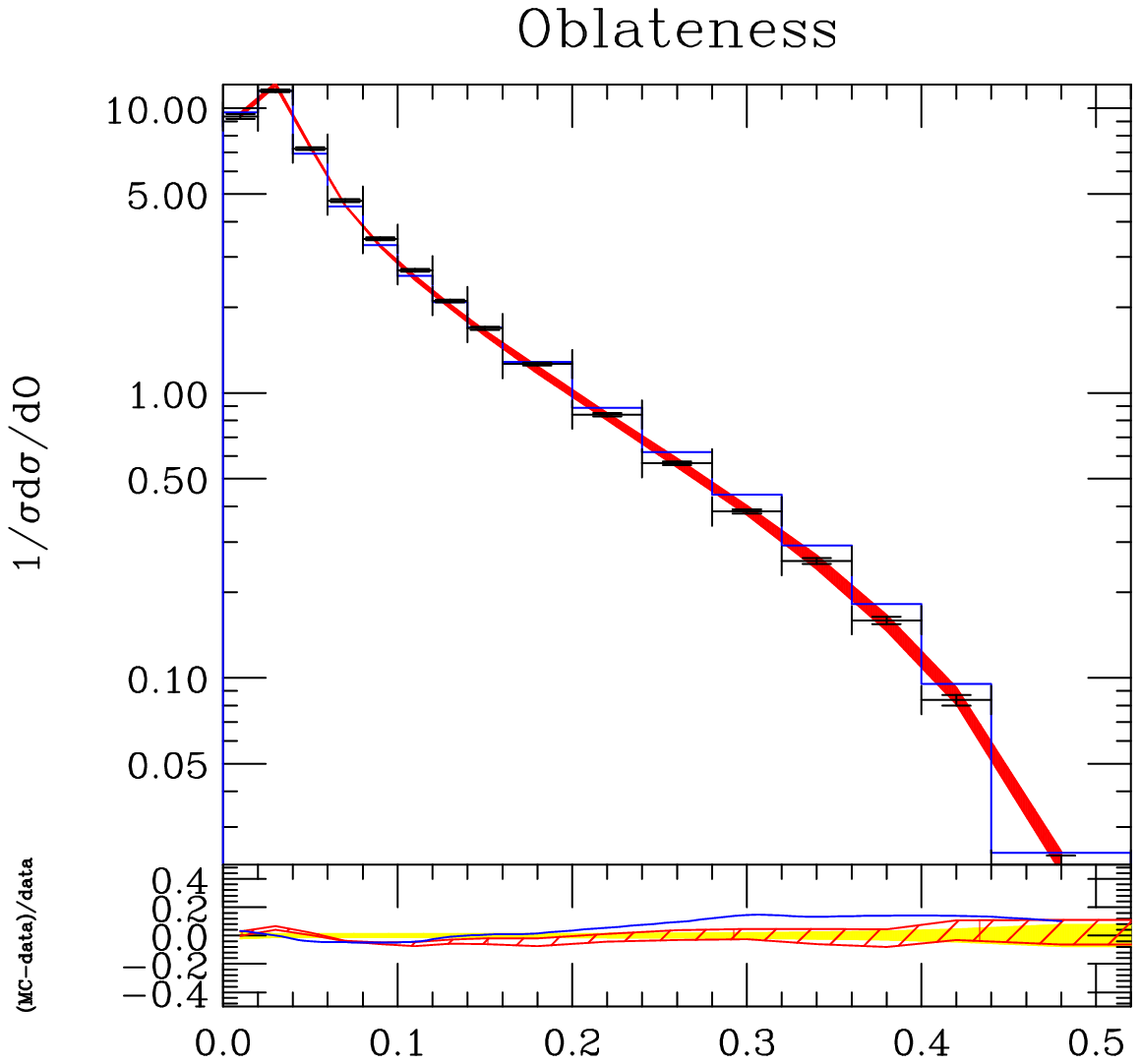} \\
 \vspace{3mm}
 \includegraphics[width=0.4\textwidth,angle=0]{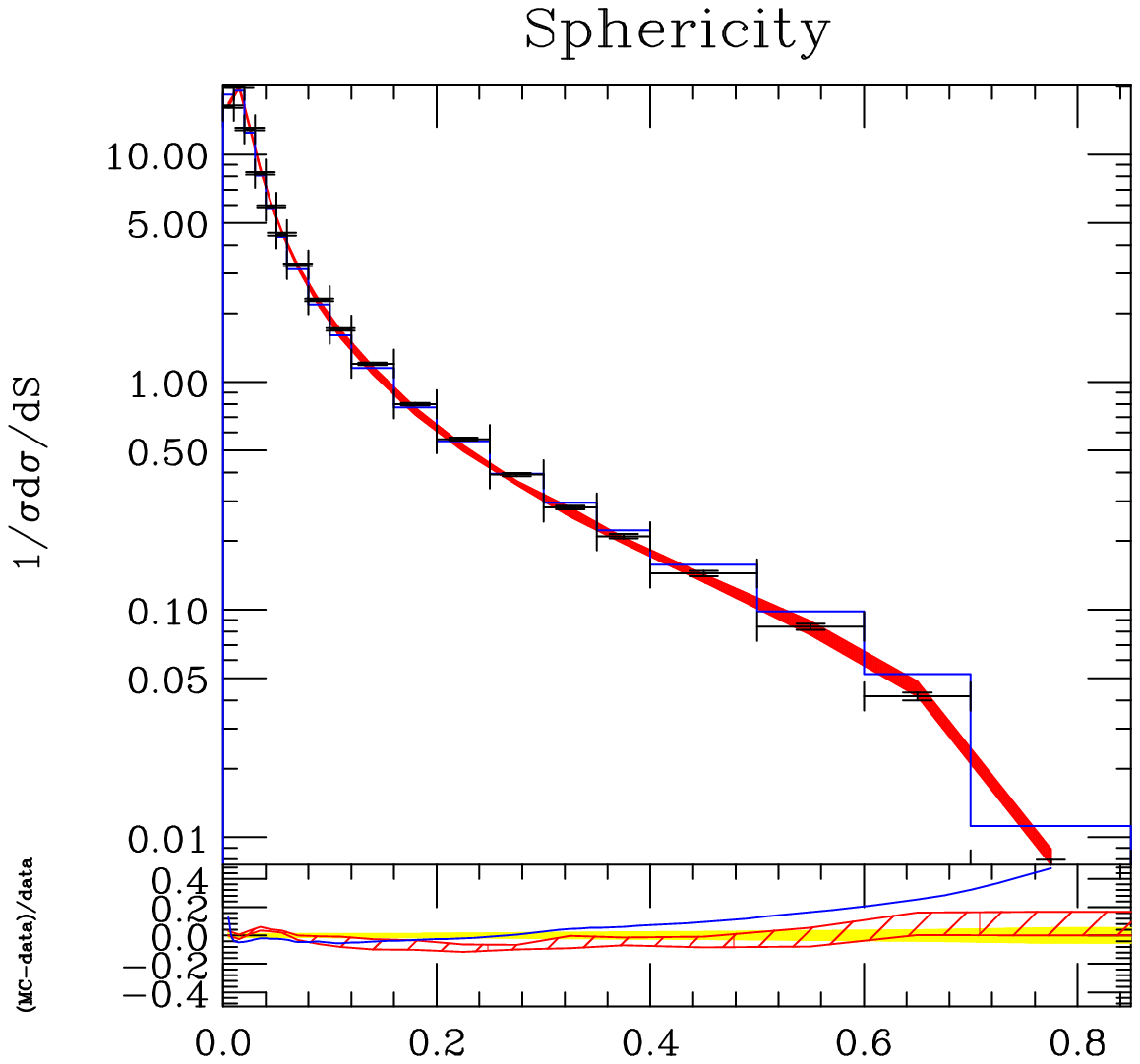}\hspace{3mm}
\includegraphics[width=0.4\textwidth,angle=0]{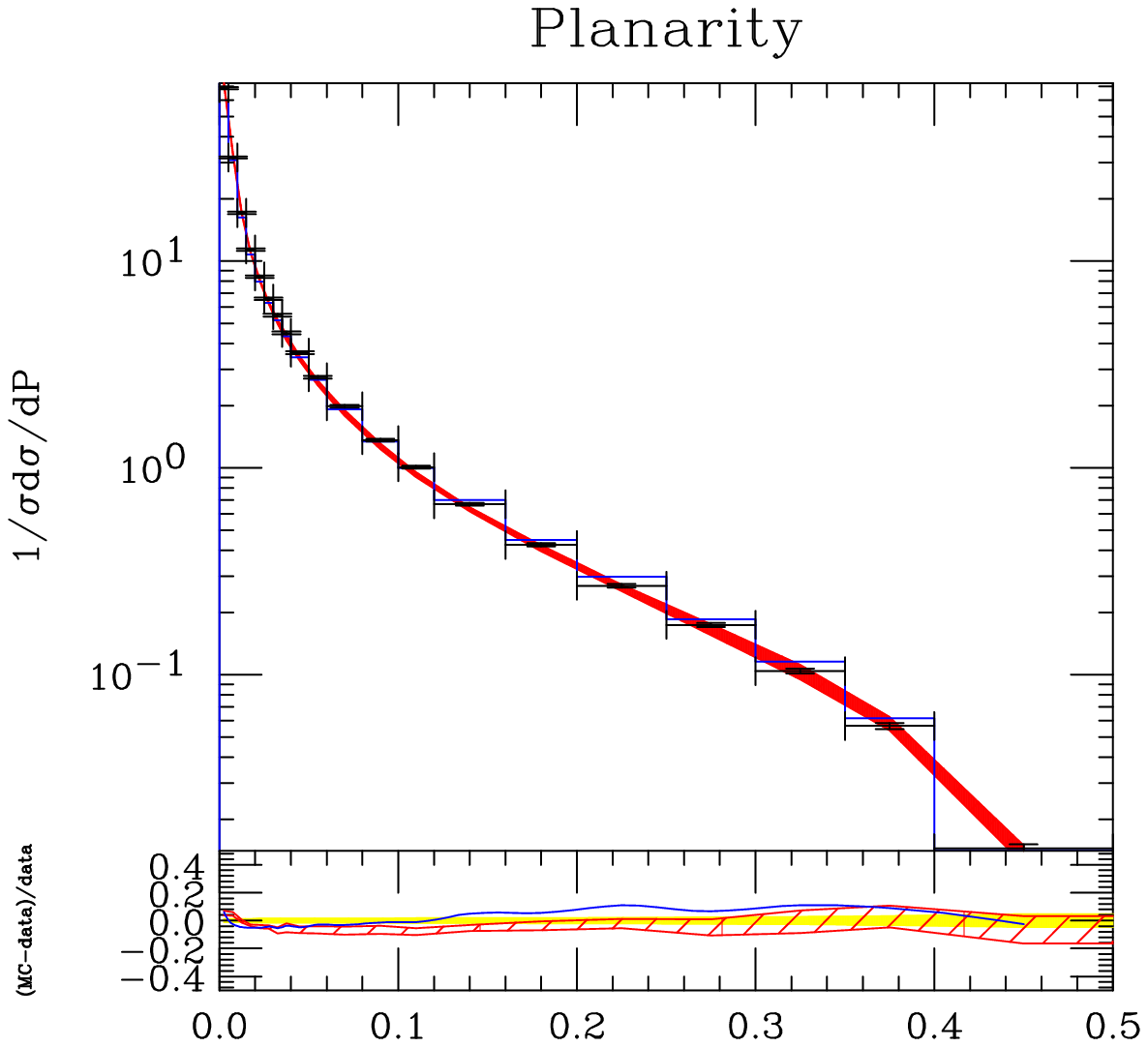} \\
\end{centering}
\caption{Distributions of the event shape variables thrust, oblateness, sphericity and planarity 
for \mbox{$e^{+}e^{-}\to\mathrm{hadrons}$} at a centre-of-mass energy of $\sqrt{s}=91.2\mathrm{\,GeV}$ in comparison 
to LEP data (black)\cite{Abreu:1996na}. 
The red band gives the variation of the distributions of the CKKW implementation with 
merging scales choices of $y_{_{MS}}=10^{-2}$ and $y_{_{MS}}=5\times10^{-3}$ 
in the Durham  and LUCLUS jet measures.  
The blue histogram gives the distributions of the default \HWPP\ parton shower with matrix element corrections.
The lower panel shows the ratio of the difference between simulation and data to the data in comparison
to the error bounds of the data (yellow region). }
\label{fig:LEPEvent}
\end{figure}

\begin{figure}
\begin{centering}
\includegraphics[width=0.4\textwidth,angle=0]{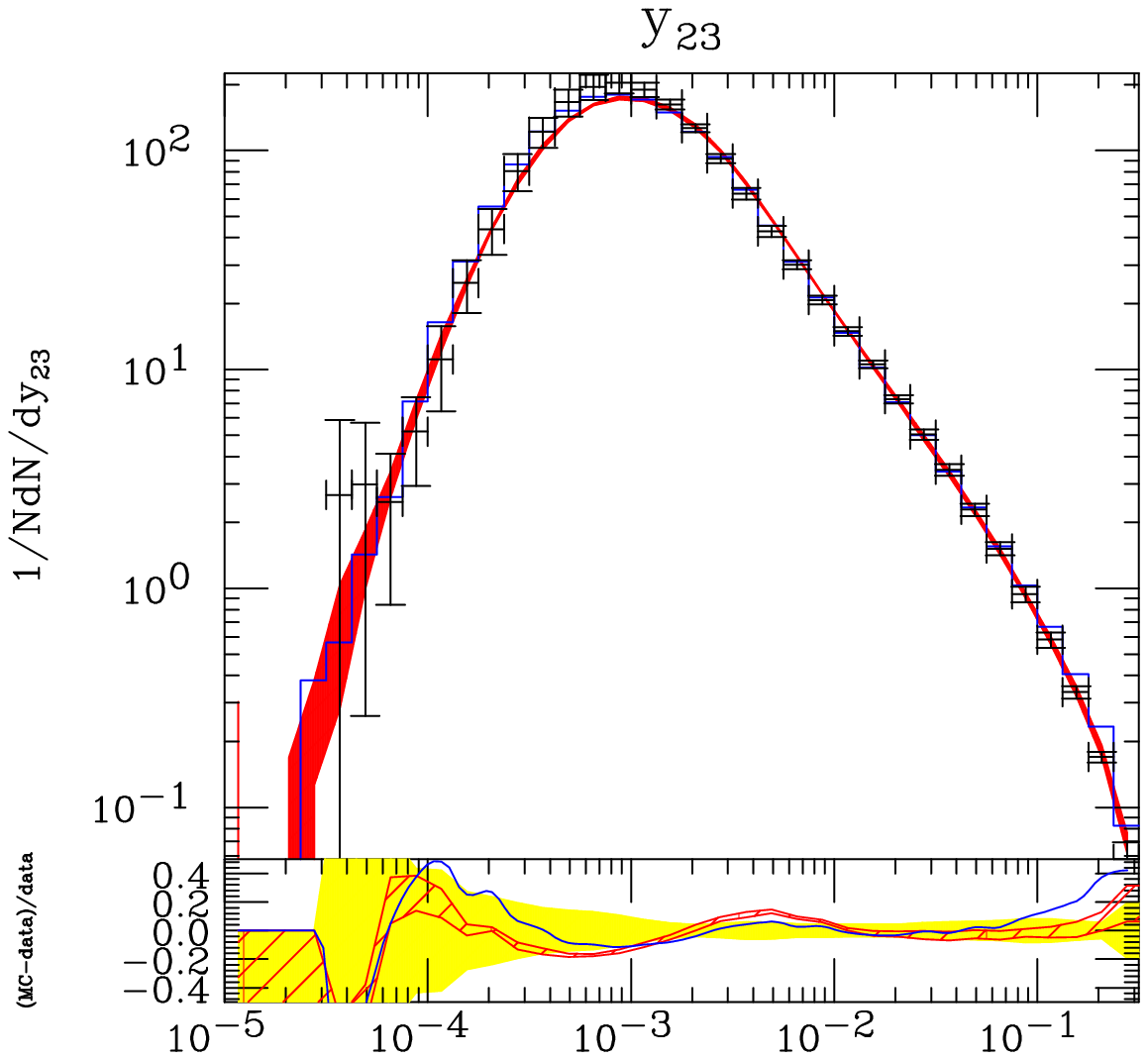}\hspace{3mm}
\includegraphics[width=0.4\textwidth,angle=0]{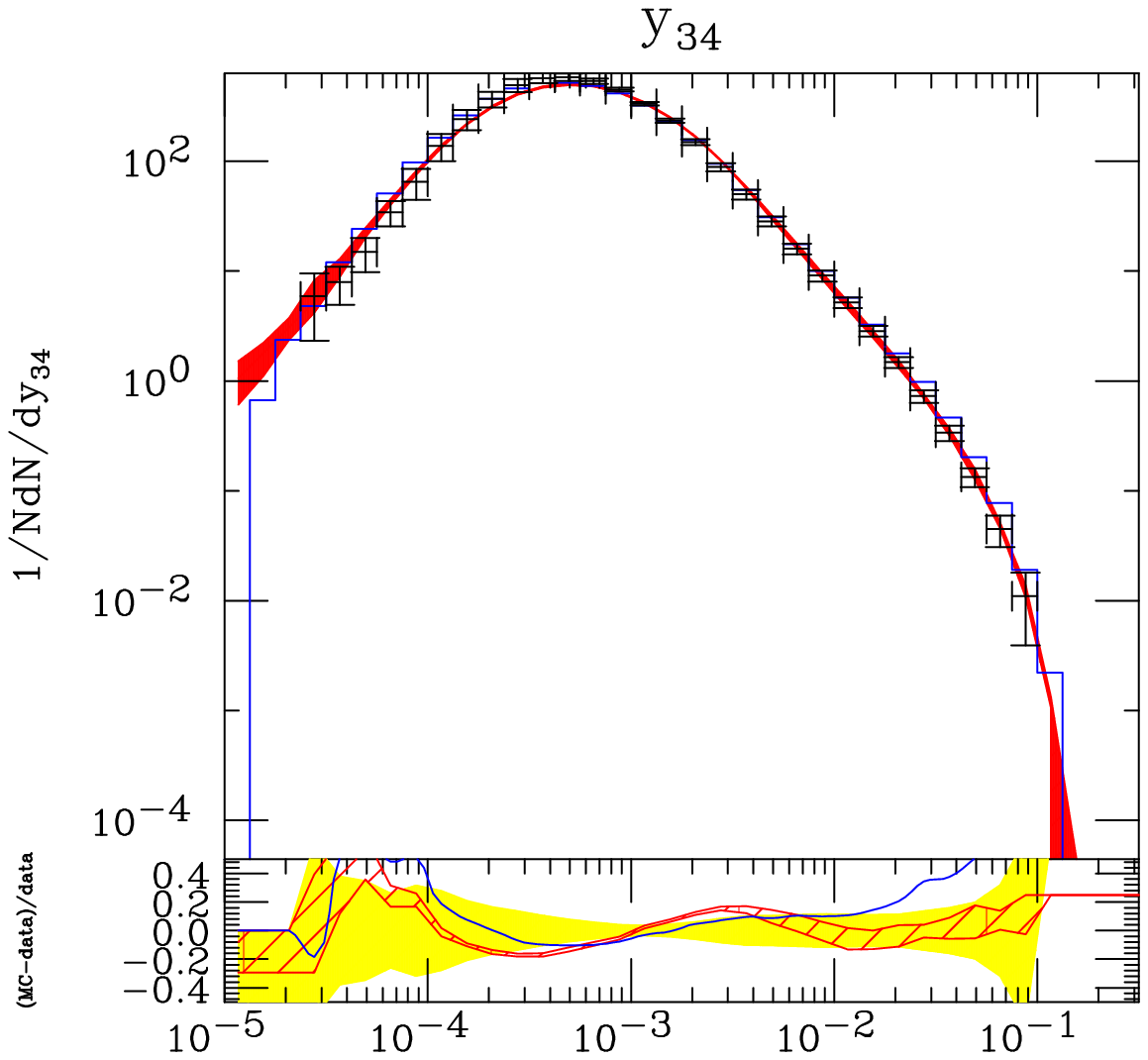} \\
 \vspace{3mm}
 \includegraphics[width=0.4\textwidth,angle=0]{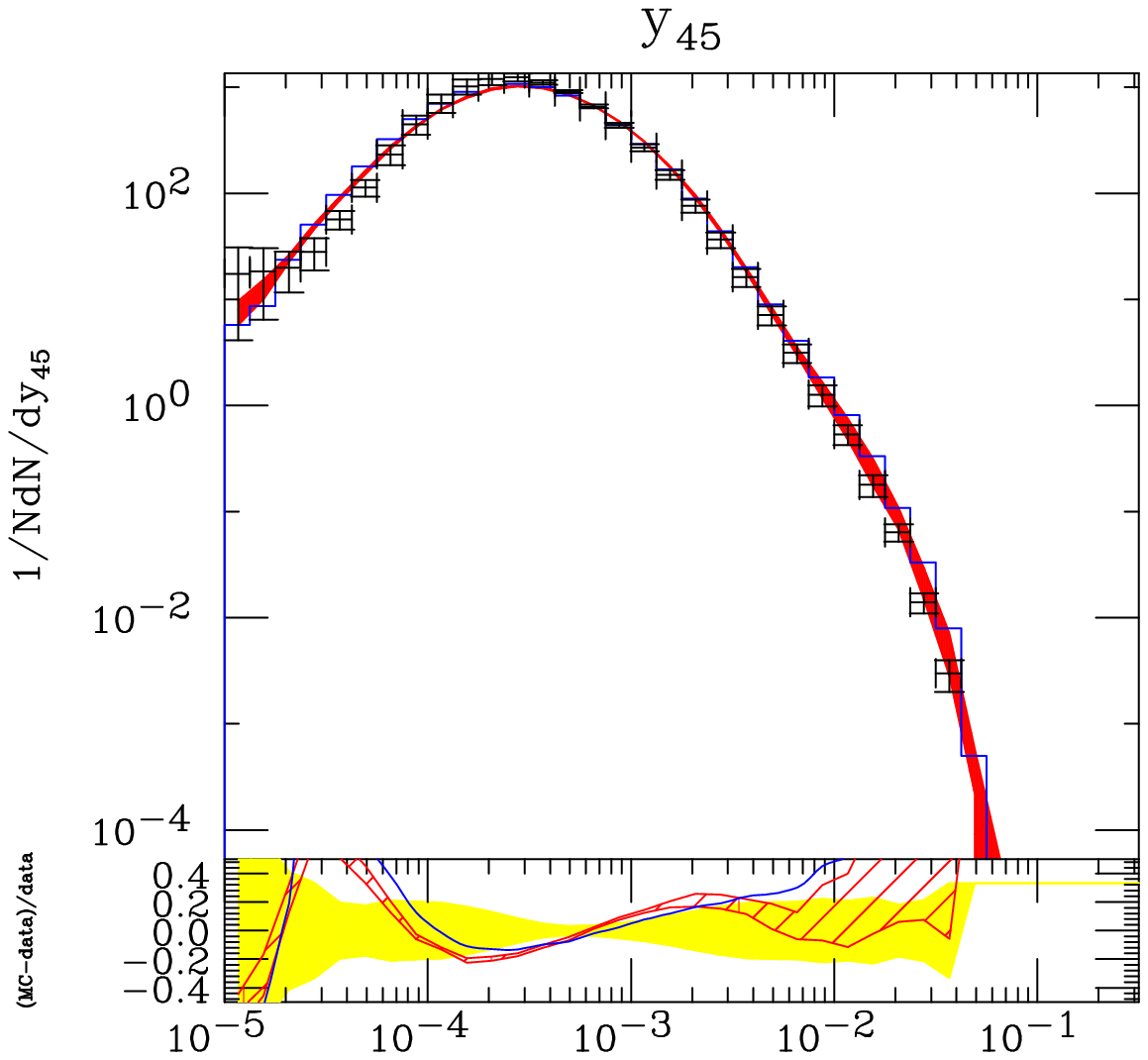} \\
\end{centering}
\caption{Distributions of the scale at which three, four and five jets are resolved in the Durham jet measure for \mbox{$e^{+}e^{-}\to\mathrm{hadrons}$} at a centre-of-mass energy of $\sqrt{s}=91.2\mathrm{\,GeV}$ in comparison to 
LEP data\protect\cite{Pfeifenschneider:1999rz}.  The colours of the lines are the same as those in Fig.~\protect\ref{fig:LEPEvent}. }
\label{fig:LEPJet}
\end{figure}

\begin{figure}
\begin{centering}
\includegraphics[width=0.4\textwidth,angle=0]{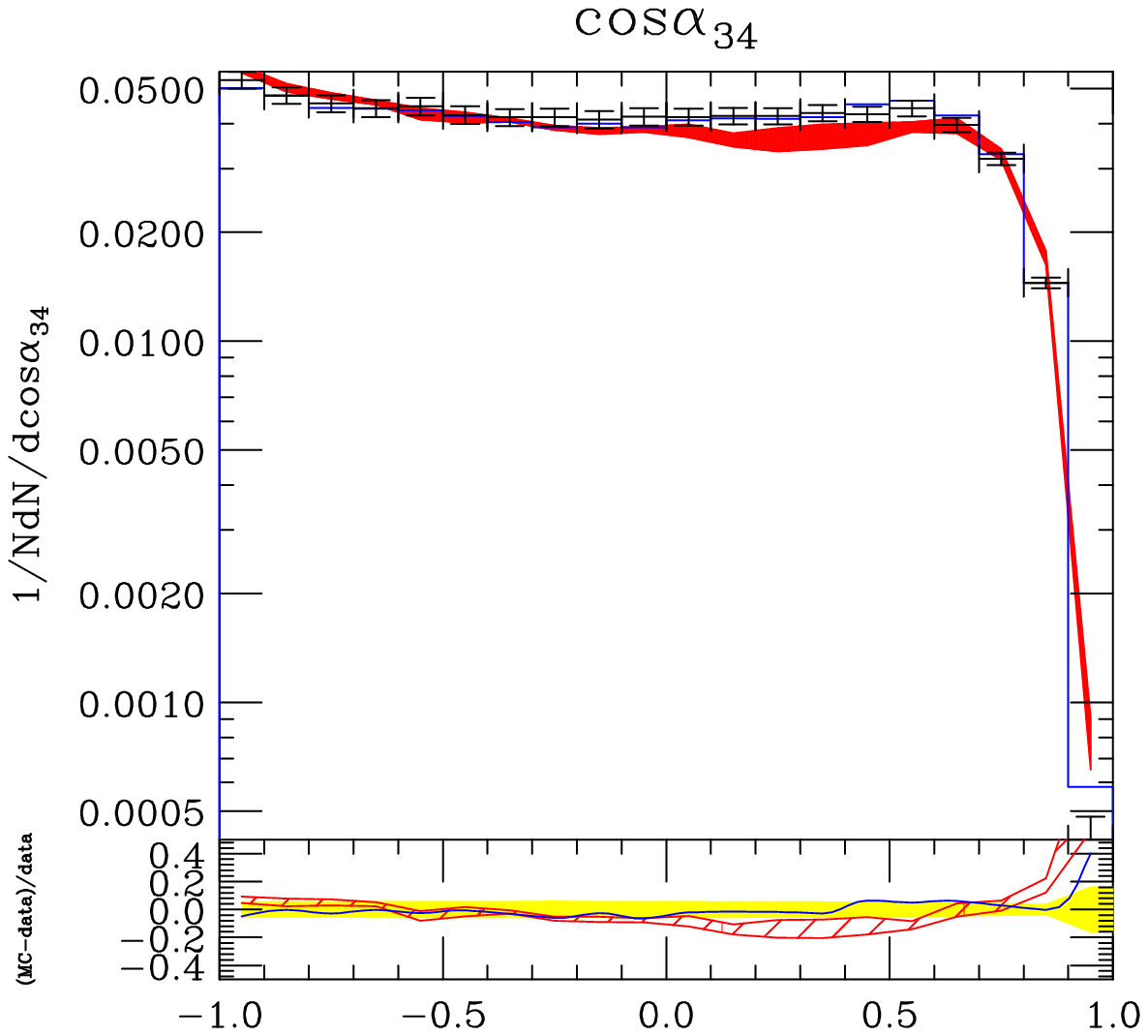}\hspace{3mm}
\includegraphics[width=0.4\textwidth,angle=0]{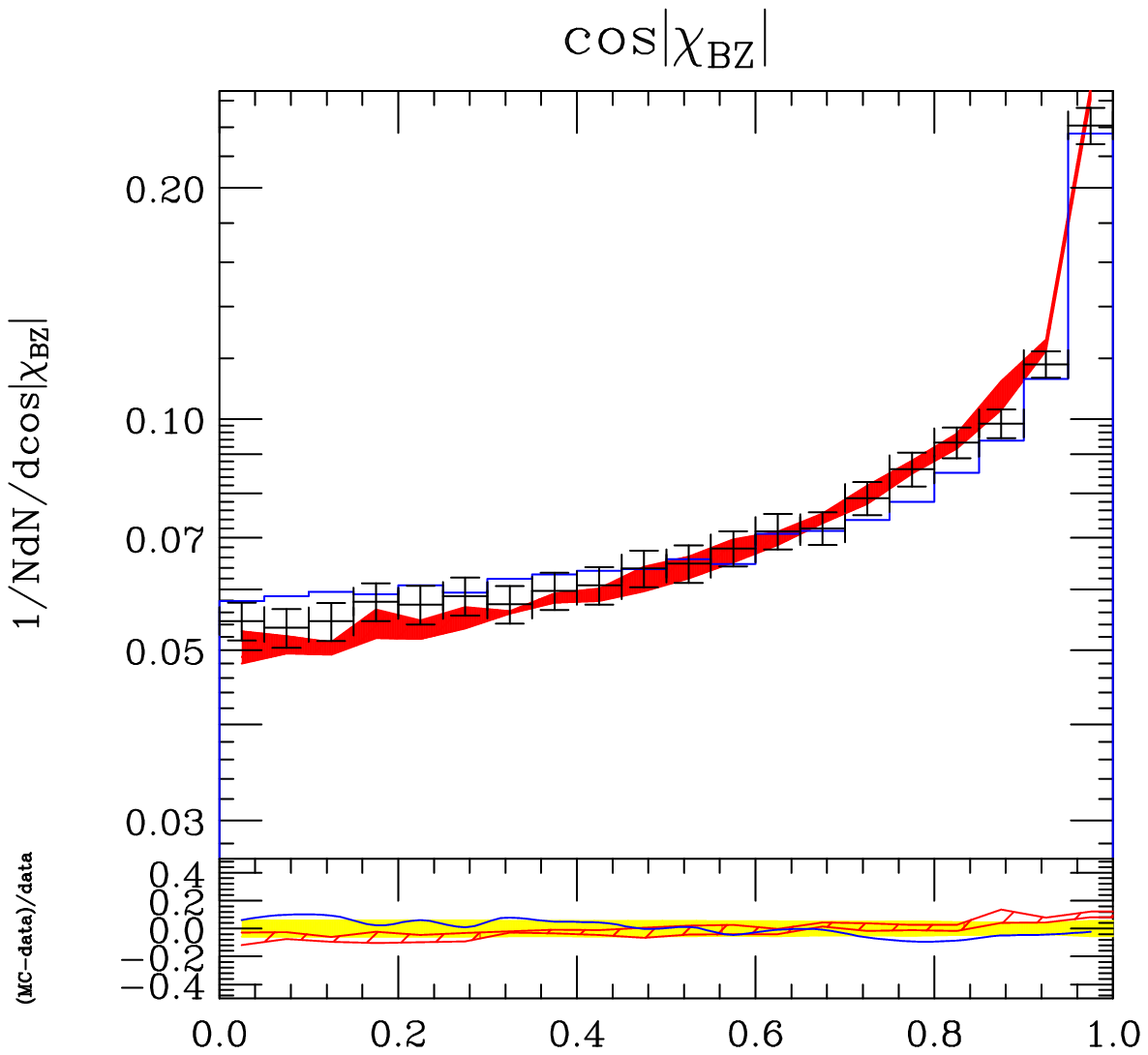} \\
 \vspace{3mm}
 \includegraphics[width=0.4\textwidth,angle=0]{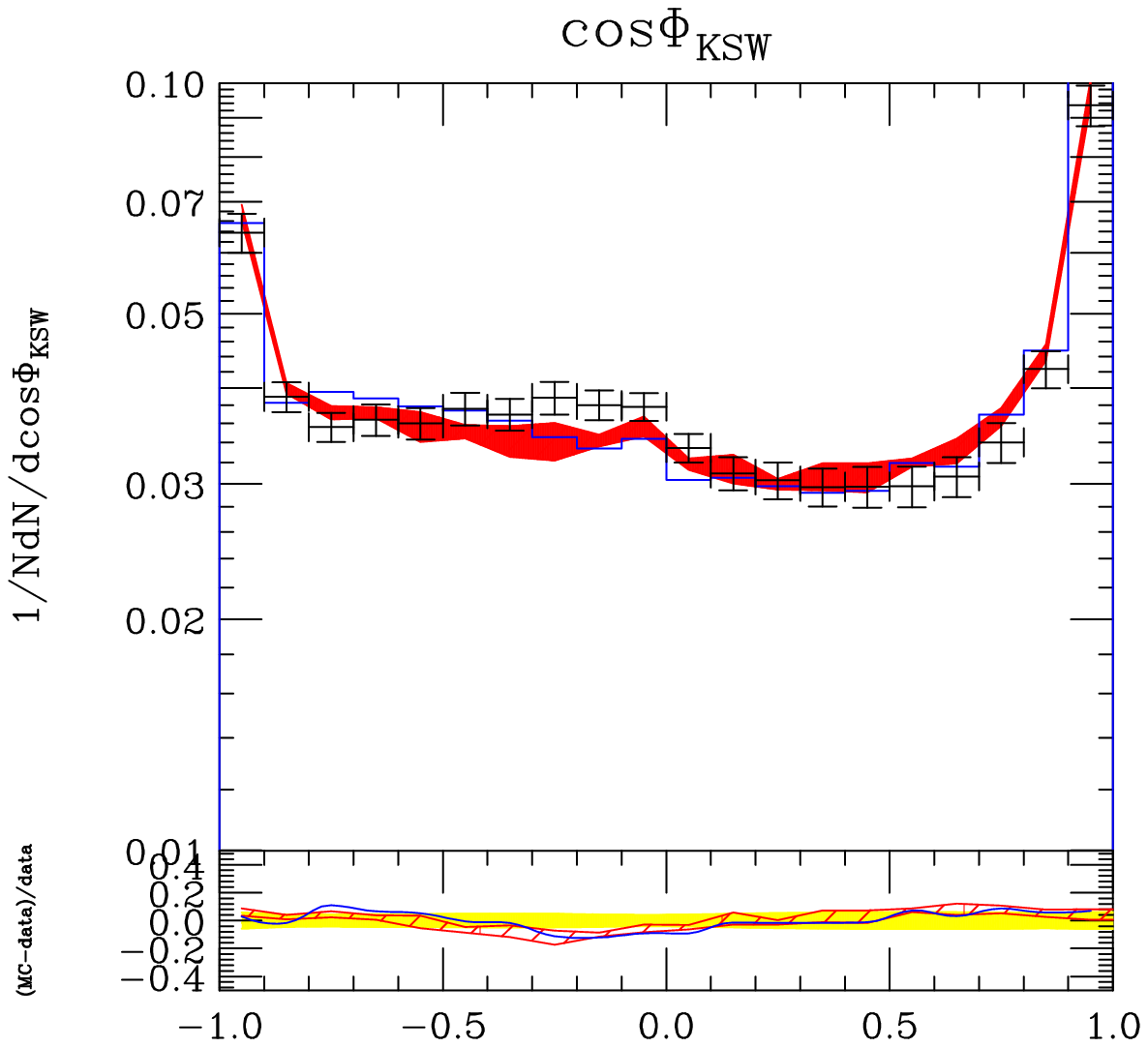}\hspace{3mm}
\includegraphics[width=0.4\textwidth,angle=0]{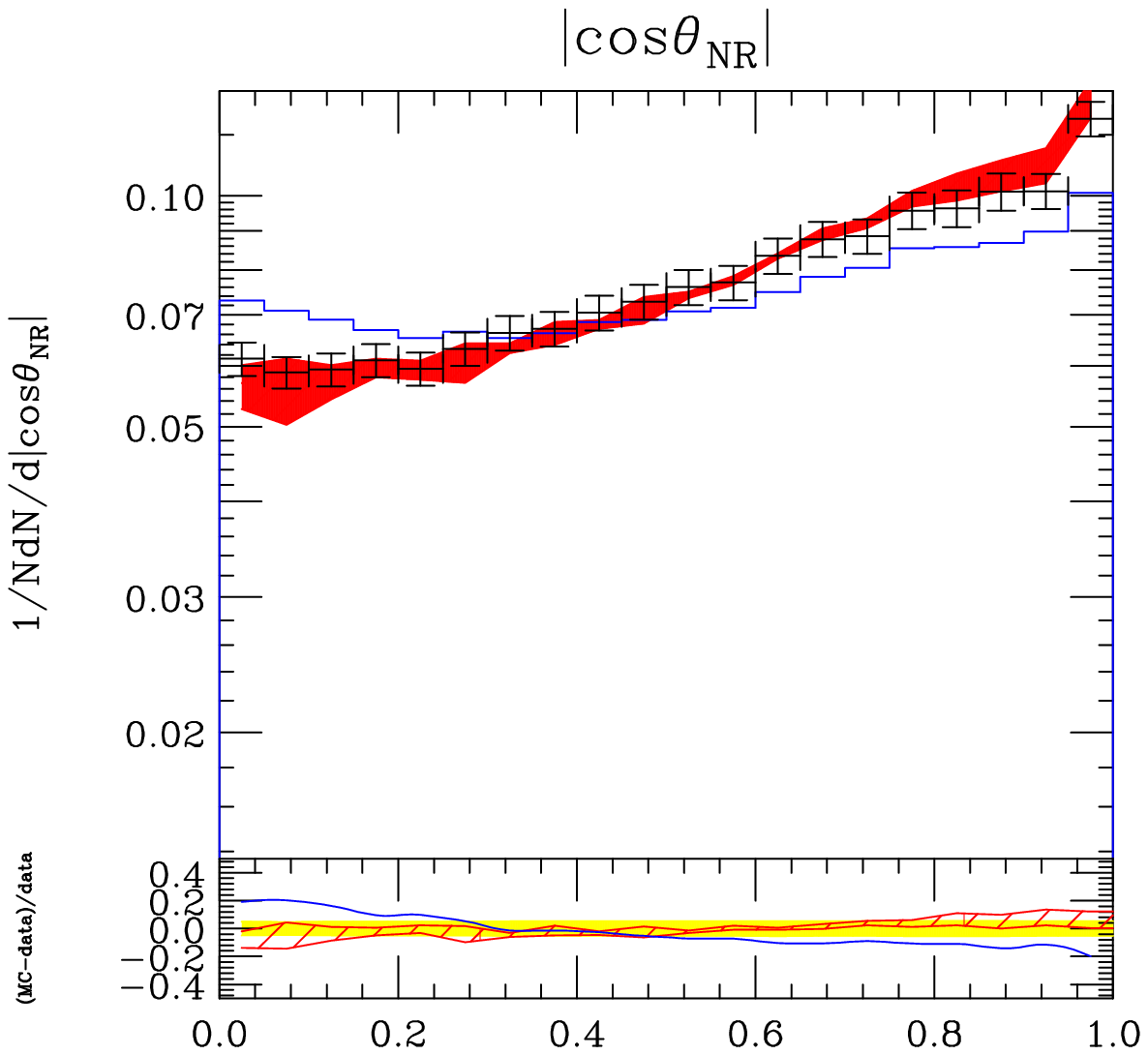} \\
\end{centering}
\caption{Distributions of four-jet angles for \mbox{$e^{+}e^{-}\to\mathrm{hadrons}$} at a centre-of-mass energy 
  of $\sqrt{s}=91.2\mathrm{\,GeV}$ in comparison to 
  LEP data\cite{Heister:2002tq}.  Figures (a)-(d) give the angle between the lowest energy jets $\alpha_{34}$, 
  the Bengtsson-Zerwas angle\protect\cite{ Bengtsson } $\chi_{BZ}$, the Korner-Sielshotlz-Willrodt\protect\cite{ Korner } $\Phi_{KSW}$ and 
  the Nachtmann-Reiter angle\protect\cite{ Nachtmann } $\theta_{NR}$.  The colours of the lines are the same as those in Fig.~\protect\ref{fig:LEPEvent}. }
\label{fig:LEPFourJet}
\end{figure}
\TABLE{
  \begin{tabular}{|l|c|c|}
    \hline
    Observable    & Hw+ME $\chi^{2}/\mathrm{d.o.f}$  & CKKW $\chi^{2}/\mathrm{d.o.f}$  \\ \hline
    Thrust        & 23.48    & 10.62 \\
    Sphericity    &  5.638   & 0.580 \\
    Oblateness    &  2.450   &  0.339 \\
    Planarity     &   1.249  &  1.211 \\        
    $y_{23}$      & 2.400     &  0.867 \\ 
    $y_{34}$      & 1.887     &  1.026 \\ 
    $y_{45}$      & 4.571     &  2.018 \\ 
    $\cos{\alpha_{34}}$ & 0.569     &  3.301 \\   
    $\cos{\chi_{BZ}}$            & 1.002    &   0.775 \\  
    $\cos{\Phi_{KSW}}$          & 1.469    &   1.337 \\    
    $\cos{\theta_{NR}}$           & 4.509    &   0.702 \\ 
    \hline
  \end{tabular}
  \caption{A comparison of the $\chi^{2}$ per degree of freedom for event shape observables in \mbox{$e^{+}e^{-}\to\mathrm{hadrons}$} with default \HWPP, with matrix element corrections, and the CKKW implementation, with merging scale set to $y_{_{MS}}=10^{-2}$ in the Durham jet measure.}
  \label{Chis}
}
\section{Conclusions}
A modified version of the CKKW algorithm has been implemented in \HWPP\ for the process 
\mbox{$e^{+}e^{-}\to\mathrm{hadrons}$}.  The modified algorithm uses truncated showers in order to 
provide smooth merging between the \HWPP\ angular-ordered parton shower and a set of 
transverse-momentum-ordered emissions defined by inverting the \HWPP\ momentum reconstruction procedure on a 
samples of parton momenta generated according to exact tree-level matrix elements.

The truncated shower was found to result in a smooth merging between parton shower and matrix element regions 
of phase space with parton level distributions appearing free of discontinuities around the 
merging scale and relatively insensitive to changes in the merging scale. 

A full tune of the \HWPP\ free parameters was performed for the CKKW implementation with a merging 
scale of $y_{_{MS}}=10^{-2}$ in the Durham jet measure.  This was found to give a good description
of LEP data, demonstrating a significant improvement over the results from the default \HWPP\
parton shower with matrix element corrections applied.

The results show a comparable level of merging scale dependence and agreement with LEP data
to that found in Ref.~\cite{Hoeche:2009rj}, in which a similar CKKW merging approach was performed
with a transverse-momentum-ordered dipole shower.

\label{conclusion}

\acknowledgments

We are grateful to all the other members of the \HWPP\ collaboration
for valuable discussions. We acknowledge the use of the UK Grid for 
Particle Physics\cite{gridPP} in tuning the \HWPP\ shower and hadronization
parameters. This work was supported
by the Science and Technology Facilities Council, formerly the Particle
Physics and Astronomy Research Council, the European Union Marie Curie
Research Training Network MCnet under contract MRTN-CT-2006-035606.
Keith\,Hamilton acknowledges support from the Belgian Interuniversity
Attraction Pole, PAI, P6/11.

\end{document}